\documentclass[preprint,showpacs,preprintnumbers,amsmath,amssymb,
tightenlines]{revtex4}
\usepackage{graphicx,color}
\usepackage{subfigure}
\def\wt{\widetilde}
\def\SphiK{S_{\phi K}}
\def\SpsiK{S_{\psi K}}
\def\Spsiphi{S_{\psi\phi}}
\def\CphiK{C_{\phi K}}

\def\bbbar{B^0$--$\bar B^0}
\def\bsbsbar{B_s^0$--$\bar B_s^0}
\def\betaphiK{\beta_{\phi K}}
\def\betapsiK{\beta_{\psi K}}
\def\BtophiKs{B\to\phi K_S}
\def\BtopsiKs{B\to\psi K_S}

\def\bbar{\bar B^0}
\def\psiK{\psi K}
\def\All{A_{\ell\ell}}
\def\Acp{A_{\rm CP}^{b\to s\gamma}}
\def\gev{\,\mbox{GeV}}
\def\gluino{\widetilde{g}}
\def\beq{\begin{equation}}
\def\eeq{\end{equation}}

\begin{document}

%\preprint{KAIST-TH 2002/24 \\ MCTP-02-58}
\preprint{\begin{tabular}{l}
\hbox to\hsize{December, 2002 \hfill KAIST-TH 2002/24}\\
\hbox to\hsize{hep-ph/0212092 \hfill MCTP-02-65}\\
\hbox to\hsize{\hfill MADPH-02-1314}\\
\end{tabular} }

\title{$B \rightarrow \phi K_S$ and supersymmetry } 



\author{G.L.~Kane}
\email[]{gkane@umich.edu}
%\thanks{}
\affiliation{Michigan Center for Theoretical Physics, University of
Michigan\\ Ann Arbor, MI 48109, USA}

\author{P.~Ko}
\email[]{pko@muon.kaist.ac.kr}
%\thanks{}
\altaffiliation{On leave of absence from 
Department of Physics, KAIST, Daejon 305-701, Korea}
\affiliation{Michigan Center for Theoretical Physics, University of
Michigan\\ Ann Arbor, MI 48109, USA}

\author{C.~Kolda}
\email[]{ckolda@nd.edu}
%\thanks{}
\affiliation{Department of Physics, University of Notre Dame \\ 
Notre Dame, IN 46556, USA}

\author{Jae-hyeon Park}
\email[]{jhpark@muon.kaist.ac.kr}
\affiliation{Department of Physics, KAIST \\ Daejon 305-701, Korea}

\author{Haibin Wang}
\email[]{haibinw@umich.edu}
%\thanks{}
\affiliation{Michigan Center for Theoretical Physics, University of
Michigan\\ Ann Arbor, MI 48109, USA}

\author{Lian-Tao Wang}
\email[]{liantaow@pheno.physics.wisc.edu}
%\thanks{}
\affiliation{Department of Physics, University of Wisconsin  \\
Madison, WI 53706, USA}


%\date{\today}

\begin{abstract}
The rare decay $\BtophiKs$ is a well-known probe of 
physics beyond the Standard Model because it arises only through
loop effects yet has the same time-dependent CP asymmetry as $\BtopsiKs$.
Motivated by recent data suggesting new physics in $\BtophiKs$,
we look to supersymmetry for possible explanations, including
contributions mediated by gluino loops and by Higgs bosons.
Chirality-preserving $LL$ and $RR$ gluino  
contributions are generically small, 
unless gluinos and squarks masses are close to  the current lower bounds.
Higgs contributions are also too small to explain a large asymmetry
if we impose the current upper limit on 
$B( B_s \rightarrow \mu^+ \mu^- )$.
On the other hand, chirality-flipping $LR$ and $RL$ gluino contributions 
can provide sizable effects and while remaining consistent
with related results in $\BtopsiKs$, $\Delta M_s$, 
$B\rightarrow X_s \gamma$ and other processes. 
We discuss how the $LR$ and $RL$ insertions can be distinguished 
using other observables, and we provide a string-based model and 
other estimates to show that the needed sizes of mass insertions are 
reasonable.
\end{abstract}

\pacs{12.60.Jv, 11.30.Er}
%\keywords{}

\maketitle

\section{Introduction}

$\BtophiKs$ is a powerful testing ground for new physics. 
Because it does not occur at tree level in the standard model (SM),
but only via loop contributions, this decay is very sensitive to
possible new physics contributions to the quark level process
$b\rightarrow s s \bar{s}$.
%an opportunity not shared by most charmless $B$ decays. 
Within the SM, it is dominated by the
QCD (and electroweak) penguin diagrams with a top quark in the loop
(see Fig.~\ref{figone}). 
Therefore the time dependent CP asymmetries are essentially due
to $\bbbar$ mixing and thus are
the same as those in $\BtopsiKs$: $\sin 2 \betaphiK
\simeq \sin 2 \betapsiK + O(\lambda^2)$~\cite{worah}. %}

Recently the  $\BtophiKs$ decay was observed by both BaBar 
and Belle at the branching ratio of $\sim (8-9) \times 10^{-6}$ (see
Table~\ref{table1}). 
This is in accord with theoretical predictions based on various factorization
approximations, albeit there are considerable uncertainties 
in theoretical predictions. Being a penguin dominated process, the 
theoretical prediction  for the branching ratio of $\BtophiKs$ 
is very sensitive to the so-called effective number of colors $N_c^{\rm eff}$, 
which has been widely used in the old factorization 
approximations~\cite{akl1}.  
Varying $N_c^{\rm eff}$ from 2 to $\infty$, the predicted 
branching ratio lies in the range $13 \times 10^{-6}$ and 
$0.4 \times 10^{-6}$~\cite{hycheng1}. If one uses the improved QCD 
factorization method developed by Beneke, Buchalla, 
Neubert and Sachrajda (BBNS)~\cite{bbns} (where $N_c^{\rm eff} = N_c = 3$), 
one obtains $B( B\to \phi K^0) \simeq 5\times 10^{-6}$, 
somewhat lower than the
current world average despite large experimental uncertainties. 
Another approach based on perturbative QCD yields the branching ratio 
of this decay at the level of $10 \times 10^{-6}$ within the SM~\cite{keum}.
On the other hand, the CP asymmetries in $\BtophiKs$ are less
model dependent, since they are ratios of branching ratios for $B^0$ and 
$\bbar$ decays.  Large theoretical uncertainties mostly cancel out 
between  the numerator and the denominator (see, {\it e.g.}, 
Ref.~\cite{akl2}). 

In general, the time dependent CP asymmetry in $B^0 ( \bbar)
\rightarrow  \phi K_S$ (or for any common CP eigenstates into which both 
$B^0$ and $\bbar$ can decay) measures two independent pieces of 
information~\cite{nir} : 
\begin{eqnarray}
{\cal A}_{\phi K} (t) & \equiv &
{{\Gamma (\bbar_{\rm phys} (t) \rightarrow \phi K_S ) - 
 \Gamma (      B^0_{\rm phys} (t) \rightarrow \phi K_S )   } \over
{\Gamma (\bbar_{\rm phys} (t) \rightarrow \phi K_S ) + 
 \Gamma (      B^0_{\rm phys} (t) \rightarrow \phi K_S )   }}
\nonumber  \\
& = & - C_{\phi K} \cos ( \Delta M t )
  + S_{\phi K} \sin ( \Delta M t ),
\end{eqnarray}
where $C_{\phi K}$ and $S_{\phi K}$ are given by
\begin{equation}
C_{\phi K}  = 
{ 1 - | \lambda_{\phi K} |^2 \over 1 + | \lambda_{\phi K} |^2 } , ~~~~~
{\rm and}~~~~~
S_{\phi K}  =  
{ 2~ {\rm Im} \lambda_{\phi K} \over 1 + | \lambda_{\phi K} |^2 } ,
\end{equation}
with 
\begin{equation}
\lambda_{\phi K} \equiv - e^{ - 2 i (\beta + \theta_d )} 
{\bar{A} ( \bbar\to\phi K_S) \over 
A (\BtophiKs) }.
\end{equation}
The angles $\beta$ and $\theta_d$ represent the SM and any
new physics contributions to the 
$\bbbar$ mixing angle; we will assume the latter to be small.
$A (\BtophiKs)$ is the amplitude for the nonleptonic $B$ 
decay of interest. If there are several independent channels relevant to the
same final states with weak phases $\varphi_m$ and strong phases $\delta_m$ 
(with $m=1,2,\ldots$ labeling different channels), the decay amplitude
$A$ for a particle is given by
\begin{equation}
A = \sum_m a_m e^{+i \varphi_m} e^{+ i \delta_m },
\end{equation}
whereas the decay amplitude $\bar A$ for the antiparticle is given by
\begin{equation}
\overline{A} = \sum_m a_m e^{-i \varphi_m} e^{+ i \delta_m }.
\end{equation}
Here $a_m >0$ is the modulus of the amplitude of the $m$-th channel.
Within the SM, the amplitude is real to a good accuracy so that
$\lambda_{\phi K} =  - e^{-2i \beta} = \lambda_{\psiK}$.  
This remains true if any new physics contributions are real relative to the
SM decay amplitude for $\BtophiKs$. Therefore one needs a 
new CP violating phase(s) $\varphi_m$ if there exists any significant 
deviation of $\SphiK$ from $\SpsiK$.  
For example, if there is a new physics contribution  to the decay with
amplitude whose modulus is $a_{\rm NP}$, weak phase is
$\varphi$ and strong phase is $\delta$, then
\begin{equation}
\lambda_{\phi K} = - e^{-2i ( \beta + \theta_d )}
{ 1 + r e^{i ( \delta - \varphi )} \over 1 + r e^{i ( \delta + \varphi )}}
\,\,\stackrel{r\gg1}{\longrightarrow}\, -e^{-2i(\beta+\theta_d+\varphi)},
\label{Lambda}
\end{equation}
where $r \equiv a_{\rm NP} / a_{\rm SM}$.  
In general, the strong phase $\delta$ will affect both $\SphiK$ and 
$\CphiK$. However if the new physics contributions dominate over the
SM ones, the strong phase effectively drops out, an approach that is
commonly (and mistakenly) used in the literature even when $r$ is not large.
We will comment further on the nature and source of the strong phases
in Section~\ref{gluinosec}.

\begin{table}[t]%[H] add [H] placement to break table across pages
\caption{ CP averaged branching ratios, 
$\SphiK$ and $\CphiK$ in 
$\BtophiKs$ from BaBar~\cite{babar,lp03} and Belle~\cite{belle,lp03}
and the SM predictions in the BBNS approach.}
\begin{ruledtabular}
\begin{tabular}{ccccc}
Observable & BaBar & Belle & Average & SM prediction 
\\
$Br$ (in $10^{-6}$) & $8.1\,{}^{+3.1}_{-2.5}\pm 0.8$ & 
$ 8.7\,{}^{+3.8}_{-3.0} \pm 1.5 $ & $8.4\,{}^{+2.5}_{-2.1}$ & 
$\simeq 5$ (see text)
\\
$S_{\phi K_S}$ & $0.45\pm 0.43\pm 0.07 $& 
$-0.96\pm 0.50\,{}^{+0.09}_{-0.11}$ & $-0.13\pm 0.33$ & $0.736\pm 0.049$
\\
$C_{\phi K_S}$ & $-0.80 \pm 0.38 \pm 0.12$ & 
$0.56 \pm 0.41 \pm 0.16$ & $-0.19 \pm 0.30$ & $-0.008$ 
\end{tabular}
\end{ruledtabular}
\label{table1}
\end{table}

As described before, $\SphiK \simeq \sin 2 \betapsiK = 
\SpsiK$ within the SM: 
\begin{equation}
\SphiK = \SpsiK = 0.736\pm 0.049, \quad\mbox{and}\quad
\CphiK = -0.008.
\end{equation}
That is, the time dependent CP asymmetry in $\BtophiKs$  
should be 
essentially the same as that in $\BtopsiKs$. 
However the BaBar and the Belle collaborations both report a deviation 
from the SM prediction for $\SphiK$.
As summarized in Table~\ref{table1}, the Belle value for $\SphiK$ is
$3.3\sigma$ away from the SM prediction,
while the Babar value is within $1\sigma$ of predictions. (Previously,
both collaborations had reported values inconsistent with the SM
prediction by about $2.7\sigma$. The average of the two values remains
$2.7\sigma$ away from the SM.) 
This result, which may be an indication of new physics contributions to 
$\BtophiKs$, has generated a wave of activity, much of it on SUSY
contributions to
$\BtophiKs$~\cite{causse,others,lucas,ourprl,murayama,masiero02,
kagan02,cheng,more}.
The direct CP asymmetry in $\BtophiKs$ is also reported by 
both collaborations (see Table~\ref{table1}). 
However, their two values have large errors.
Measurements of both the direct and indirect CP violation are in
disagreement between the two experiments, 
so no firm conclusions can be drawn at present.

However in this paper, 
we wish to entertain the possibility that there is indeed a 
deviation of $\sin 2\betaphiK$ from  $\sin 2 \betapsiK$,  
and that it originates from supersymmetry (SUSY) effects. 
More specifically, we consider the $\BtophiKs$ decay 
within several classes of general SUSY models with $R$-parity conservation.  
We will study two interesting classes of modifications to  
$b\rightarrow s s \bar{s}$ within SUSY models: 
\begin{itemize}
\item Gluino-mediated $b\rightarrow s q \bar{q}$  with
$q = u,d,s,c,b$: Such operators are 
induced by flavor mixings in the down-squark sector.
We consider all possible combinations of $LL$, $LR$, $RL$ and $RR$
mixings in the $\wt{b}$--$\wt{s}$ squark sector.
Such contributions do not distinguish among 
flavors of light quarks due to the flavor independence of SUSY QCD.  
Therefore other decays such as $B\rightarrow K \pi, 
\eta^{(}{}'{}^{)} K$ could be affected as well.
\item Higgs-mediated $b\rightarrow s s \bar{s}$ in the large $\tan\beta $ 
limit ($\propto \tan^3 \beta$ at the amplitude level): This mechanism is 
important only for $b\rightarrow s s \bar{s}$, and not for $s c \bar{c}$, 
$s u \bar{u}$ or $s d \bar{d}$ transitions, which makes it an attractive  
possibility. It would affect $\phi K$ and $\eta^{(}{}'{}^{)} K$ 
modes but not $K\pi$ modes. 
\end{itemize} 
In this article, we carefully analyze the effects of these two mechanisms
on $\BtophiKs$ and related observables.
More specifically, we consider 
\begin{itemize}
\item $\BtophiKs$ : its branching ratio and the time dependent
CP asymmetries, $\SphiK$ and $\CphiK$;
\item Correlations with $B\rightarrow X_s \gamma$, both its branching
fraction and its direct CP asymmetry;
\item Correlation of the Higgs-mediated $b\rightarrow s s \bar{s}$ 
transition with $B_s \rightarrow \mu^+ \mu^-$ which has been, and 
is being, searched for at the Tevatron; 
\item Correlations with $\bsbsbar$ mixing coming from SUSY
contributions to $\Delta M_{B_s}$, and with
the dilepton charge asymmetries $\All$ 
and the time-dependent CP asymmetry in 
$B_s \rightarrow \psi \phi$ (which is proportional to the phase of the
mixing).
\end{itemize}
The new phase in the $b\rightarrow s s \bar{s}$ will affect other 
CP-violating observables in calculable ways, and our explanations can be
tested by measuring other quantities as we will discuss.

Before proceeding, we comment on the following concern: why should there
be a large deviation in $\BtophiKs$ but not in related decay
modes such as 
$B \rightarrow \eta' K_S, \pi K,  K^+ K^- K_S$?
Unlike $\BtophiKs$, these decays have SM contributions 
at tree level %and one loop levels, 
while the SUSY contributions are loop-suppressed.
Therefore it is reasonable that only $\BtophiKs$, which is 
already one loop suppressed in the SM, is modified by a significant amount. 
Because considerable hadronic physics is involved, we will not make
any more precise
statement here, but we do not think any contradiction of our 
analysis of $\phi K_S$ is implied by the data. 
The unexpectedly large branching 
ratio for $\eta' K_S$  is also consistent with this view.

We also note that for $B \rightarrow D^{*+} D^{-*}$ there also is a 
$2.7 \sigma$ deviation between the SM prediction and the data~\cite{ddbar}. 
This decay is dominated by the tree level $b\rightarrow d c \bar{c}$ 
transition, but its amplitude is suppressed by a factor of 
$\lambda = \sin \theta_c \approx 0.22$ relative to that of 
$B\rightarrow J/\psi K$.
Therefore new physics contibutions at one loop level might have a chance to
compete with the SM contribution to $B_d \rightarrow D^{*+} D^{-*}$.
Possible SUSY contributions involve $b\rightarrow d q \bar{q}$ 
(with $q=u,d,s,c$) so that the relevant mass insertion parameter is 
$( \delta_{AB}^d )_{13}$. 
This parameter is independent of $( \delta_{AB}^d )_{23}$ which affects
$B\rightarrow \phi K$.  One could perform a similar study as presented below
for the $B_d \rightarrow D^{*+} D^{-*}$, but we do not pursue that here.

In the following, we will deduce that $LL$ (and $RR$) insertions in
gluino penguins generally provide contributions 
too small to cause an observable deviation between $\SphiK$ and 
$\SpsiK$,  unless gluino and squark masses are close to
the current lower bounds.
We will also find that Higgs-mediated $b\to ss\bar s$ is not
sufficient to explain the data once we impose the existing CDF limit
on $B_{d,s}\to\mu\mu$.
However, we find that the down-sector $LR$ and $RL$ insertions 
in gluino penguins can in fact  
explain a sizable deviation in a way consistent with all other data.

\section{$\Delta B = 1$ effective Hamiltonian}

All approaches to exclusive $B \to M_1 M_2$ decays (where
$M_1$ and $M_2$ are light mesons) use factorization methods at
various levels of sophistication.
One starts from the $\Delta B = 1$
effective Hamiltonian at the renormalization scale $\mu \sim m_b$, 
which can be obtained from the underlying ultraviolet physics by
integrating out heavy particles, with the effects of hard gluon 
taken into
account by the renormalization-group-improved perturbation theory
(RG-improved PQCD). 

The effective Hamiltonian for $B\rightarrow \phi K$ in the SM  
can be written as~\cite{bbns}
\begin{equation}
H_{\rm eff} = \frac{G_F}{\sqrt{2}}\sum_{p=u,c} 
\lambda_p^{(s)} \left[ C_1 O_1^p + C_2 O_2^p + \sum_{i=3}^{10} 
C_i ( \mu ) O_i ( \mu ) + C_{7 \gamma} O_{7\gamma} + C_{8g} O_{8 g} \right]
+ {\rm H.c.},
\end{equation}
where $\lambda_p = V_{ps}^* V_{pb}$ with $p = u,c$ are CKM
factors, and $\lambda_u + \lambda_c + \lambda_t = 0$ due to the unitarity 
of the CKM matrix. 
The operators $O_i$ are all those relevant to $\Delta S = 1$ hadronic
decays of a $b$ quark:
\begin{equation}
\begin{array}{lcl}
O_1^p=(\bar pb)_{V-A}(\bar sp)_{V-A} & &
O_2^p=(\bar p_\alpha b_\beta)_{V-A} (\bar s_\beta p_\alpha)_{V-A} \\
O_3=(\bar s b)_{V-A} \sum_q (\bar qq)_{V-A} &&
O_4=(\bar s_\alpha b_\beta)_{V-A} \sum_q (\bar q_\beta q_\alpha)_{V-A} \\
O_5=(\bar sb)_{V-A}\sum_q(\bar qq)_{V+A} &&
O_6=(\bar s_\alpha b_\beta)_{V-A}\sum_q(\bar q_\beta q_\alpha)_{V+A} \\
O_7=(\bar sb)_{V-A}\sum_q\frac32 e_q(\bar qq)_{V+A} &&
O_8=(\bar s_\alpha b_\beta)_{V-A}
        \sum_q\frac32 e_q(\bar q_\beta q_\alpha)_{V+A} \\
O_9=(\bar sb)_{V-A}\sum_q\frac32 e_q(\bar qq)_{V-A} &&
O_{10}=(\bar s_\alpha b_\beta)_{V-A}
        \sum_q\frac32 e_q(\bar q_\beta q_\alpha)_{V-A} \\
O_{7\gamma} = -\frac{e}{8\pi^2}\, m_b\bar s\,\sigma_{\mu\nu}(1+\gamma_5)
F^{\mu\nu} b & \phantom{2+2=4} &
O_{8g} = -\frac{g_s}{8\pi^2}\, m_b\bar s\,\sigma_{\mu\nu}(1+\gamma_5)
G^{\mu\nu} b 
\end{array}
\end{equation}
where $\alpha,\beta$ are color indices.
The operators $O_{1,2}$ are charged current operators relevant at
next-to-leading order; $O_{3,4,5,6}$ are generated by gluonic penguins
at leading order; $O_{7,8,9,10}$ are electroweak penguin operators, also
generated at leading order; $O_{7\gamma,8g}$ are the magnetic and
chromomagnetic transition operators, also generated at leading order.
The Wilson coefficients $C_i$ contain all the 
relevant information regarding the short-distance physics (and possible 
new physics effects, if any). For simplicity, we will ignore the electroweak 
penguin operators $O_{7\ldots10}$ whose contributions to $\BtophiKs$ are 
roughly $\sim 10 \%$ within the  SM. 
The expressions for $C_{i}$'s within the SM 
can be found in Ref.~\cite{buraslec}.

The most difficult task is evaluating the matrix element of the
above effective Hamiltonian between the initial $| B \rangle$
state and the final $\langle M_1 M_2 |$ state. 
In this work, we adopt the BBNS approach~\cite{bbns} to estimate 
the hadronic amplitude for $\BtophiKs$. 
It is inevitable that our results will depend on the factorization
scheme chosen. The direct CP asymmetry $\CphiK$ is particularly 
dependent
on the method we use for evaluating the hadronic matrix element, 
more so than $\SphiK$. 
For example, $\CphiK=0$ for naive factorization 
without one loop corrections to the matrix elements of four-quark operators.
Including those corrections, or going to the BBNS approach, can lead to very
large asymmetries. On the other hand, $\SphiK$ can be large and
negative in either scheme. We will discuss some of these unavoidable
scheme dependencies again at the end of Section~\ref{uncertainties}.

\section{Gluino-mediated FCNCs}
\label{gluinosec}

\begin{figure}
\includegraphics[width=16cm]{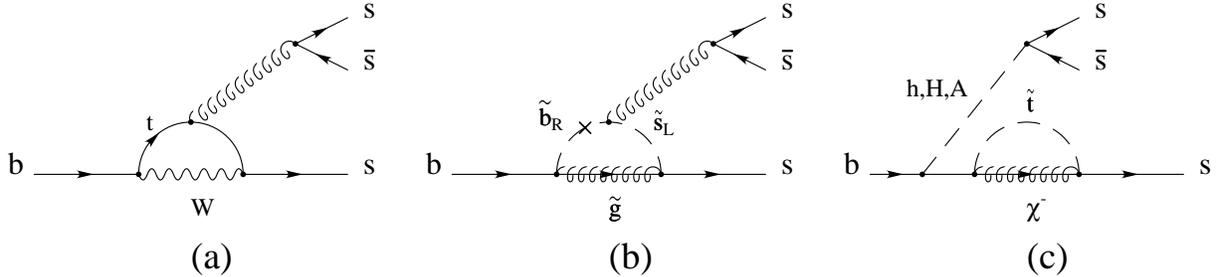}
\caption{
To illustrate different contributions, we show some Feynman diagrams 
relevant to $b\rightarrow s s \bar{s}$: (a) the SM,
(b) the gluino mediated and (c) $\tan\beta-$enhanced Higgs mediated 
contributions.
}
\label{figone}
\end{figure}

In the MSSM, supersymmetric versions of the SM contributions to
$\BtophiKs$ exist. For example, the $W$-$t$ loop of
Fig.~\ref{figone}(a) is accompanied a new $\wt{W}$-$\wt{t}$ loop. The
flavor-changing in these diagrams is intrinsically tied to the usual,
CKM-induced flavor changing of the SM. If that were the only new
source of flavor physics, we would say the model is minimally flavor
violating (see Ref.~\cite{mfv} for a consistent definition of minimal
flavor violation in two-Higgs doublet models such as the MSSM). 
However such a model will not generate large corrections to
$\SphiK$ or $\CphiK$ and so cannot describe the experimental data. (We
will return to a more-minimal flavor violation when we discuss the
Higgs-mediated contributions in Section~\ref{higgssec}.)

But the general MSSM is not minimally flavor violating.
For a generic MSSM a new source of flavor violation is introduced by 
the squark mass matrices, which usually cannot be diagonalized in the
same basis as the quark mass matrices.
This means gluinos (and other gauginos) 
will have flavor-changing couplings to quarks and squarks, which
implies FCNCs which are mediated by gluinos and thus have strong 
interaction strengths.
In order to analyse the phenomenology of these couplings,
it is helpful to rotate the effects
so that they occur in squark propagators rather than in couplings, and to 
parametrize them in terms of dimensionless parameters.
We work in the usual mass insertion approximation (MIA)~\cite{mia}, 
where the flavor mixing $j \rightarrow i$ in the down type squarks 
associated with $\tilde{q}_B$ and $\tilde{q}_A$ are parametrized by 
$( \delta_{AB}^d )_{ij}$. More explicitly, 
\[ 
( \delta_{LL}^d )_{ij} = \left( V_L^{d \dagger} M_{Q}^2 V_L^d \right)_{ij}  
/ \tilde{m}^2, ~~~
 ( \delta_{RR}^d )_{ij} = \left( V_R^{d \dagger} M_{D}^2 V_R^d \right)_{ij}
/ \tilde{m}^2, ~~~
( \delta_{LR}^d )_{ij} = \left( V_L^{d T} M_{LR}^2 V_R^d \right)_{ij} 
/ \tilde{m}^2, ~~~
\] 
in the super CKM basis where the quark mass matrices are diagonalized by
$V_L^d$ and $V_R^d$, and the squark mass matrices are rotated in the same
way. Here $ M_{Q}^2$, $M_{D}^2$ and $M_{LR}^2$ are squark mass matrices, 
and $\tilde{m}$ is the average squark mass.
Then, the gluino box/penguin diagrams generate the QCD penguin operators,
$O_{3\ldots6}$ and 
$\widetilde{O}_{3,..,6} \equiv O_{3\ldots6} (L \leftrightarrow R)$. 
Since the decay $B_d \rightarrow \phi K_S$ is dominated by the SM QCD penguin 
operators (an example is shown in Fig.~\ref{figone}(a)), 
the gluino-mediated QCD 
penguins such as that in Fig.~\ref{figone}(b) may be significant 
if the gluinos and squarks are relatively light. 
Similar studies were carried out by Lunghi and Wyler~\cite{lunghi} and 
Ciuchini and Silvestrini~\cite{lucas} (and in the context of a SUSY GUT 
by Moroi~\cite{moroi} and Causse~\cite{causse}). 
Our approach extends these existing papers in the following respects: 
\begin{itemize}
\item We use the recent BBNS approach to evaluate the 
hadronic matrix element 
for $\BtophiKs$ ({\it i.e.}, $a_{\phi K}^{\rm eff}$ in Eq.~\ref{master}), 
which is important for calculating direct CP violation $\CphiK$, 
unlike other papers (except for Ref.~\cite{lucas}).
\item We consider the $LR$, $RL$, $LL$ and $RR$ insertions, whereas Ciuchini
{\it et al.} consider only the $LL$ insertion, and Moroi and Causse consider 
mainly the $RR$ insertion. We will find that the $LR$, $RL$ contributions 
can dominate, while the $LL$, $RR$ insertions are too small to affect 
$B\rightarrow \phi K$ significantly unless the gluino 
and squark masses are close to the current lower bounds.
\item We also study the Higgs-mediated contribution 
(Fig.~\ref{figone}(c)), the
amplitude of which can be enhanced by $\tan^3 \beta$ for large $\tan\beta$.
Naively this contribution would be a natural flavor-violating candidate to
provide a non-SM enhancement of the $s \bar{s}$ final state. 
\end{itemize}

Note that the gluino mediated QCD penguin operators are not enhanced 
in the large $\tan\beta$ region, unlike the Higgs-mediated contributions.
But the gluino-mediated diagrams contribute to $b\rightarrow s q\bar{q}$ 
for all $q = u,d,s,c,b$, independent of flavor.
Therefore one should check that these new sources of flavor changing
do not contribute 
too much to other charmless $B$ decays such as $B\rightarrow \pi\pi, \pi K$, 
etc. In this paper we will not deal with this problem directly. But we
note that there
are known mechanisms by which this can be achieved. As an example,
consider a scenario in which 
both $LL$ and $RR$ operators contribute to $\BtophiKs$ with nearly
equal amplitudes.
Because vector mesons ($\phi, \rho, \ldots$) and pseudoscalar mesons 
($\pi, K,\eta^{(')}$) have opposite parity, the gluino loop effects appear
as $( \delta_{LL}^d )_{23} + ( \delta_{RR}^d )_{23}$ for the $VP$ modes and 
$( \delta_{LL}^d )_{23} - ( \delta_{RR}^d )_{23}$  for the $PP$ modes, 
respectively. So one can suppress the gluino contributions to the $PP$ 
modes by simply assuming $( \delta_{LL}^d )_{23} = ( \delta_{RR}^d )_{23}$,
even if $LL$ and $RR$ contributions are sizable~\cite{kagan02}.  

There are additional issues that arise in comparing $\BtophiKs$ to
$B\to \pi\pi$ and others. For example,
there are multiple diagrams contributing to processes such as
$B\to\pi\pi$, unlike the case of $\BtophiKs$. This makes the inclusion
of SUSY effects in these processes either irrelevant or highly
uncertain. We will take an indirect approach to this problem
by requiring that the total branching ratio of
$\BtophiKs$ is consistent with observation after including SUSY
effects. By demanding that the SUSY contributions to $\BtophiKs$ are
of the same order or less than the SM contributions, we can
safely assume that the SUSY contributions to related processes are
also close to experiment. Given the uncertainties inherent in
the BBNS or any other factorization scheme we feel that this is an
appropriate approach. However it also means that the size of the
strong phase will play an important role in our calculations, another
reason for using the best technology available today, namely BBNS
factorization. 

The relevant Wilson coefficients due to the gluino box/penguin loop diagrams 
involving the $LL$ and $LR$ insertions are  
given (at the scale $\mu \sim m_W$) by~\cite{ko01} 
\begin{eqnarray}
   C_3^{\rm SUSY} & = & 
   -\frac{\alpha_s^2}{ 2 \sqrt{2} G_F \wt{m}^2 \lambda_{t} }
   \left(-\frac{1}{9} B_1(x)-\frac{5}{9} B_2(x)-\frac{1}{18} P_1(x)
   -\frac{1}{2} P_2(x) \right) 
   \left(\delta^d_{LL}\right)_{23} \nonumber \\
   C_4^{\rm SUSY} & = & 
   -\frac{\alpha_s^2}{  2 \sqrt{2} G_F \wt{m}^2 \lambda_{t}}
   \left(-\frac{7}{3} B_1(x)+\frac{1}{3} B_2(x)+\frac{1}{6} P_1(x)
   +\frac{3}{2} P_2(x) \right) 
   \left(\delta^d_{LL}\right)_{23} \nonumber \\
   C_5^{\rm SUSY} & = & 
   -\frac{\alpha_s^2}{ 2 \sqrt{2} G_F \wt{m}^2 \lambda_{t} }
   \left(\frac{10}{9} B_1(x)+\frac{1}{18} B_2(x)-\frac{1}{18} P_1(x)
   -\frac{1}{2} P_2(x) \right) 
   \left(\delta^d_{LL}\right)_{23} \nonumber \\
   C_6^{\rm SUSY} & = & 
   -\frac{\alpha_s^2}{ 2 \sqrt{2} G_F \wt{m}^2 \lambda_{t} }
   \left(-\frac{2}{3} B_1(x)+\frac{7}{6} B_2(x)+\frac{1}{6} P_1(x)
   +\frac{3}{2} P_2(x) \right) \left(\delta^d_{LL}\right)_{23} 
    \nonumber \\
   C_{7 \gamma}^{\rm SUSY} & = &  
   {8 \pi \alpha_s \over 9 \sqrt{2} G_F \tilde{m}^2 \lambda_t }
   \left[ ( \delta_{LL}^d )_{23}  M_4 (x) 
    - ( \delta_{LR}^d )_{23} \left( {m_{\tilde{g}} \over m_b} \right) 
   4 B_1 (x) \right],
    \nonumber  \\
   C_{8 g}^{\rm SUSY} & = & - {2 \pi \alpha_s \over 
   \sqrt{2} G_F \tilde{m}^2 \lambda_t }
   \left[ ( \delta_{LL}^d )_{23} \left( {3\over 2} M_3 (x) - 
   {1\over 6} M_4 (x) \right) \right.
   \nonumber  \\
   & & \left.
   + ( \delta_{LR}^d )_{23} \left( {m_{\tilde{g}} \over m_b} \right) 
   ~{1\over 6}~\left( 4 B_1 (x) - 9 x^{-1} B_2 (x) \right) \right], 
\end{eqnarray}
where $x \equiv ( m_{\tilde{g}} / \tilde{m} )^2$, and the loop functions
can be found in Ref.~\cite{ko01}.  
In the presence of the $RR$ and $RL$ insertions, 
we have additional operators $\widetilde{O}_{i=3\ldots6,7\gamma,8g}$
that are obtained by $L\leftrightarrow R$ in the operators in the SM. The 
associated Wilson coefficients $\widetilde{C}_{i=3\ldots6,7\gamma,8g}$
are given by the similar expressions as above
with the replacement $L\leftrightarrow R$. The remaining coefficients are
either dominated by their SM contributions ($C_{1,2}$) or are
electroweak penguins ($C_{7\ldots10}$) and therefore small.

A remark is in order concerning the above expressions. 
These Wilson coefficients differ  from 
the 1996 results in Gabbiani {\it et al.}~\cite{masiero96}
in {\it (i)}\/ the signs of the $(\delta_{LR}^d )_{23}$ contributions in 
$C_{7\gamma}$ and $C_{8g}$, and {\it (ii)}\/ 
the loop function $9 x B_2 (x)$ in 
$C_{8g}$ should be corrected to $9 x^{-1} B_2 (x)$ as given 
above~\footnote{
There are typos in the overall signs of $C_{7\gamma}$ and 
$C_{8g}$ in Eqs.~(44) and (46) in Ref.~\protect\cite{ko01}.    
We also verified this by explicit calculations.}.
These differences are not important for numerical results, however. 

The decay rate for $\BtophiKs$ is given by
\begin{equation}
\Gamma (\BtophiKs) = {G_F^2 f_\phi^2 m_B^3 \over 32 \pi} 
~( F_1^{B\rightarrow K} )^2 
\left| a_{\phi K}^{\rm eff} \right|^2 ~\lambda^{3/2} 
( 1, m_\phi^2 / m_B^2, m_K^2 / m_B^2 ) 
\label{master}
\end{equation}
where 
\begin{equation}
a_{\phi K}^{\rm eff} = 
\sum_{p=u,c} \lambda_p \left[ ( a_3 + a_4^p + a_5 ) 
- {1\over 2}~( a_7 + a_9 + a_{10}^p ) \right]  
\end{equation}
and $\lambda$ is the magnitude of the $\phi$ 3-momentum relative to
$m_B$.
In the numerical analysis, we use $f_{\phi} = 237$ MeV and 
$ F_1^{B\rightarrow K}  = 0.38$.  
The $a_i$'s are given in terms of the Wilson coefficients $C_i$ times
various nonperturbative hadronic parameters:
\begin{eqnarray*}
a_3 &=& C_3 + \frac{C_4}{N_c} + \frac{C_4}{N_c} 
\frac{C_F \alpha_s}{4\pi N_c}
\left(N_c V_\phi + 4\pi^2 H_{\phi K}\right) \\
a_4^p &=& C_4 + \frac{C_3}{N_c} + \frac{C_3}{N_c} 
\frac{C_F \alpha_s}{4\pi N_c}
\left(N_c V_{\phi} + 4\pi^2 H_{\phi K}\right) +\frac{C_F\alpha_s}{4\pi}
\frac{P^p_{\phi,2}}{N_c} \\
a_5 &=& C_5 + \frac{C_6}{N_c} - \frac{C_6}{N_c} 
\frac{C_F \alpha_s}{4\pi N_c}
\left(N_c V'_\phi + 4\pi^2 H'_{\phi K}\right) %\\
\end{eqnarray*}
where
$C_F=(N_c^2-1)/2N_c$ and $N_c=3$. The various $V^({}'{}^)$, 
$H^({}'{}^)$ and $P$ terms
represent the nonperturbative hadronic parameters at the heart of the
BBNS calculation and are described in Ref.~\cite{bbns}.

Besides generating the amplitudes necessary for calculating physical
rates, the BBNS approach also provides us with the appropriate strong
phase. We will digress for a moment to talk about the importance of
this piece in the calculation. 

\begin{figure}
\includegraphics[width=8cm]{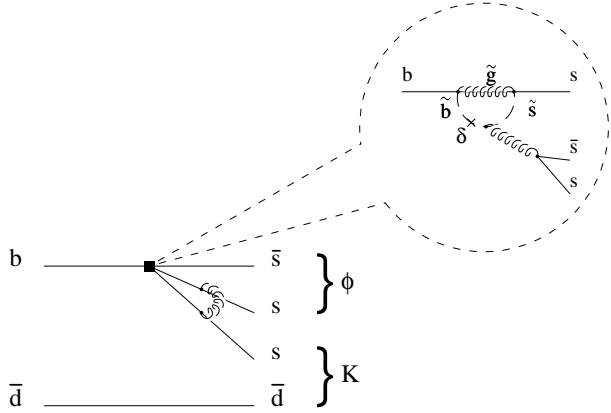}
\caption{
A typical final state interaction diagram which generates a {\it strong}\/
phase in $\BtophiKs$. At long distances, the {\it weak}\/ phase is contained
in the 4-fermion vertex shown as a small square. At short distances,
that vertex resolves to SUSY diagrams like those studied in this paper.
}
\label{strong}
\end{figure}

In order to have nonzero CP asymmetries, we need at least two independent
amplitudes with different weak and strong phases. In the SUSY models 
we are considering,
the weak phases reside in the complex mass insertion parameters, 
the $\delta$'s, 
and appear in the Wilson coefficients $C_{3,...,6}$ and $C_{8g}$: see  
Fig.~\ref{figone}(a)--(c). 
These weak phases are odd under a CP transformation. 
On the other hand, the CP-even 
strong phases arise from scatterings among the final
state particles. At the parton level, they are generated by gluon 
exchange between various quarks participating in the $B$ decay processes. 
In the BBNS approach
there are four classes of diagrams that can generate strong phases: 
vertex corrections, penguins, hard scattering with spectators and 
annihilation diagrams. We show a vertex correction diagram in
Fig.~\ref{strong} for 
illustration, referring the reader to the BBNS papers~\cite{bbns} for 
the details. The strong phases are then encoded in the 
$V^{(}{}'{}^{)}$, $H^({}'{}^)$ and $P$ terms that appear in the
definitions of the $a_i$ parameters above. 

Before the BBNS approach, the strong phase was generated by the so-called 
Bander-Soni-Silverman (BSS) mechanism, in which the one loop diagrams 
involving the 4-quark operators develop a strong phase consistent with
unitarity. However, there is an ambiguity in this approach regarding the
momentum flow into the quark loop, and one usually made the {\it ad
hoc}\/ assumption that the average
$q^2$ through the virtual gluon is $\langle q^2 \rangle \simeq m_b^2 / 4$.
This ambiguity disappears in the BBNS 
approach, where one has a definite relation between the loop momenta of the 
quark and antiquark and the meson lightcone wavefunction. In particular,
the contributions of the $b\rightarrow s g$ operators $O_{8g}$ and 
$\wt{O}_{8g}$ are obtained in an unambiguous way at least in the leading 
order.

\section{$B\to X_s\gamma$}

It has long been claimed that the most 
stringent bounds on the  $( \delta_{AB}^d )_{23}$ insertion comes from 
$b\to s\gamma$ (or $B\rightarrow X_s \gamma$). 
Because the measured value is close to the 
SM prediction, there is little room for new physics lurking into 
$b\rightarrow s$ transition. This statement is
particularly true for $LR$ mixing because the SUSY contributions
interfere with the SM ones at the amplitude level (in $C_{7\gamma}$
and $C_{8g}$) and further because the SUSY contributions pick up 
an $m_{\widetilde g}/m_b$ enhancement relative to the SM.
The $LL$ insertion also contributes to $C_{7\gamma,8g}$ but lacks the
gluino mass enhancement. Thus $LL$ insertions can be very large while
remaining consistent with $b\to s\gamma$, while $LR$ insertions are
constrained to be small, of ${\cal O}(10^{-2})$.

The contribution of the $RL$ and $RR$ insertions to $C_{7\gamma,8g}$ are
suppressed by $m_s/m_b$ where $m_s$ is the strange quark mass and so
we ignore them. However
they make unsuppressed contributions to $\widetilde C_{7\gamma,8g}$,
but because these operators do not appear in the SM, their contributions
cannot interfere with the SM, nor can they generate CP violation. Thus
the constraint from $B\to X_s\gamma$ on $RL$ and $RR$ operators
will always be somewhat weaker than that on $LR$ and $LL$ operators
respectively.  

We will impose a rather generous bound for $B\rightarrow X_s \gamma$,
$$
2.0 \times 10^{-4} < B ( B\rightarrow X_s \gamma ) < 4.5 \times 10^{-4},
$$
in order to take into account the theoretical uncertainties in the NLO
calculation of the SUSY contributions.
(We have also checked that our results for $\SphiK$ do not
change very much if we narrow the allowed window for the 
$B\rightarrow X_s \gamma$ branching ratio. It will be relatively
straightforward to see the effect of narrowing the window in the
figures we present in Section~\ref{sec:numbers}.) 
The CP-averaged branching ratio for $B\rightarrow X_s \gamma$ in the leading
log approximation is given by
\begin{equation}
{B ( B \rightarrow X_s \gamma ) \over B( B \rightarrow X_c e \nu ) } 
= \left| {V_{ts}^* V_{tb} \over V_{cb}} \right|^2 ~{6 \alpha \over \pi f(z) }~
\left[ | C_{7 \gamma} ( m_b ) |^2 + 
| \widetilde{C}_{7 \gamma} ( m_b ) |^2  \right]. 
\end{equation} 
where $f(z) = 1 - 8 z + 8 z^3 - z^4 - 12 z^2 \log z$ is the phase space factor
for the $b\rightarrow c$ semileptonic decays and $\alpha^{-1} = 137.036$.
Neglecting the RG running between the heavy SUSY particles and the top quark 
mass scales, we get the following relations :
\begin{eqnarray}
C_{7 \gamma} ( m_b ) & \simeq & - 0.31 + 0.67 ~C_{7 \gamma}^{\rm SUSY} 
( m_W ) + 0.09 ~C_{8 g}^{\rm SUSY} ( m_W ), 
\nonumber   \\
C_{8 g} ( m_b ) & \simeq & - 0.15 + 0.70~ C_{8 g}^{\rm SUSY} ( m_W ). 
\end{eqnarray}

In most of our analysis of $B\to X_s\gamma$ we will assume that the only
contributions are those of the SM plus the gluino-mediated SUSY
loops. It is well known that the charged Higgs and chargino sectors in
the MSSM may contribute to $B\to X_s\gamma$ with strengths equal to or
greater than the SM piece. We will ignore this possibility except in
Section~\ref{RLdom}, where we will consider a particularly interesting
limit in which all but the gluino loops cancel out among themselves.

A new CP-violating phase in $( \delta^d_{AB} )_{23}$ will also 
generate CP violation  in $B\rightarrow X_s \gamma$. 
In order to have a nonvanishing direct CP asymmetry, one needs at least 
two independent amplitudes with different strong (CP-even) and weak (CP-odd) 
phases.  In $B\rightarrow X_s \gamma$, strong phases are provided by
quark and gluon loop diagrams, whereas weak phases are provided 
by the CKM angles and $(\delta_{AB}^d )_{23}$. 
For conventional models with the same operator basis as in the SM, 
the resulting direct CP asymmetry in 
$B\rightarrow X_s \gamma$ can be written as~\cite{ali,kn}
\begin{equation}
\Acp({\rm in} ~\% ) = 
{1\over |C_{7\gamma}|^2}~\left[ 1.23 ~{\rm Im} 
\left( C_2 C_{7\gamma}^* \right) 
- 9.52 ~{\rm Im} \left( C_{8g} C_{7\gamma}^* \right) 
+ 0.10 ~{\rm Im} \left(C_2 C_{8g}^* \right) \right].
\label{eq:acp}
\end{equation}
Since SUSY contributions to $C_2$ are negligible, we use 
$C_2 ( m_b )\simeq C_2^{\rm SM} ( m_b ) \simeq 1.11$. 
In the presence of operators $\widetilde{O}_i$ with opposite chirality,
Eq.~(\ref{eq:acp}) is modified according to Ref.~\cite{kn}.
Within the SM, the predicted CP asymmetry is less than $\sim 0.5\%$;
any larger asymmetry would be a clear indication of new physics. 

What we currently know about $\Acp$ comes from CLEO~\cite{cleo}:
\begin{equation}
\Acp = (-7.9 \pm 10.8 \pm 2.2)
(  1.0 \pm 0.030) \%.
\label{Acp}
\end{equation}
This is clearly not a very strong constraint at present, but should
become significantly more important in the near future.

Despite the importance of $B\rightarrow X_s \gamma$ for constraining
$( \delta_{AB}^d )_{23}$, we will  find in the following sections that
the $B\rightarrow \phi K$ branching ratio itself provides a 
constraint on the $LR$ and $RL$ insertions which is every bit 
as strong as $B\rightarrow X_s \gamma$.

\section{$\bsbsbar$ mixing}

Gluino-mediated box diagrams will also affect 
$\bsbsbar$ mixing, which  is phenomenologically very important.
The most general effective Hamiltonian for $\bsbsbar$ mixing
($\Delta B = 2$) and its SUSY contributions have been calculated in 
Refs.~\cite{masiero2002,kkp} and we do not repeat that discussion here.
Following the explicit formulation of 
Ref.~\cite{kkp}, we calculate $\Delta M_s$ and the phase of the $\bsbsbar$ 
mixing, $\beta_s$.  For the $B_{d,s}$ meson decay constants, 
we have assumed $f_{B_d} = 200 \pm 30$ MeV, and 
\[
{f_{B_s}^2 B_{B_s} \over f_{B_d}^2 B_{B_d}} = 1.16 \pm 0.05.
\]
Currently, only the experimental 
lower limit $\Delta M_{B_s} > 14.9~{\rm ps}^{-1}$ 
is known~\cite{bs-exp}. 
Also we assume that the SUSY effects on $\bbbar$ mixing 
are reasonably small~\cite{kkp} ({\it i.e.}, $\theta_d$ in
Eq.~\ref{Lambda} is nearly zero).  
According to the analysis by Ko {\it et al.}~\cite{kkp}, 
the CKM angle $\gamma$ can still be in the range between 
$-60^{\circ}$ and $60^{\circ}$ for the $LL$ insertion
$( \delta_{LL}^d )_{13} \neq 0$ 
without any conflict with the measured 
$\Delta M$ and $\sin 2 \beta_{\psi K_S}$. Then 
these effects will only appear in
$\lambda_{\phi K}$ as $2 (\beta + \theta_d)$, which is fixed to be  
$\sin 2 \beta_{\psi K} = 0.734 \pm 0.054 $ by data.
Furthermore, the SM contributions  
to $\BtophiKs$ or $\bsbsbar$ mixing are  
independent of the CKM angle $\gamma$, so that we %can safely neglect 
don't have to worry about 
the SUSY contribution to $\bbbar$ mixing.

In our model, the CP violating phase in the mass insertion parameters  
$( \delta_{AB}^d )_{23}$  contributes not only to $\SphiK$, as 
discussed before, but also to the imaginary part of the 
$\bsbsbar$ mixing, $M_{12} ( B_s)$, which is real 
within the SM to a very good approximation.  Thus there should be 
some correlation between $\SphiK$ and 
$2 \beta_s \equiv - \mbox{Arg}\, [M_{12} ( B_s )]$.
The observable $\Spsiphi=\sin 2\beta_s$ plays the same role in
$B_s\to\psi\phi$ as $\SphiK$ does in $\BtophiKs$, and thus we may
expect large deviations in the time-dependent decays of $B_s$ into 
$\psi\phi$. The SM predicts $\beta_s$ vanishes (to experimental
accuracy), so any measurement of a non-zero $\beta_s$ would be a sign
of new physics.

The new phase in $\bsbsbar$ mixing will also appear in the 
dilepton charge asymmetry of $B$ decays 
that can be measured at hadron colliders 
(see, {\it e.g.}, \cite{randall}):
\begin{equation}
\All \equiv \frac{N(\ell^+\ell^+)-N(\ell^-\ell^-)}
{N(\ell^+\ell^+)+N(\ell^-\ell^-)}=
\frac{N(BB) - N(\bar{B}\bar{B})}{N(BB) + N(\bar{B}\bar{B})}
\simeq {\rm Im} (\Gamma_{12} / M_{12} ).
\end{equation}
In the SM, the phases of $M_{12}$ and $\Gamma_{12}$ are almost the same, 
and $\All^{\rm SM} \simeq 10^{-4}$ ($10^{-3}$ for $B_d$).
But in the presence of SUSY, 
a much larger dilepton charge asymmetry may be possible.
Though SUSY is not expected to provide a large correction to
$\Gamma_{12}$ (which is dominated by large SM contributions), it can
strongly affect $M_{12}$. In such a case,
the dilepton charge asymmetry could be approximated as
\begin{equation}
\All \simeq {\rm Im} \left( { \Gamma_{12}^{\rm SM} \over
M_{12}^{\rm SM} + M_{12}^{\rm SUSY} } \right).
\end{equation}
The possible range of $\All$ in a classes of SUSY models were 
studied in Refs.~\cite{randall,kkp}. We will follow the formulation of
Ref.~\cite{kkp} in our analysis. We will demand that all solutions
obey the lower bound $\Delta M_s>14.9\,\mbox{ps}^{-1}$, and for those
that do, present their prediction for $\Delta M_s$.

\section{Numerical Analyses of Gluino Mediation}
\label{sec:numbers}

Now we come to the heart of our analysis. We would like to consider
whether gluino-induced flavor changing can explain the $\BtophiKs$
data while remaining consistent with all other known constraints. And
if so, we wish to identify other observables that can be used to check
our interpretation of the data. We will consider each possible mass
mixing $(\delta^d_{AB})_{23}$ for $AB=LL,LR,RL,RR$ one at a time. The
technique assumes that only one insertion dominates the new
physics. However, where relevant, we will explain how a second, subdominant
insertion could change some of the correlating observables.

In all of the studies that follow, we will scan over the modulus and
phase of the flavor-changing mass insertions 
($|\mbox{Re,Im}\,(\delta^d_{AB})_{23}|\leq1$), 
calculating all observables for a
fixed gluino and (universal) squark mass of $400\gev$. We will then
vary the ratio $x=m^2_{\widetilde g}/\widetilde m^2$ away from unity
(keeping $m_{\widetilde g}$ fixed) and finally, at the end of the
section, discuss the effects of varying the gluino mass. We will keep
only those points which are consistent with the branching ratio for
$B\to X_s\gamma$, and which have $\Delta
M_s>14.9\,\mbox{ps}^{-1}$. Both of these constraints were discussed in
previous sections. We will also require that the branching ratio for
$\BtophiKs$ be less than $1.6\times 10^{-5}$, as calculated in the
BBNS approach. This value is twice the observed branching ratio, but
we feel that the uncertainties inherent in the BBNS factorization
scheme may be large enough to merit this window.

\subsection{The $LL$ and $RR$ insertions} 

We will begin by considering the 
the $LL$ and $RR$ insertions, though assuming that one or the other
appears dominantly. Since the physics effects of these two operators
are almost identical, we will couch our discussion in the language of
the $LL$ insertion and specifically point out those few places 
where the $RR$ analysis diverges.

\begin{figure}
\subfigure[Allowed region for the $LL$ insertion]
{\includegraphics[width=5.5cm]{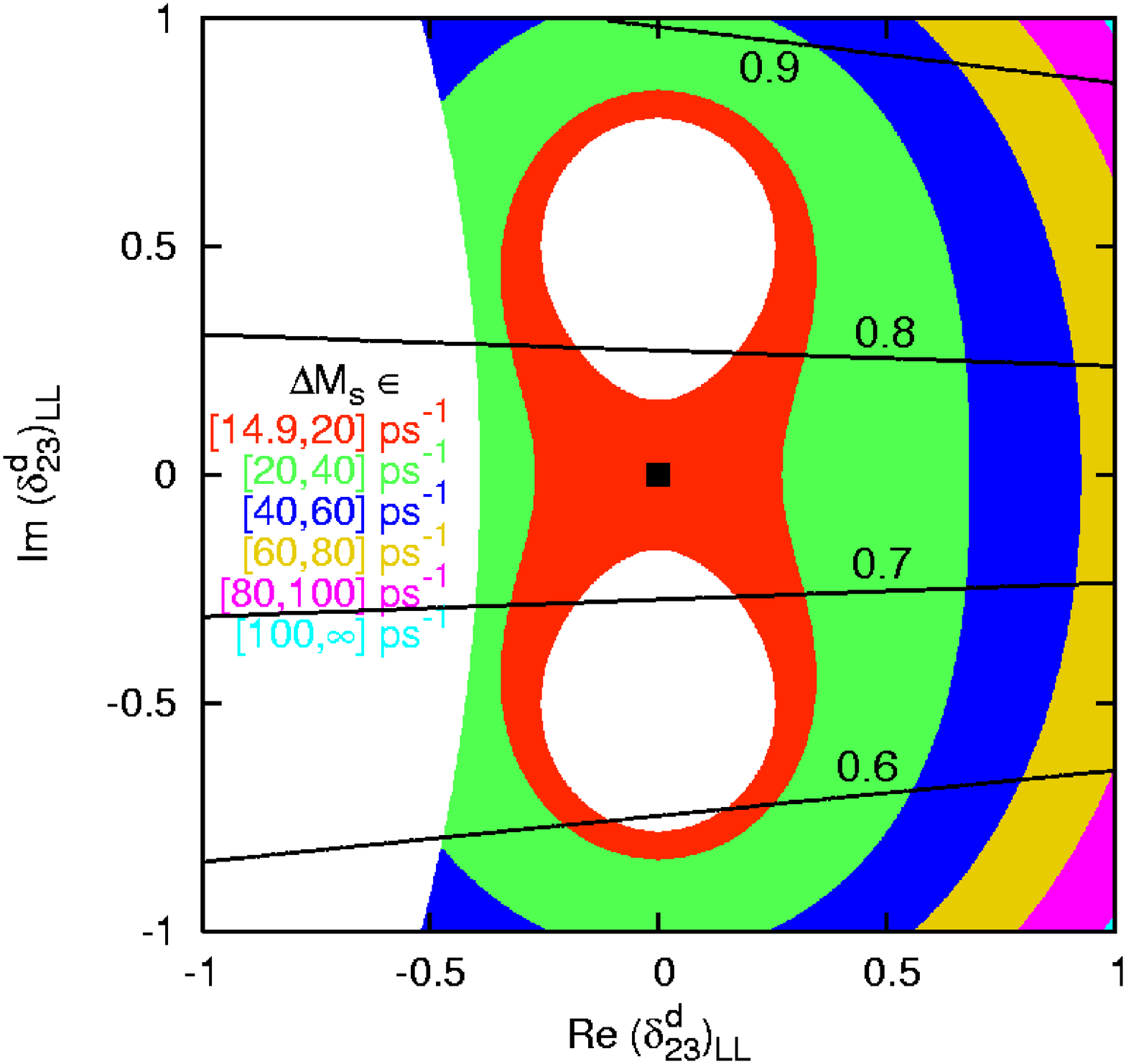}}
\subfigure[$\SphiK$ vs.\ $B ( \BtophiKs )$]
{\includegraphics[width=5.5cm]
{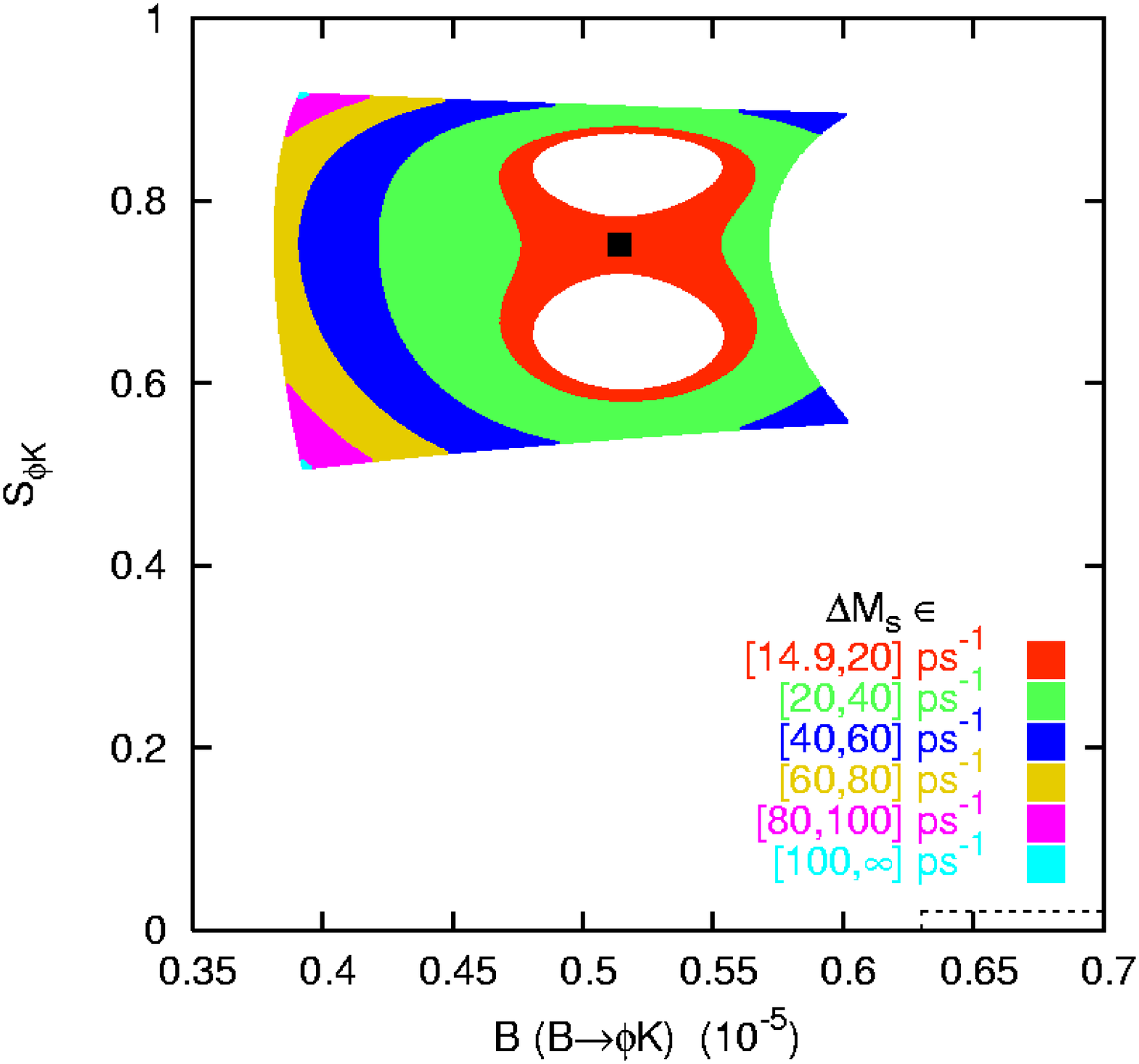}}
\subfigure[$\SphiK$ vs.\ $\CphiK$]
{\includegraphics[width=5.5cm]%
{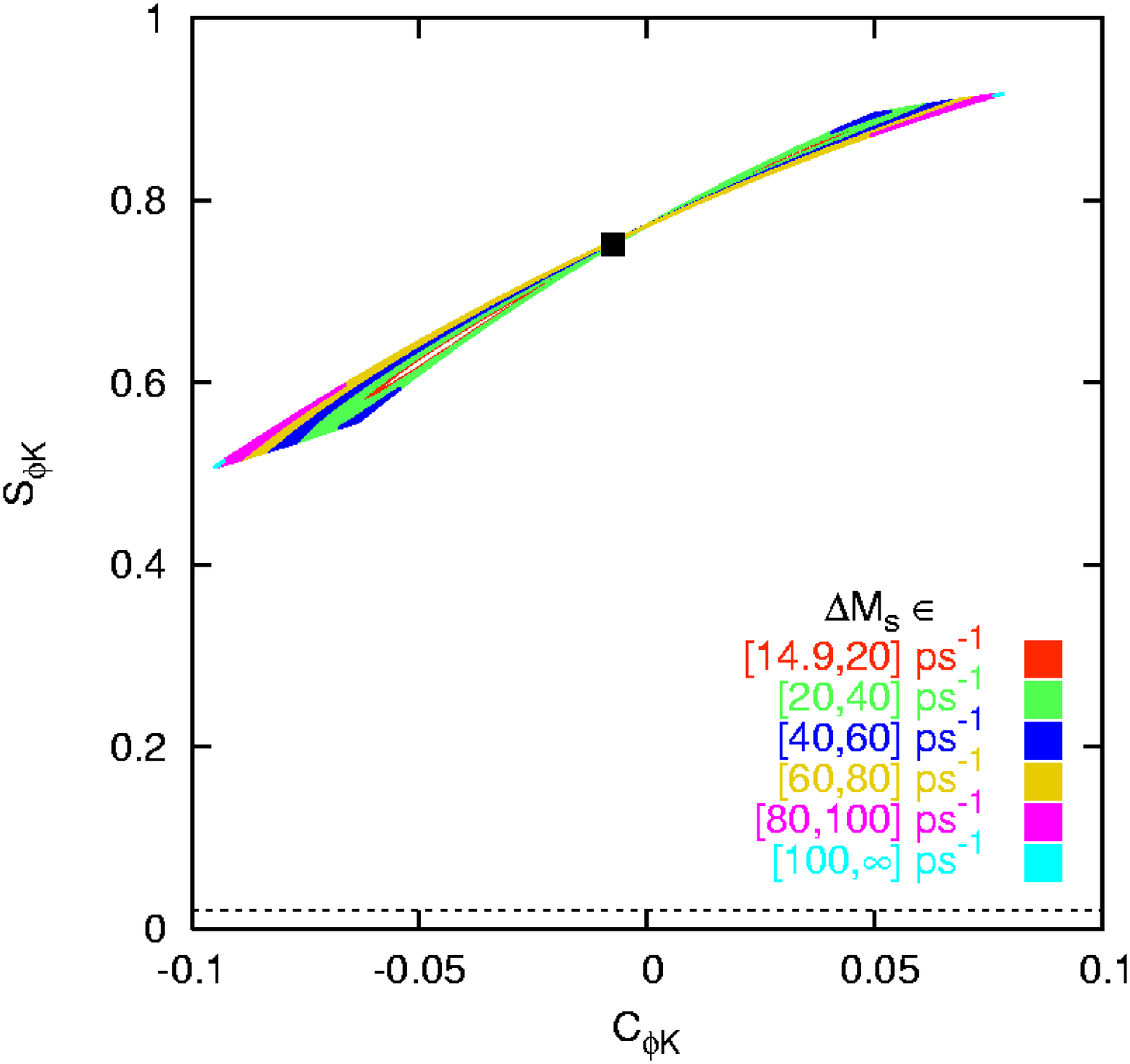}}
\subfigure[$\SphiK$ vs.\ $\Acp$]
{\includegraphics[width=5.5cm]
{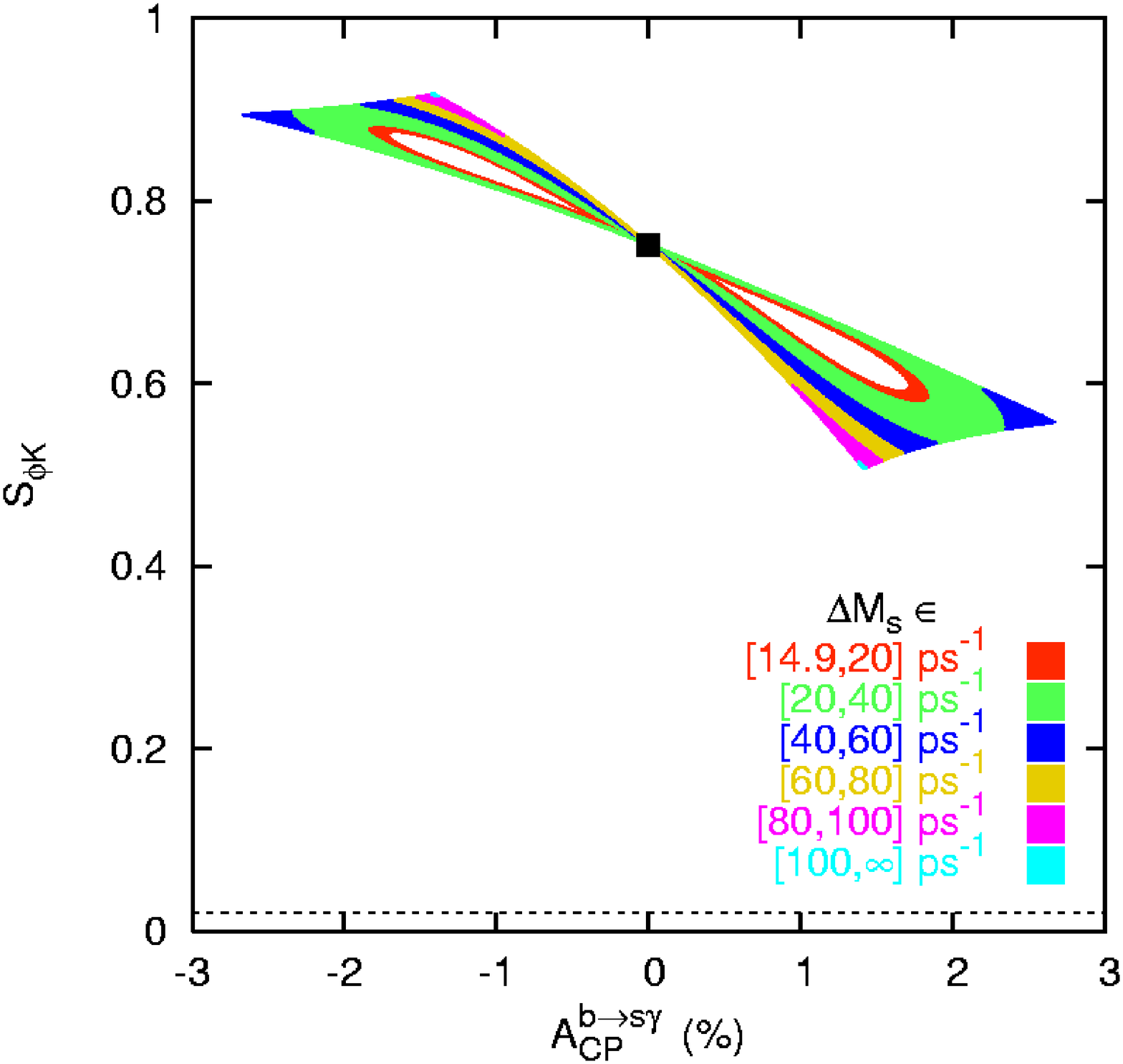}}
\subfigure[$\SphiK$ vs.\ $\All$]
{\includegraphics[width=5.5cm]
{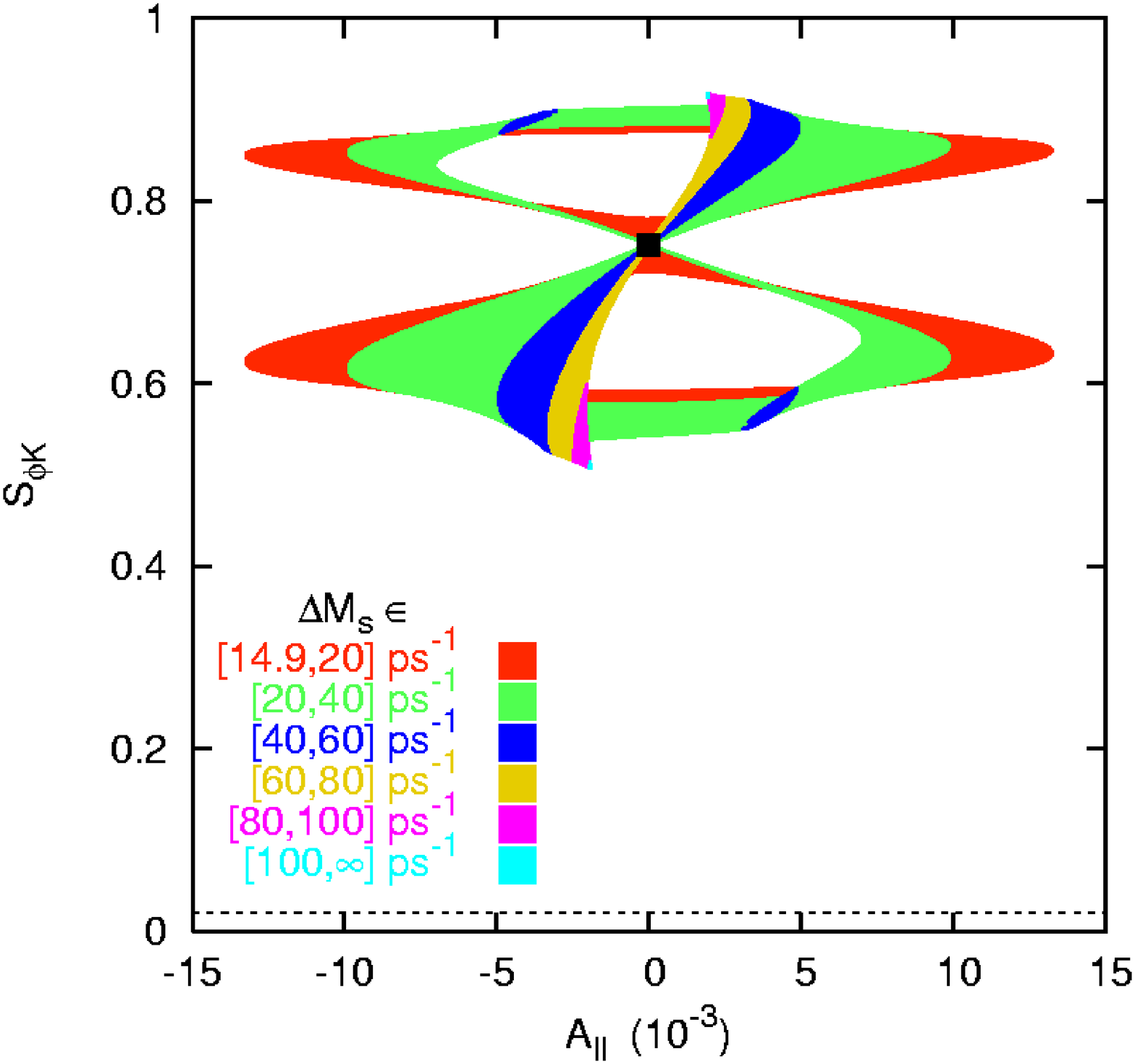}}
\subfigure[$\SphiK$ vs.\ $\sin 2 \beta_s$]
{\includegraphics[width=5.5cm]
{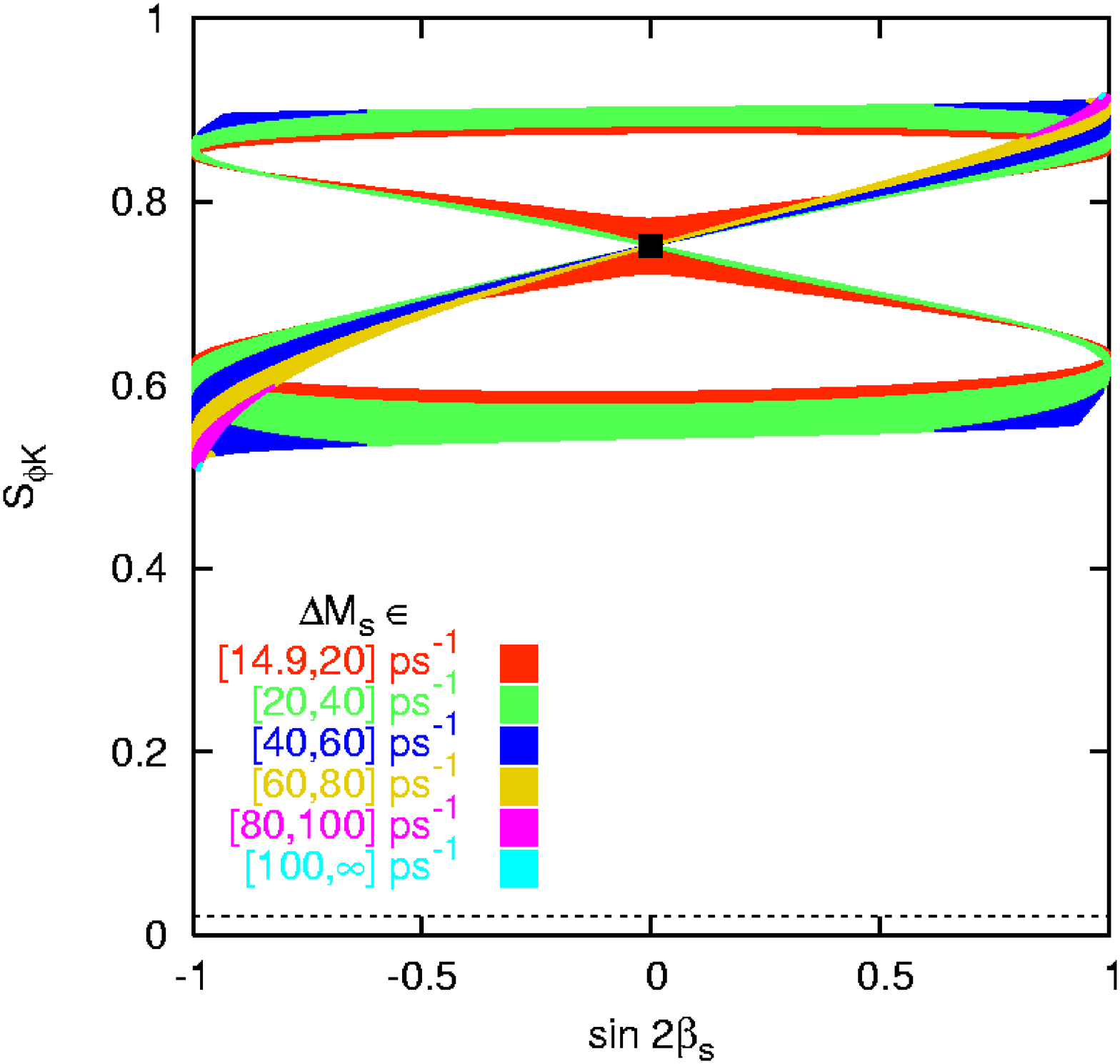}}
\caption{
(a) The allowed region in the $( {\rm Re}\, ( \delta_{LL}^d )_{23}, 
     {\rm Im}\, ( \delta_{LL}^d )_{23} )$ plane, and the correlations between
(b) $\SphiK$ and $B ( \BtophiKs )$, 
(c) $\SphiK$ and $\CphiK$, 
(d) $\SphiK$ and $\Acp$,
(e) $\SphiK$ and $\All$,  and
(f) $\SphiK$ and $\sin 2 \beta_s$ for
$m_{\tilde{g}} = 400$ GeV and $x=1$. 
Different shades are used for different ranges of $\Delta M_s$. 
The small black boxes represent the SM prediction.
}
\label{LLfig}
\end{figure}

A non-zero $LL$ mass insertion $(\delta_{LL}^d)_{23}$ generates
the QCD penguin operators $O_{3,...,6}$ and the (chromo)magnetic operators
$O_{7\gamma,8g}$. The $B\to X_s\gamma$ and $\Delta M_s$ constraints
remove two distinct regions in the $\{\mbox{Re}\,(\delta^d_{LL})_{23},
\mbox{Im}\,(\delta^d_{LL})_{23}\}$ plane. The constraint on the
branching ratio of $b\to s\gamma$ removes all points with 
$\mbox{Re}\,(\delta^d_{LL})_{23}< -0.5$. The constraint that $\Delta
M_s> 14.9\,\mbox{ps}^{-1}$ removes points where
$|\mbox{Re}\,(\delta^d_{LL})_{23}|< 0.25$ and 
$0.25< |\mbox{Im}\,(\delta^d_{LL})_{23}|< 0.75$. 
See Fig.~{\ref{LLfig}}(a).  The resulting $\Delta M_s$ 
is indicated by the different colors. 
For $x=1$, $\Delta M_s$ can be as large as
$\sim 100$ ps$^{-1}$. 

However, nowhere in the plane are we able to get substantial shifts in
$\SphiK$. In particular, for $m_{\widetilde g}=\widetilde m=400\gev$
we find no points with $\SphiK<0.5$. (We will comment shortly about
the effect of changing the SUSY masses.) The lack of impact on
$\SphiK$ arises in part because of 
substantial cancellations between   
$C_{3\dots6}^{\rm SUSY}$ and $C_{8g}^{\rm SUSY}$. This result is consistent
with that obtained by Lunghi and Wyler~\cite{lunghi}.
If we lower the common squark mass to $x = 3$, the asymmetries become 
somewhat smaller and $\Delta M_s$ can be as large as a few tens of ps$^{-1}$.
On the other hand, for a heavier squark of $x=0.5$,  $\Delta M_s$ can be
even larger than 100$\,$ps$^{-1}$,  which is beyond the reach of 
hadron colliders. 

(On the other hand, 
if we lower the gluino mass down to $250$ GeV with $x=1$,  
$\SphiK$ can shift down as low as $0.05$, but this is possible only 
if the gluino and squark masses are close their current lower limits
of 200 to $250\gev$. See the Section~\ref{gluino} for further discussion.) 

Furthermore there is very little effect on $\CphiK$ and $\Acp$, the
former varying only
between $\pm0.1$, the latter between $\pm3\%$ (see the figure).
However there are one-to-one correlations
among $\SphiK$, $\CphiK$ and $\Acp$
for an $LL$ insertion, so measuring these
correlations would provide strong evidence for the model. However the
experimental precision achievable in the near future will make this a
very difficult task.

The lack of signal in $\BtophiKs$ and $B\to X_s\gamma$ 
does not mean that an $LL$ insertion
is without signature. It is clear from Fig.~{\ref{LLfig}}(e) that there can
be sizable effects in the dilepton charge asymmetry $\All$,  
with asymmetries
as large as an order of magnitude above the SM. 
But the largest effect is in the $B_s$ sector,
specifically $\bsbsbar$ mixing and the 
CP-violating phase $\beta_s$ measured in $B_s\to\psi\phi$. Examining
Figs.~{\ref{LLfig}}(a),(e) and (f), it is remarkable that SUSY could
drive $\Delta M_s$ to values as large as $100\,\mbox{ps}^{-1}$, and
could shift the CP asymmetry to any value
$|\sin2\beta_s|\leq1$. 
Thus an $LL$ insertion may be better probed in $B_s$ decays
than in $B_d$ decays such as $\BtophiKs$.

In summary we conclude that the effects of the $LL$ insertion on
$\BtophiKs$ are insignificant unless
the squarks and gluino are very close to their current experimental
limit. For squarks and gluinos with masses of several hundred
GeV or more, the $LL$ insertion alone is inadequate for
explaining the current data on $\SphiK$. However, the
$LL$ insertion plays its strongest role in the $B_s$ system, where it
can contribute sizably to $\bsbsbar$ mixing
even for heavy squarks and gluinos. Thus there remains the
possibility that $\Delta M_s$ and $\sin 2\beta_s$
will be quite large while 
$\BtophiKs$ and $B\to X_s\gamma$ remain close to their SM
predictions.

The physics of the $RR$ insertion is identical in almost every way 
to that of the $LL$ insertion. The prime exception is the 
$b\to s\gamma$ constraint, which is weaker 
because the $RR$ contribution
does not interfere with the dominant SM amplitude for $B\to X_s\gamma$.
 
\begin{figure}
\subfigure[Allowed region for the $RR$ insertion]
{\includegraphics[width=5.5cm]{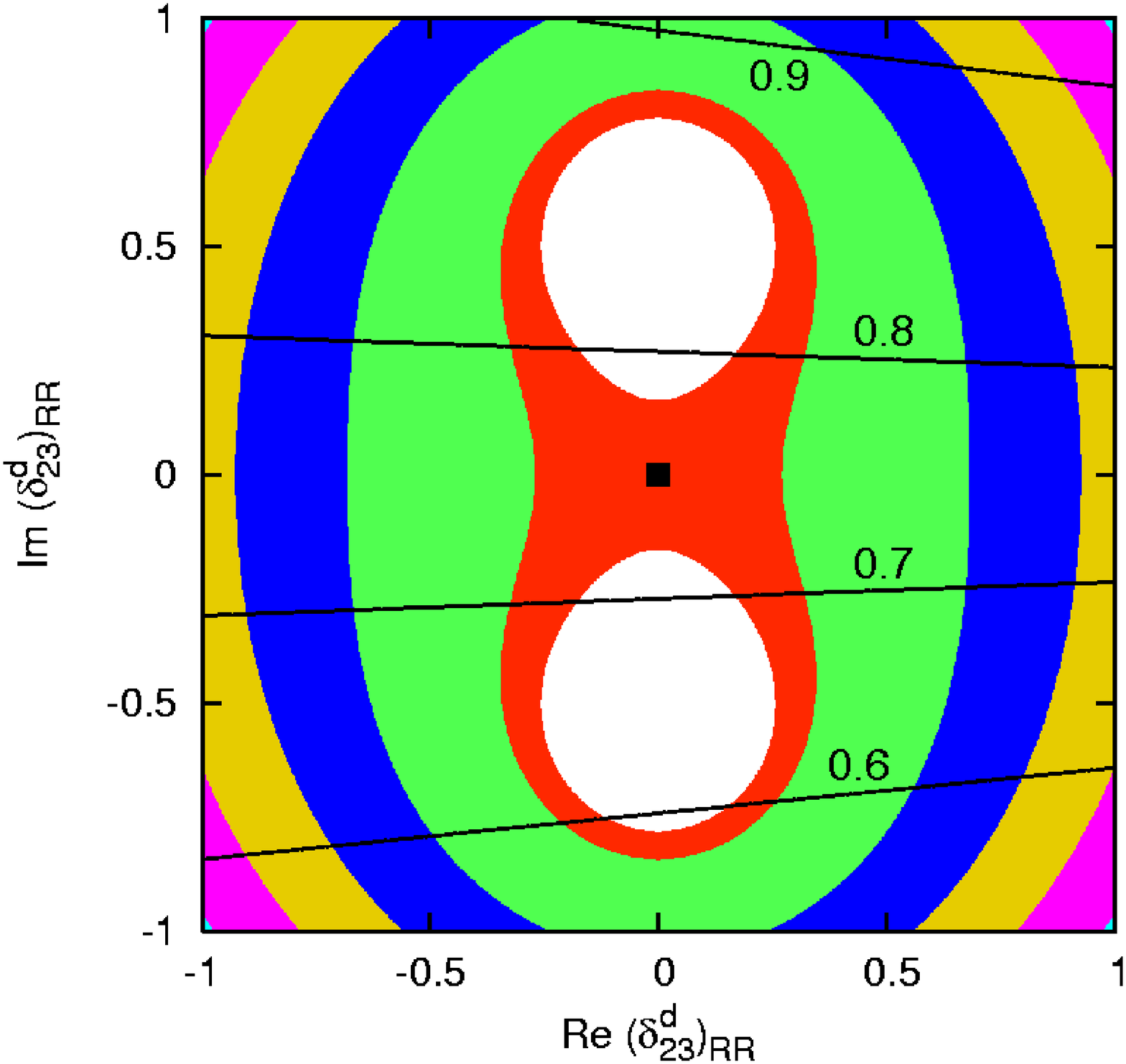}}
\subfigure[$\SphiK$ vs.\ $B ( \BtophiKs )$]
{\includegraphics[width=5.5cm]
{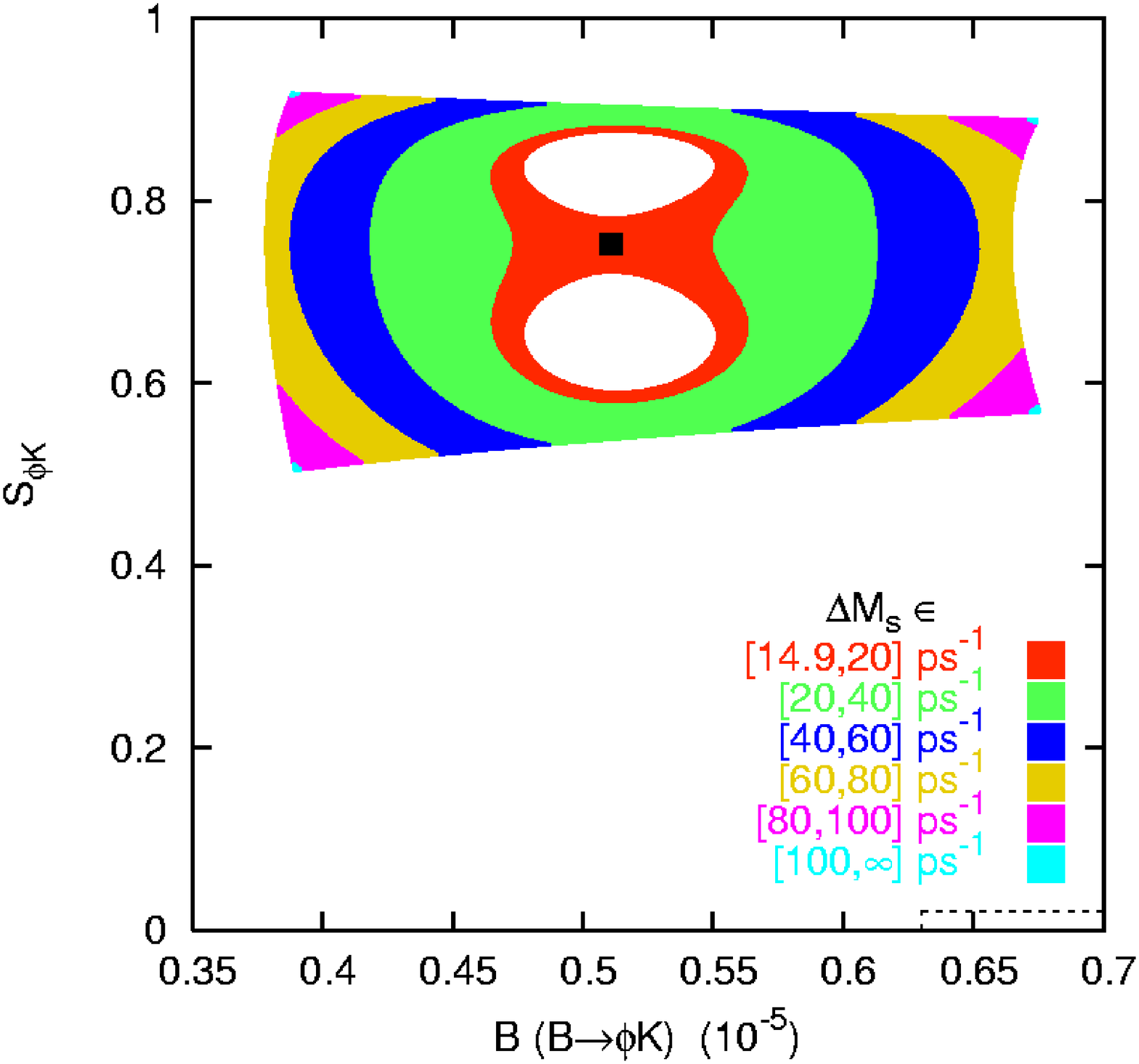}}
\subfigure[$\SphiK$ vs.\ $\CphiK$]
{\includegraphics[width=5.5cm]%
{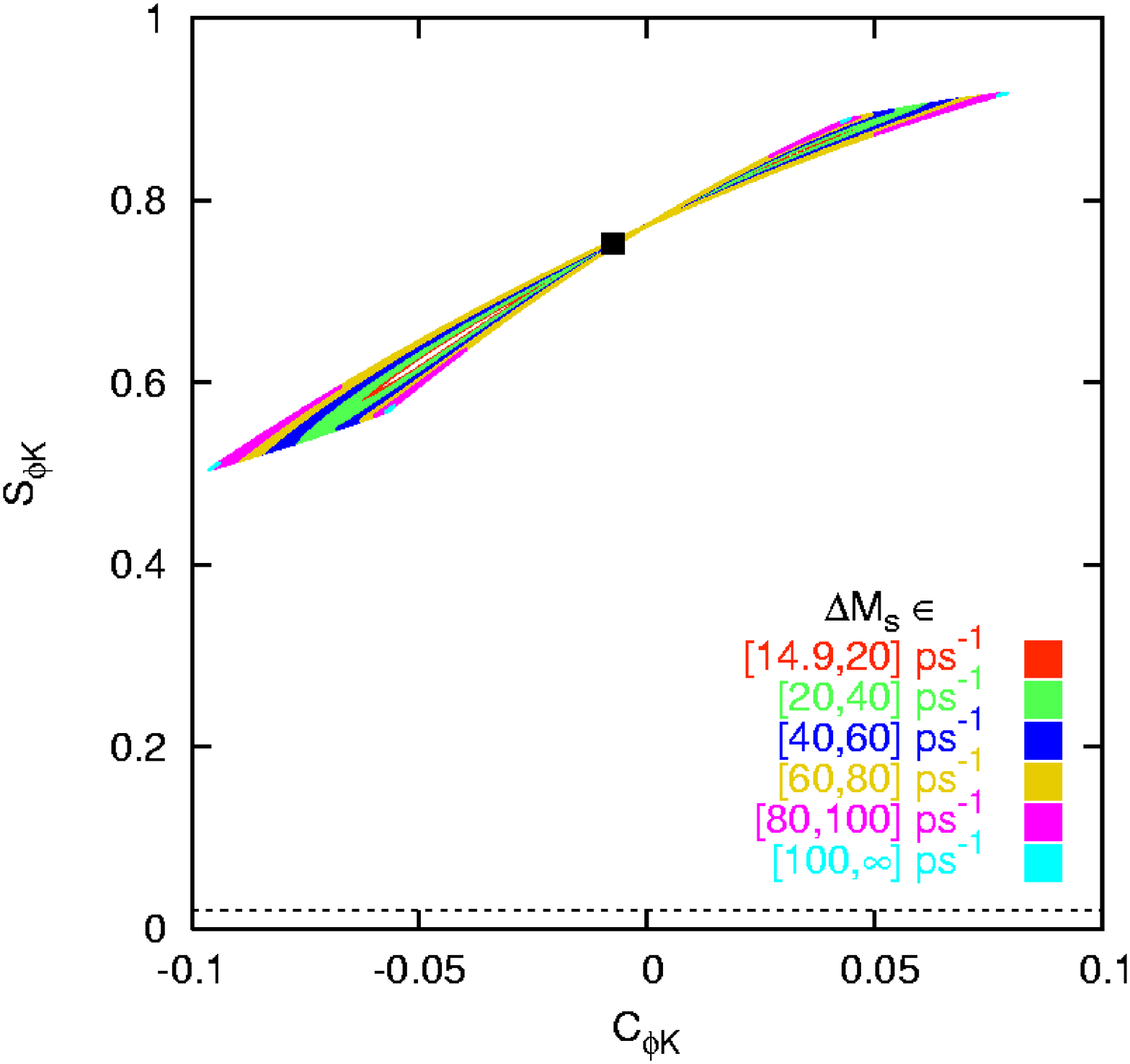}}
\subfigure[$\SphiK$ vs.\ $\sin 2 \beta_s$]
{\includegraphics[width=5.5cm]
{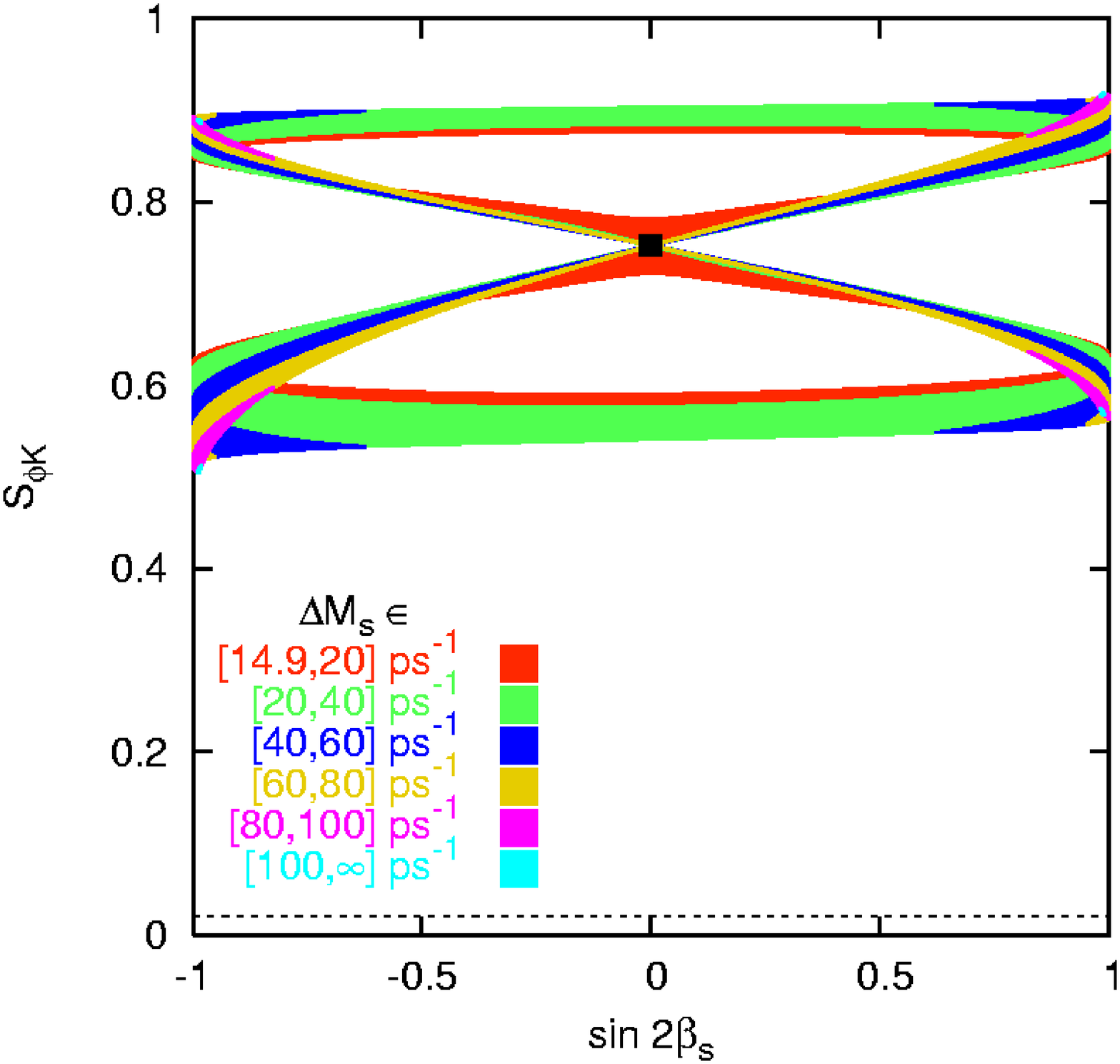}}
\caption{
(a) The allowed region in the $( {\rm Re}\, ( \delta_{RR}^d )_{23}, 
     {\rm Im}\, ( \delta_{RR}^d )_{23} )$ plane, 
and the correlations between
(b) $\SphiK$ and $B (\BtophiKs )$, 
(c) $\SphiK$ and $\CphiK$, 
(d) $\SphiK$ and $\sin 2 \beta_s$ for
$m_{\tilde{g}} = 400$ GeV and $x=1$. 
Different shades are used for different ranges of $\Delta M_s$. 
}
\label{RRfig}
\end{figure}

Therefore, the entire region of $\{\mbox{Re}\,(\delta_{RR}^d )_{23}, 
\mbox{Im}\,(\delta_{RR}^d )_{23}\}$ 
with $| (\delta_{RR}^d )_{23} | \leq 1$ is now 
allowed apart from the small areas in which $\Delta
M_s<14.9\,\mbox{ps}^{-1}$: see Fig.~{\ref{RRfig}}(a).  
In Fig.~{\ref{RRfig}}, 
we show the plots for the $RR$ insertion case with $m_{\tilde{g}}
= \tilde{m} = 400$ GeV. 
It is obvious that the $RR$ case is essentially identical to the $LL$
case in Fig.~{\ref{LLfig}}. There are two issues which are not obvious
from the figures. First, as already indicated, the $RR$ insertion
contributes very little to $B\to X_s\gamma$. Thus we have omitted a plot
of the prediction of the direct CP violation in $B\to X_s\gamma$,
$\Acp$, since it would differ little from the
SM. Secondly, for light gluinos and squarks, near their experimental
limit, it is possible to get slightly lower values of $\SphiK$ for the
$RR$ insertion than it was for the $LL$; this is because the 
$LL$ insertions required to further push down $\SphiK$ are
inconsistent with $B\to X_s\gamma$.

A few general remarks on the $LL$ and $RR$ insertions are in order.
First, we note that one requires
a large 2-3 mixing in order to obtain any observable effect in $\SphiK$.
But for such large mixing, our mass insertion approximation 
is no longer valid. One must then consider the full
vertex mixings in the $d_i - \tilde{d}_j - \tilde{g}$ couplings
with squarks in the mass eigenstates, as done in 
Ref.~\cite{murayama}. In this case, the superGIM suppression is less 
effective and the loop functions are enhanced, which could lead to
somewhat larger
effects on $\BtophiKs$. 

If one considers a large 2-3 mixing in the $LL$ or $RR$ sector, 
an important constraint from $\bbbar$ mixing, 
which is proportional to $( \delta_{AB}^d )_{13}$, may also arise.
Since the SM contribution to $\Delta M_d$ is proportional to 
$( V_{tb} V_{td}^* )^2$, the gluino loop contribution with light 
gluinos/squarks can easily dominate the SM, unless the $RR$ or 
$LL$ insertion for $b\rightarrow d$ transition is very small. 
In Ref.~\cite{kkp}, two of us discussed the allowed region for the 
$( \delta_{LL}^d )_{13}$  insertion, imposing the measured 
$\Delta M_d$, $\sin 2 \betapsiK$ and $B\rightarrow X_d \gamma$.  
It was found that one needs $( \delta_{LL}^d )_{13} \lesssim 
5 \times 10^{-2}$ 
even if we assume that the CKM angle $\gamma$ is a completely free 
parameter.  This should be compared to  $( \delta_{LL}^d )_{23} \sim 1$ 
which is needed in order for the $LL$ insertion to explain
the observed shift in $\SphiK$. 
It would be difficult, although not impossible, 
to build a flavor model where this is the case.

\subsection{The $LR$ insertion}
 
The case of an $LR$ insertion of the form $\wt{s}_L^\dagger\wt{b}_R$
is very different from that of either $LL$ or $RR$. While $LL$, $RR$
insertions can be ${\cal O}(1)$, the
parameters $( \delta_{AB}^d )_{23}$'s must be small in order to avoid
excessively large
FCNC amplitudes in charmless nonleptonic $B$ decays.
But this also implies that 
even small  $( \delta_{AB}^d )_{23}$'s can lead to 
observable effects in the $B$ sector. 
The analysis of the $LR$ insertion is particularly simple, in part
because it does not contribute to the penguin operators
$C_{3\ldots10}$. But its contributions to the (chromo)magnetic dipole
operators can be quite large because it breaks the SM chiral symmetry in
the $b$-quark sector allowing dipole operators to occur without the usual
$m_b$ suppression, replaced instead with a SUSY-breaking
$m_{\widetilde g}$ insertion. Thus, as previously noted, 
the $LR$ insertion is much more strongly constrained by 
$B\rightarrow X_s \gamma$, roughly  
$| ( \delta_{LR}^d )_{23} | \lesssim 10^{-2}$. 
However the same chiral symmetry breaking also ensures that the $LR$
contributions to $C_{8g}$ are likewise large, and so there may still
be a residual effect on $\BtophiKs$, both in its branching ratio and
CP asymmetries.
(This last effect was also noticed in Refs.~\cite{kagan,kagan02}).

\begin{figure}
\subfigure[Allowed region for the $LR$ insertion]
{\includegraphics[width=5.5cm]{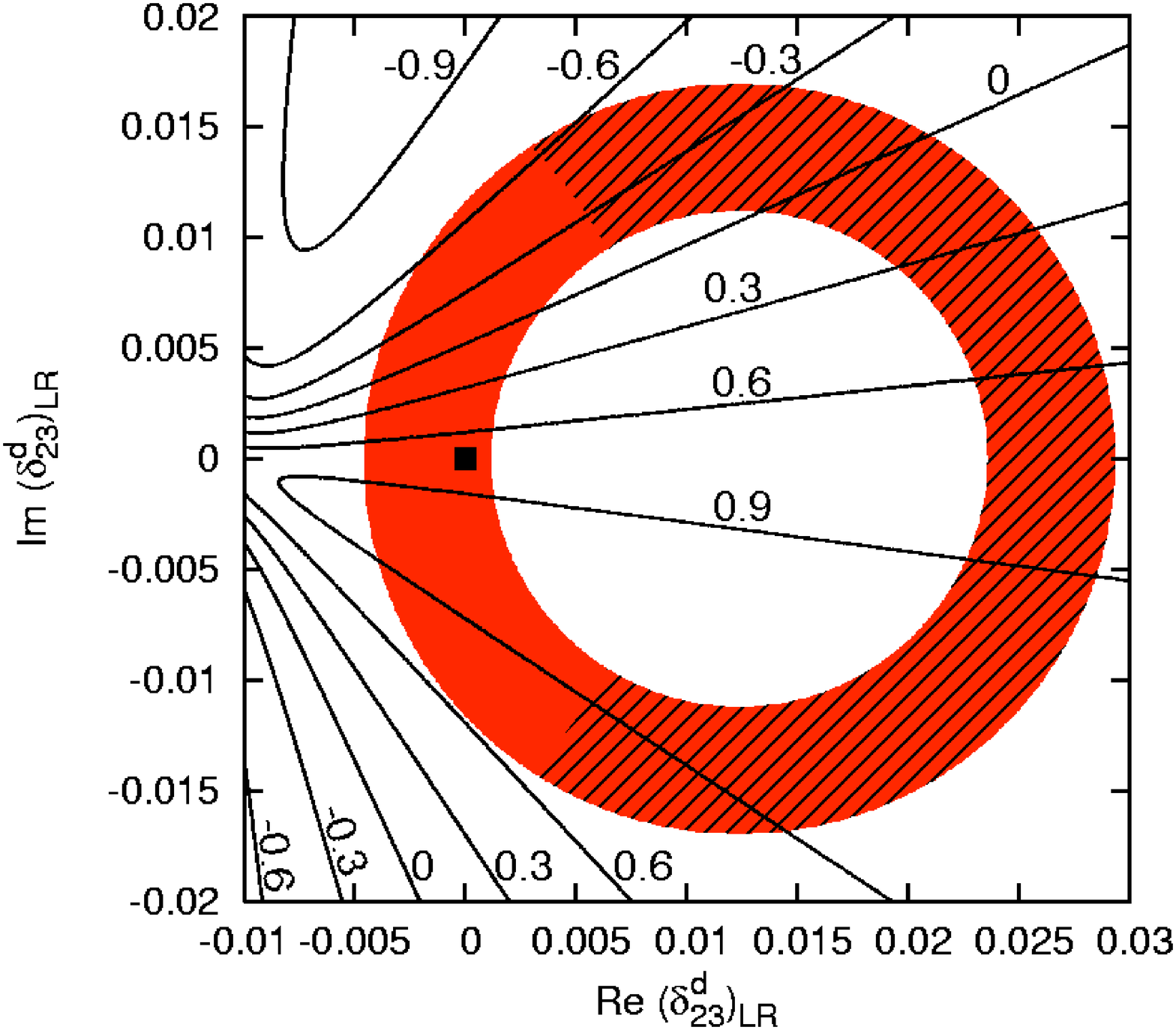}}
\subfigure[$\SphiK$ vs.\ $B ( \BtophiKs )$]
{\includegraphics[width=5.5cm]
{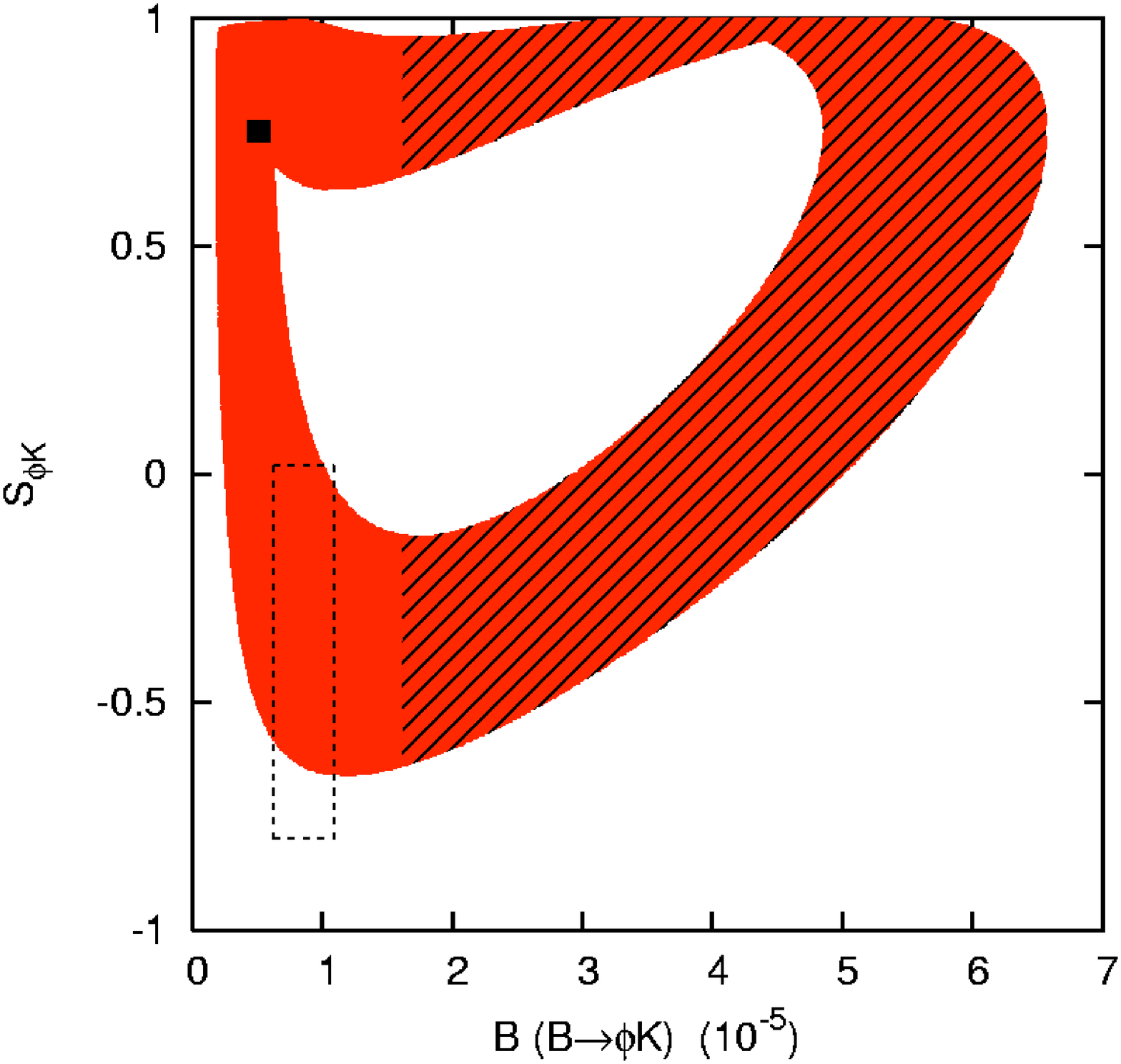}}
\subfigure[$\SphiK$ vs.\ $\CphiK$]
{\includegraphics[width=5.5cm]%
{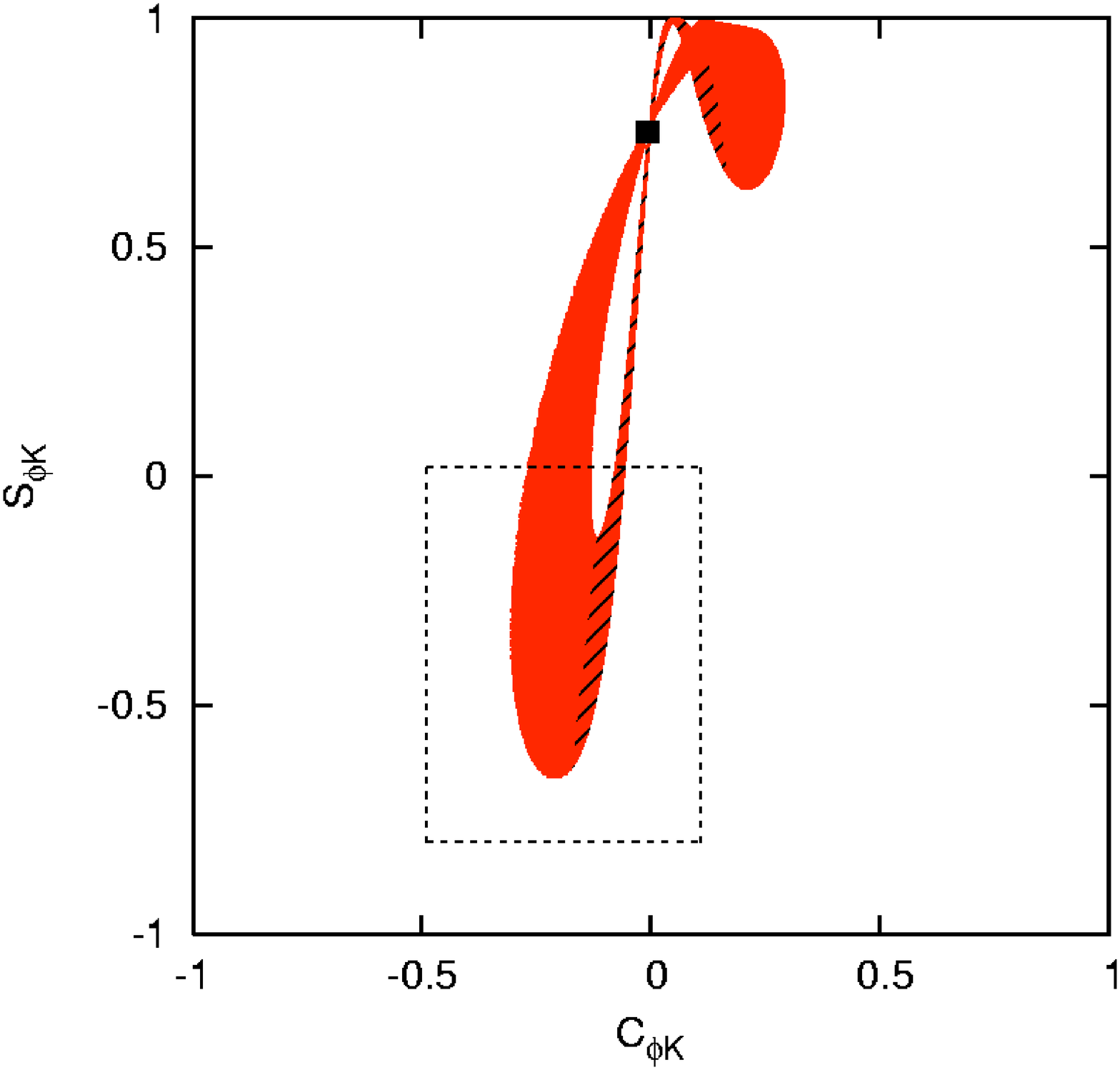}}
\subfigure[$\SphiK$ vs.\ $\Acp$]
{\includegraphics[width=5.5cm]
{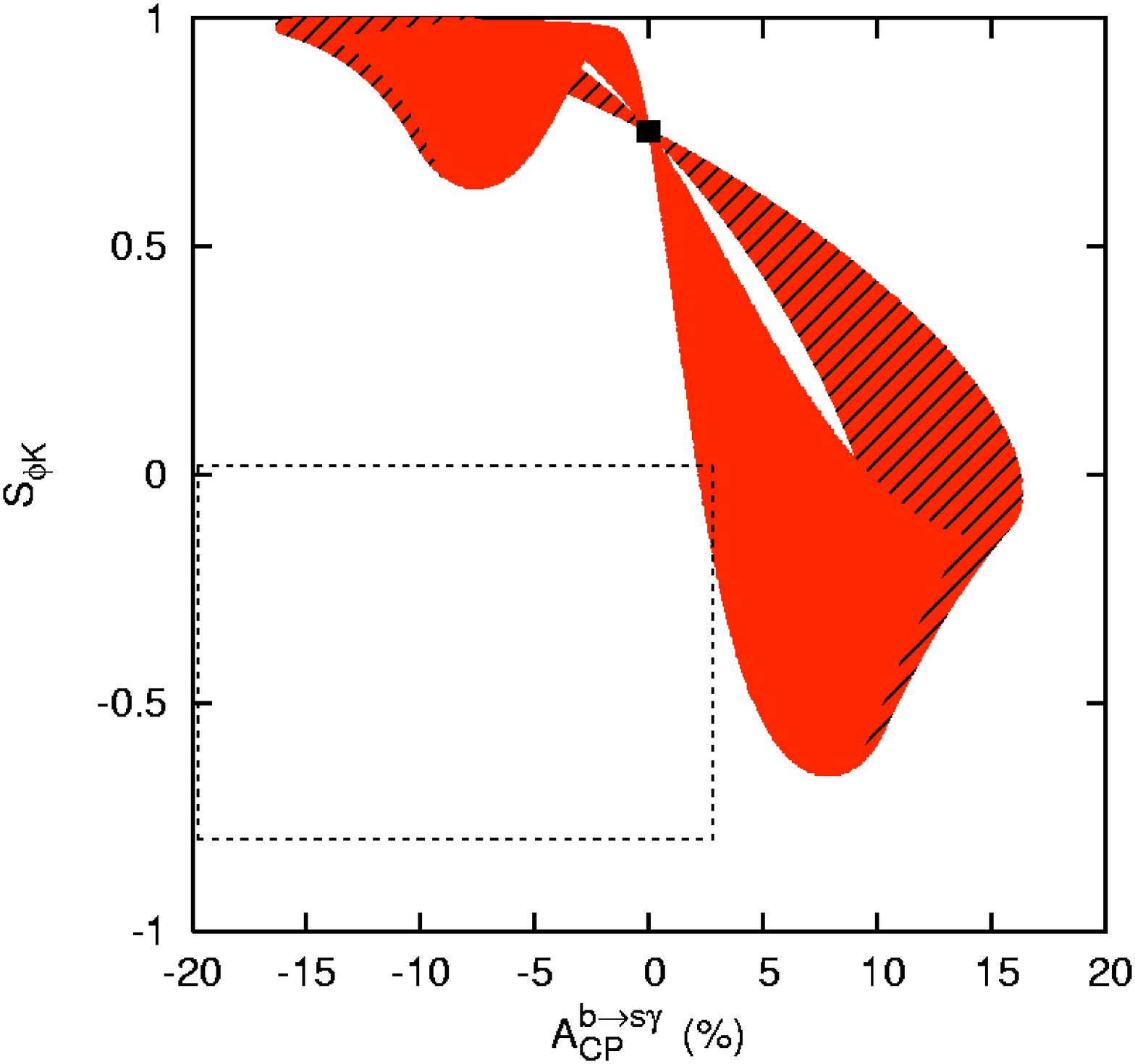}}
\subfigure[$\SphiK$ vs.\ $\All$]
{\includegraphics[width=5.5cm]
{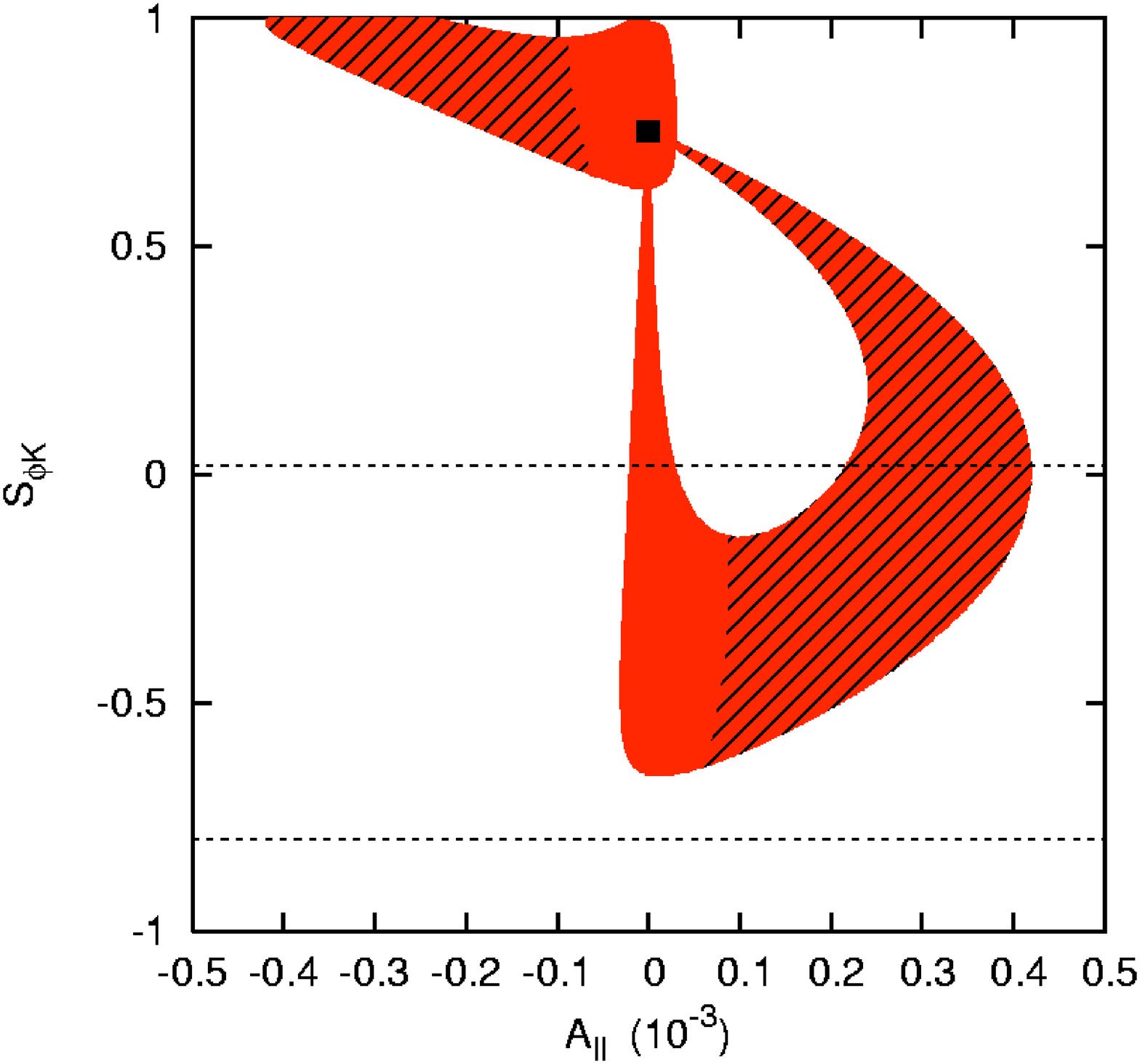}}
\subfigure[$\SphiK$ vs.\ $\sin 2 \beta_s$]
{\includegraphics[width=5.5cm]
{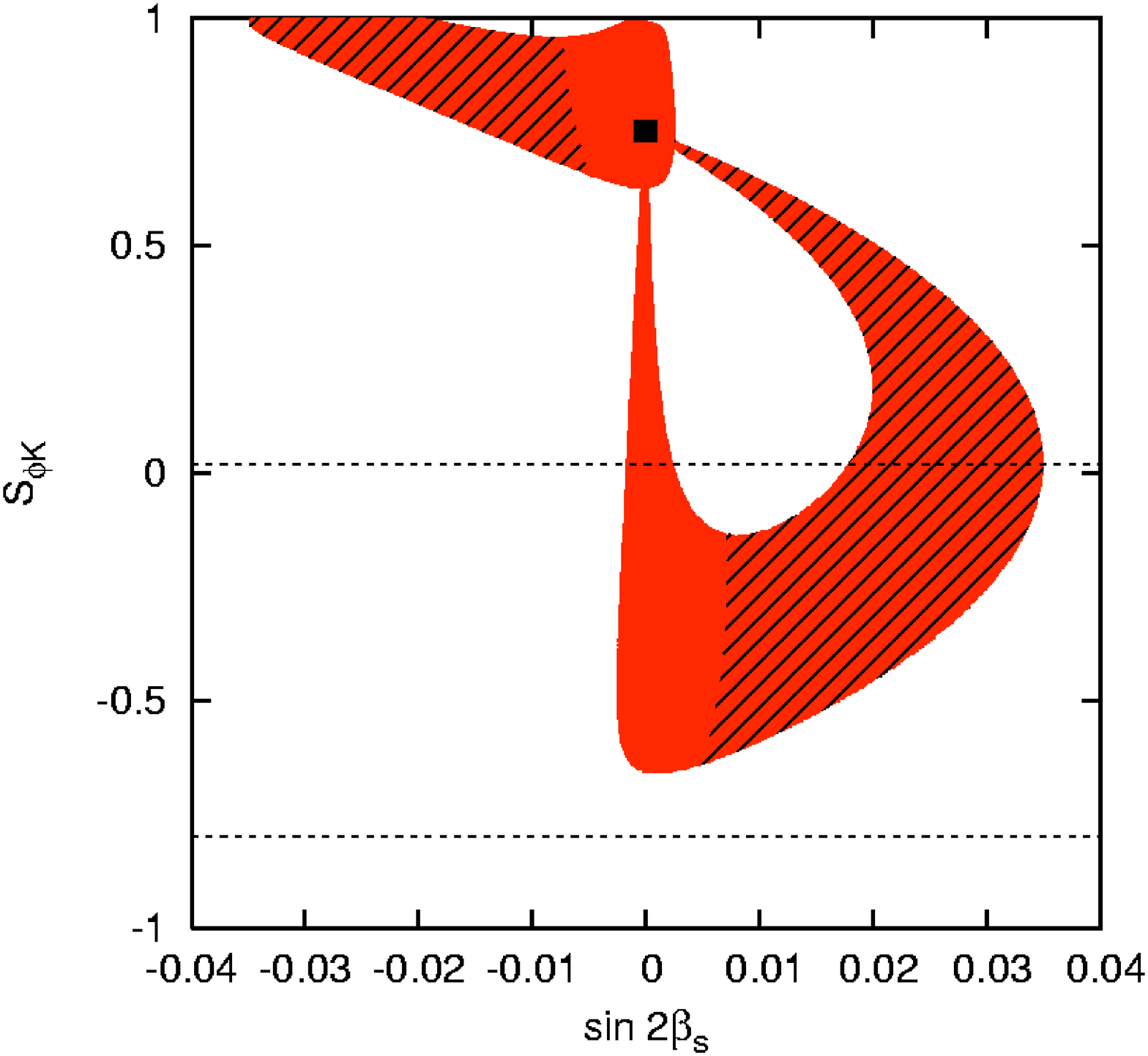}}
 \caption{
(a) The allowed region in the $( {\rm Re} \,( \delta_{LR}^d )_{23}, 
     {\rm Im} \,( \delta_{LR}^d )_{23} )$ plane, and the correlations 
between
(b) $\SphiK$ and $B (\BtophiKs)$, 
(c) $\SphiK$ and $\CphiK$, 
(d) $\SphiK$ and $\Acp$,
(e) $\SphiK$ and $\All$, and 
(f) $\SphiK$ and $\sin 2 \beta_s$ in the $LR$ insertion case for 
$m_{\tilde{g}} = 400$ GeV and $x=1$.
The dotted boxes show the $1\,\sigma$ current expermental data, and
the hatched regions represent
the region where $B (\BtophiKs ) > 1.6 \times 10^{-5}$.
 }
\label{LRfig}
\end{figure}

In Fig.~{\ref{LRfig}}(a), 
the allowed region for $\{ {\rm Re} ( \delta_{LR}^d )_{23}, 
{\rm Im} ( \delta_{LR}^d )_{23} \}$ is shown for $\tilde{m} = 400\gev$
and $x = 1$. The dark regions are consistent with the $B(B\to
X_s\gamma)$ and $\Delta M_s$ constraints, but are not constrained by
the bound on the $\BtophiKs$ branching ratio. We impose this
constraint by hatching the regions in which 
$B ( \BtophiKs) > 1.6 \times 10^{-5}$. Note that 
there are large regions of parameter space which are consistent with
$b\to s\gamma$ but not $B(\BtophiKs)$, as we advertised.
And though $(\delta^d_{LR})_{23}$ is constrained to be $\lesssim
10^{-2}$, it is still very important for  
$\BtophiKs$, affecting both its branching  ratio  
and the asymmetries $\SphiK$ and $\CphiK$ by significant amounts.
In particular, it is encouraging that the branching ratio for 
$\BtophiKs$ can easily be enhanced compared to the SM, 
moving it  closer to the experimental data. 

Fig.~{\ref{LRfig}}(b) 
indicates that the branching ratio for $\BtophiKs$ 
can in fact be greatly enhanced relative to the SM case. 
Also note that $\SphiK$ can take any value between $-0.6$ and 1, 
even after we impose $B(\BtophiKs)<16\times 10^{-6}$ as a constraint.
This is because the SUSY amplitude has a size similar to the SM amplitude
so that the resulting branching ratio is not too different from the 
SM prediction. But it has very different strong and weak phases from 
the SM amplitude so that $\SphiK$ and $\CphiK$ can vary greatly with
respect to SM predictions. 

In Fig.~{\ref{LRfig}}(c), 
we show the correlation between $\SphiK$ and $\CphiK$.
Note that we can get negative $\SphiK$, for which $\CphiK$ also 
becomes negative. 
Conversely, a positive $\CphiK$ implies that $\SphiK > 0.6$ in 
our model. Thus the current data on $\SphiK$ argues strongly in favor
of a negative $\CphiK$ if the $LR$ insertion is dominating the new
physics contributions.
This correlation between $\SphiK$ and $\CphiK$ will be 
tested at B factories in the near future.

The correlation between $\SphiK$ and the direct CP asymmetry in 
$B \rightarrow X_s \gamma$ ($\equiv \Acp$) 
is shown in Fig.~{\ref{LRfig}}(d). 
We find $\Acp$ 
becomes positive for a negative $\SphiK$, and a negative 
$\Acp$ implies that $\SphiK > 0.6$.
However the present CLEO bound on $\Acp$
(Eq.~\ref{Acp}) cannot constrain the $LR$ model very much.

In Fig.~{\ref{LRfig}}(e), 
the correlation between $\SphiK$ and $\All$ is shown. 
Note that $\All \sim 10^{-4}$ within the SM so that any appreciable amount
of $\All$ will be a clear indication of new physics in 
$\bsbsbar$ mixing.  For our model with the $LR$ insertion,  
the deviation of $\All$ from the SM prediction is small after imposing the 
$B \rightarrow X_s \gamma$ and $\BtophiKs$ branching ratio 
constraints.  Finally in Fig.~\ref{LRfig}(f), we show the correlation between 
$\SphiK$ and $\sin 2 \beta_s$ (the latter of which is zero within the SM).
Note that $\sin 2 \beta_s$  is nothing but $\Spsiphi$ in the
$B_s \to\psi \phi$ decay, and can be measured at hadron 
colliders if $\Delta M_s$ is not too large. 
However, the resulting $\sin 2\beta_s$ for an $LR$ insertion is 
too small to be observed.

If we lower the common squark mass to $x = 3$ for a fixed
$m_{\widetilde g}=400\gev$, 
the asymmetries  become somewhat smaller and $\Delta M_s$ can be in the 
range $[15.8, 16.2]$ ps$^{-1}$.
On the other hand, for a heavier squark $x=0.5$, the asymmetries can be
larger, but we still have $\Delta M_s \simeq \Delta M_s^{\rm SM}$. 
Thus the results are not particularly sensitive to the gluino and 
squark masses. As the masses get very large the SUSY contributions 
slowly decreases, and are too small to provide an explanation for a 
negative $\SphiK$.  

Before closing this subsection, let us comment further on the 
sensitivity of our results to the methods used for calculating the hadronic 
amplitude. It is clear that the direct CP asymmetry $\CphiK$ depends 
very strongly on the method used for evaluating hadronic matrix elements.
For example, one finds $\CphiK=0$ 
in naive factorization (without one loop corrections to the matrix
elements of four-quark operators), 
whereas it can take on values of ${\cal O}(1)$ in the BBNS approach.
The calculation of $\SphiK$ is also technique-dependent, though less
so. Different calculation schemes can produce values of $\SphiK$
which are different by factors of a few. For example,
$\SphiK$ can go as negative as $-0.6$ in the $LR$ insertion
case when we use the recent BBNS approach, but when the same parameter
space is studied using the techniques of Lunghi and
Wyler~\cite{lunghi}, $\SphiK$ only drops as negative as $-0.2$. We consider
the BBNS approach to be the most accurate and well-motivated
available, which is why we use it, but view the differences between
these schemes to be an unavoidable source of uncertainty at present.

\subsection{The $RL$ insertion}

\begin{figure}
\subfigure[Allowed region for the $RL$ insertion]
{\includegraphics[width=5.5cm]{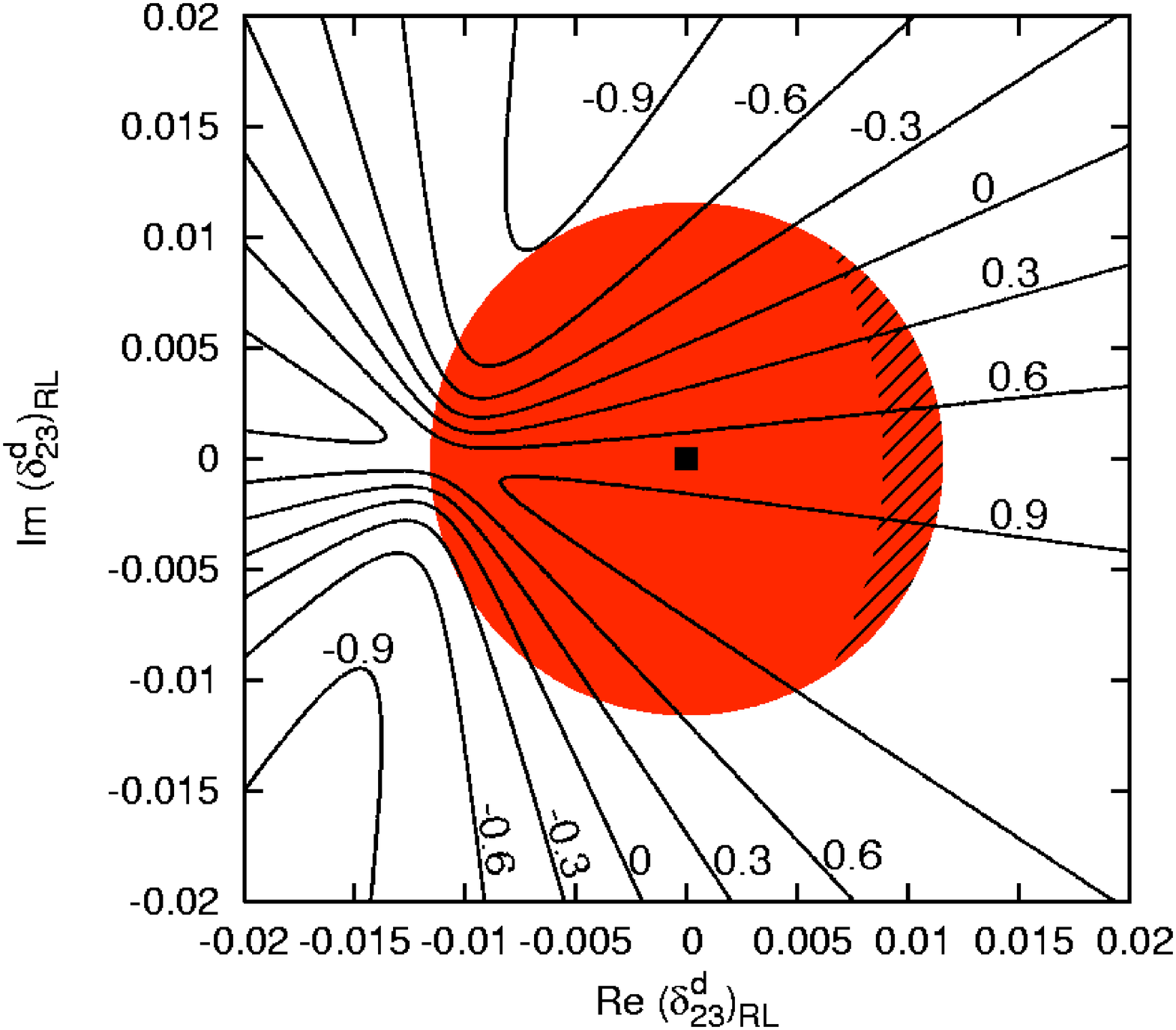}}
\subfigure[$\SphiK$ vs.\ $B ( \BtophiKs )$]
{\includegraphics[width=5.5cm]
{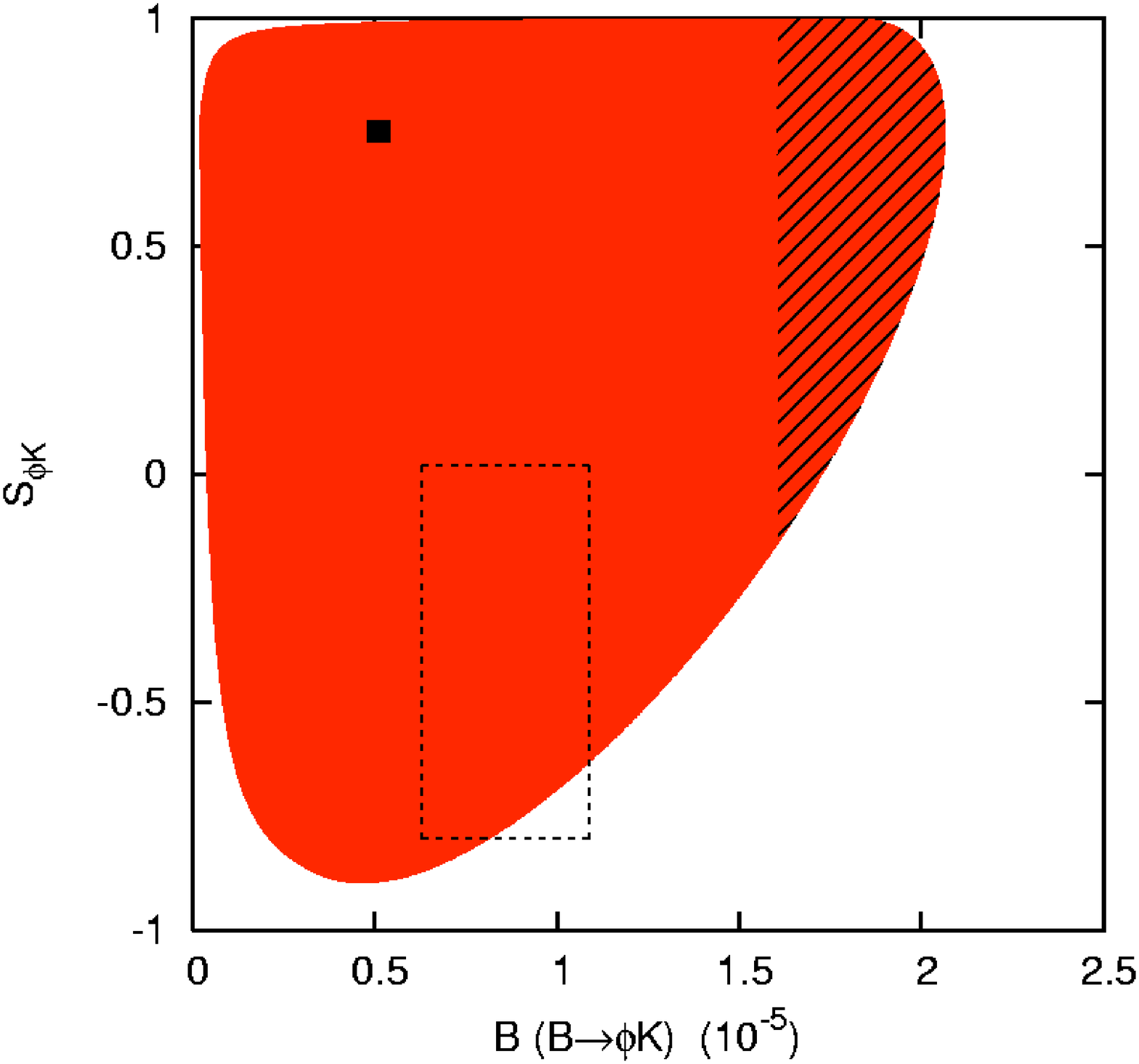}}
\subfigure[$\SphiK$ vs.\ $\CphiK$]
{\includegraphics[width=5.5cm]%
{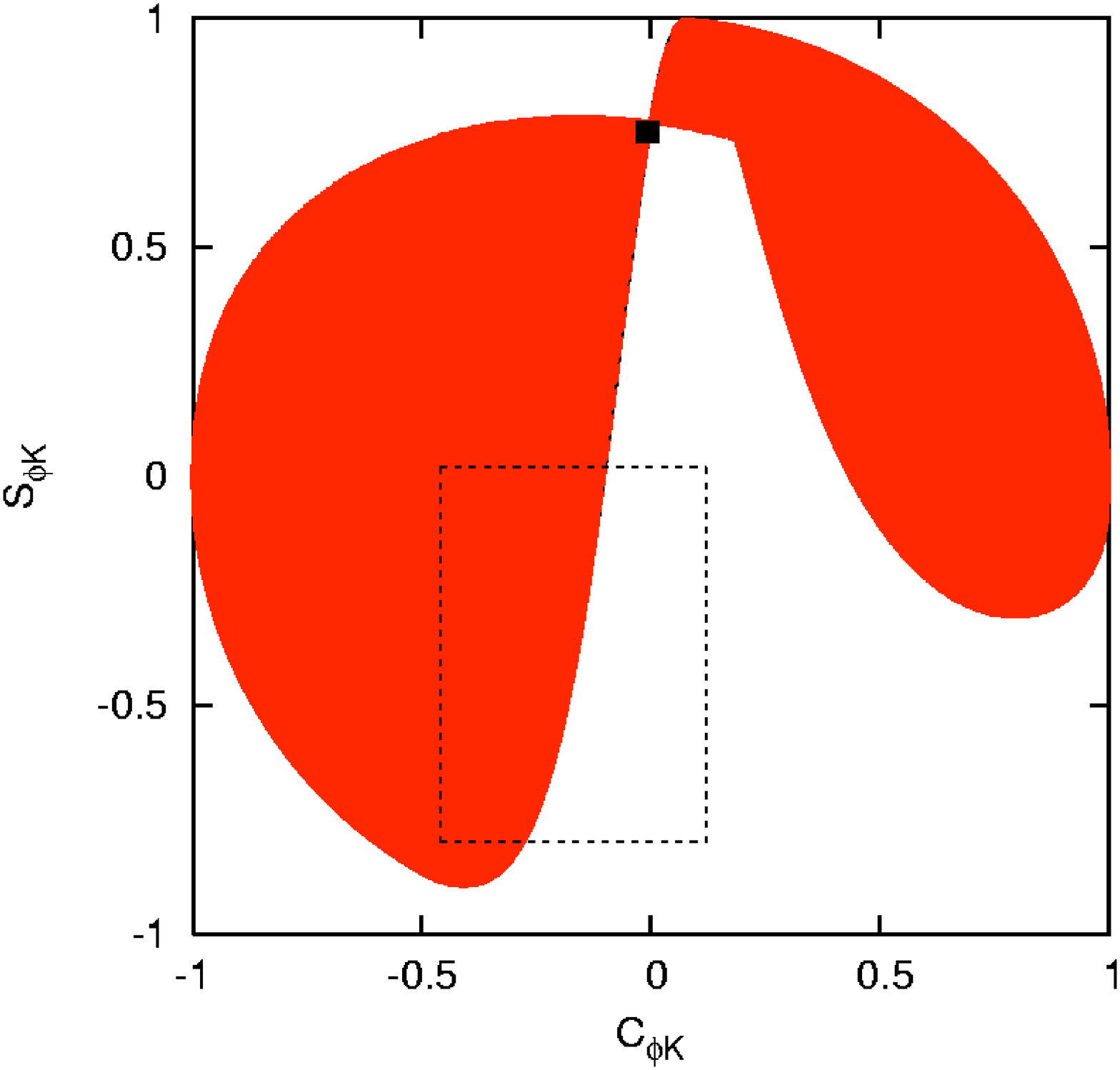}}
 \caption{
(a) The allowed region in the $( \mbox{Re}\,( \delta_{RL}^d )_{23}, 
     {\rm Im}\,( \delta_{RL}^d )_{23} )$ plane, and the correlations between
(b) $\SphiK$ and $B (\BtophiKs)$, 
(c) $\SphiK$ and $\CphiK$, and 
(d) $\SphiK$ and $\Delta M_s$ for 
$m_{\tilde{g}} = 400$ GeV and $x=1$. 
Current data at $1\,\sigma$ level is shown 
by the dotted boxes, and the hatched regions represent
the region where $B (\BtophiKs) > 1.6 \times 10^{-5}$.
}
\label{RLfig}
\end{figure}

The final case to be considered is perhaps the most unusual. An
$RL$ insertion is of the form $\wt{s}_R^\dagger\wt{b}_L$, coupling the
RH strange squark to the LH bottom squark. Our intuition from the SM
would lead us to believe that such insertions would be small compared
to $LR$ insertions, naturally suppressed by $\sim m_s/m_b$. In
SUSY models with minimal flavor violation, this is indeed the case.
However, once one moves beyond minimal flavor violation there is no
reason for the $RL$ insertion to be particularly suppressed with
respect to $LR$. Furthermore, it has a phenomenology that is quite
different than the $LR$ insertions because it does not interfere with
the SM contributions which are themselves of the $LR$ form.
(Because of this, we had to extend the BBNS approach to the case with 
opposite chirality 4-quark operators.)
A quick comparison of the results for the $RL$ insertion in
Fig.~\ref{RLfig} to those of the $LR$ in the last section should
convince the reader of this easily.

Though
the $RL$ insertion generates a wide range in $B(\BtophiKs)$ like the
$LR$ case (Fig.~\ref{RLfig}(b)), the similarities end there.
Because the $RL$ insertions do not directly interfere with the SM,
the allowed region for $(\delta^d_{23})_{RL}$ is centered around zero
(that is, only its square is relevant in most observables) as seen in
Fig.~\ref{RLfig}(a). And although both cases can easily generate
large, negative $\SphiK$ and negative $\CphiK$, only in the $LR$ case
is the correlation completely clean. As we pointed in the previous
section, an $LR$ insertion with negative $\SphiK$ implies negative
(but not very large) $\CphiK$. For the $RL$ case, negative $\SphiK$
can be accompanied by $\CphiK$ of either sign (Fig.~\ref{RLfig}(c)). 
However there is a considerable hole in the figure around $\CphiK=0$,
which may help in disentangling the structure of the insertions once
more data is to be had.

Otherwise $RL$ insertions leave very little other evidence. They do
not appreciably alter $\Delta M_s$, and they do not generate observable
$\Acp$ or $\All$. Thus for a pure $RL$ insertion, one expects to
reproduce the SM in almost all ways except in $\BtophiKs$. In fact, of
all the cases we have considered so far, the $RL$ case best fits the
current data, for precisely this reason.

There is a particularly interesting variation on the $RL$ theme that
we now turn our attention towards.

\subsubsection{A special case: the $RL$ dominance scenario}
\label{RLdom}

Thus far, in our calculations of $B\to X_s\gamma$ we have made a
simplifying assumption that can often be incorrect in realistic
models. We have assumed that the calculation of $B\to X_s\gamma$
consists entirely of a SM piece and gluino-induced
SUSY pieces. However low-energy models of SUSY usually have several
other, often large, contributions to $B\to X_s\gamma$, most particularly
from charged Higgs loops (which add to the SM terms), and from 
chargino loops (which subtract from the SM). It is well-known (see,
{\it e.g.}, Ref.~\cite{bg}) that in the supersymmetric limit, the SM, charged
Higgs and chargino contributions cancel each other exactly, predicting
a rate for $B\to X_s\gamma$ of precisely zero. If SUSY is broken but the
superpartners are relatively
light, much of that cancellation can still occur. In such a case, the
observed rate for $b\to s\gamma$ would be due almost entirely to the
gluino loops.

Ref.~\cite{kane} has argued for just such an unconventional
interpretation of $B\rightarrow X_s \gamma$. Here the
Standard Model and usual SUSY contributions to $C_{7 \gamma}$ 
approximately cancel:
$C_{7\gamma} ( m_b ) \approx C_{8 g} ( m_b) \approx 0$.  Then the 
dominant contribution to $b\to s\gamma$ 
is from the gluino-$\tilde{b}_L$-$\tilde{s}_R$
penguin in $\widetilde{C}_7$, and is proportional to $(\delta^d_{RL})_{23}$.
We will call this the $RL$-dominance scenario in $b\to s\gamma$.

\begin{figure}
\subfigure[Allowed region for the $RL$ insertion]
{\includegraphics[width=5.5cm]{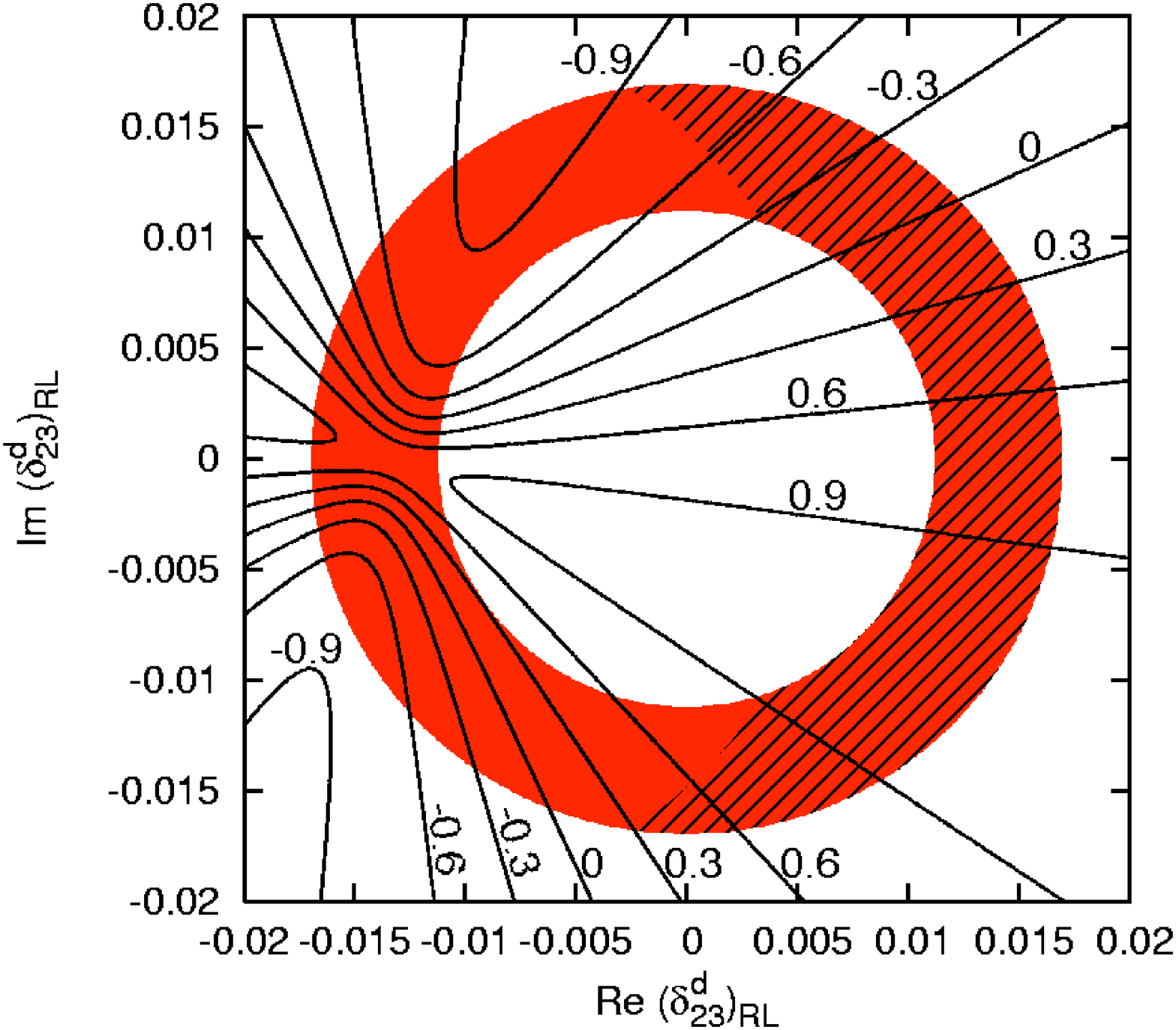}}
\subfigure[$\SphiK$ vs.\ $B ( \BtophiKs )$]
{\includegraphics[width=5.5cm]
{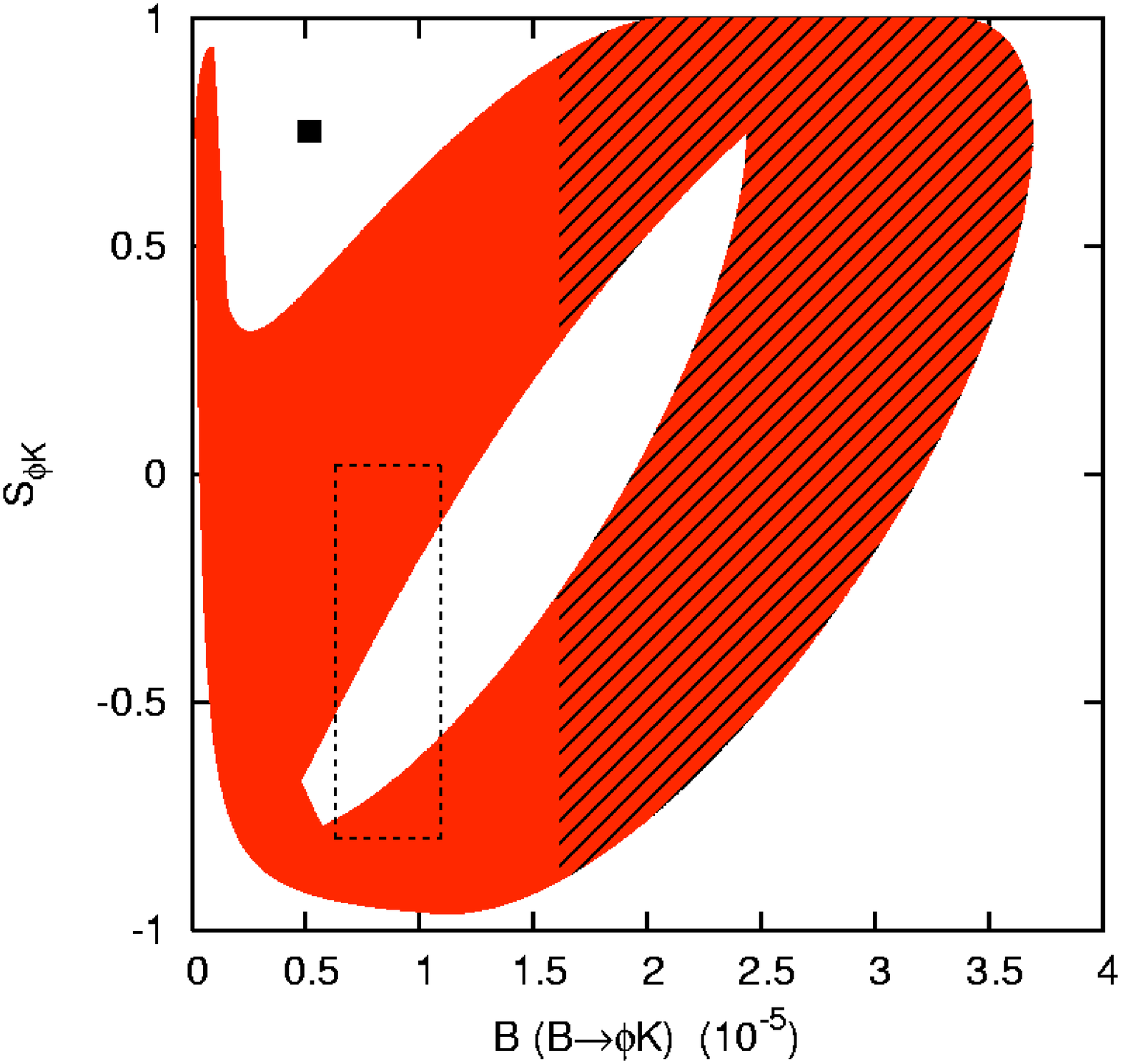}}
\subfigure[$\SphiK$ vs.\ $\CphiK$]
{\includegraphics[width=5.5cm]%
{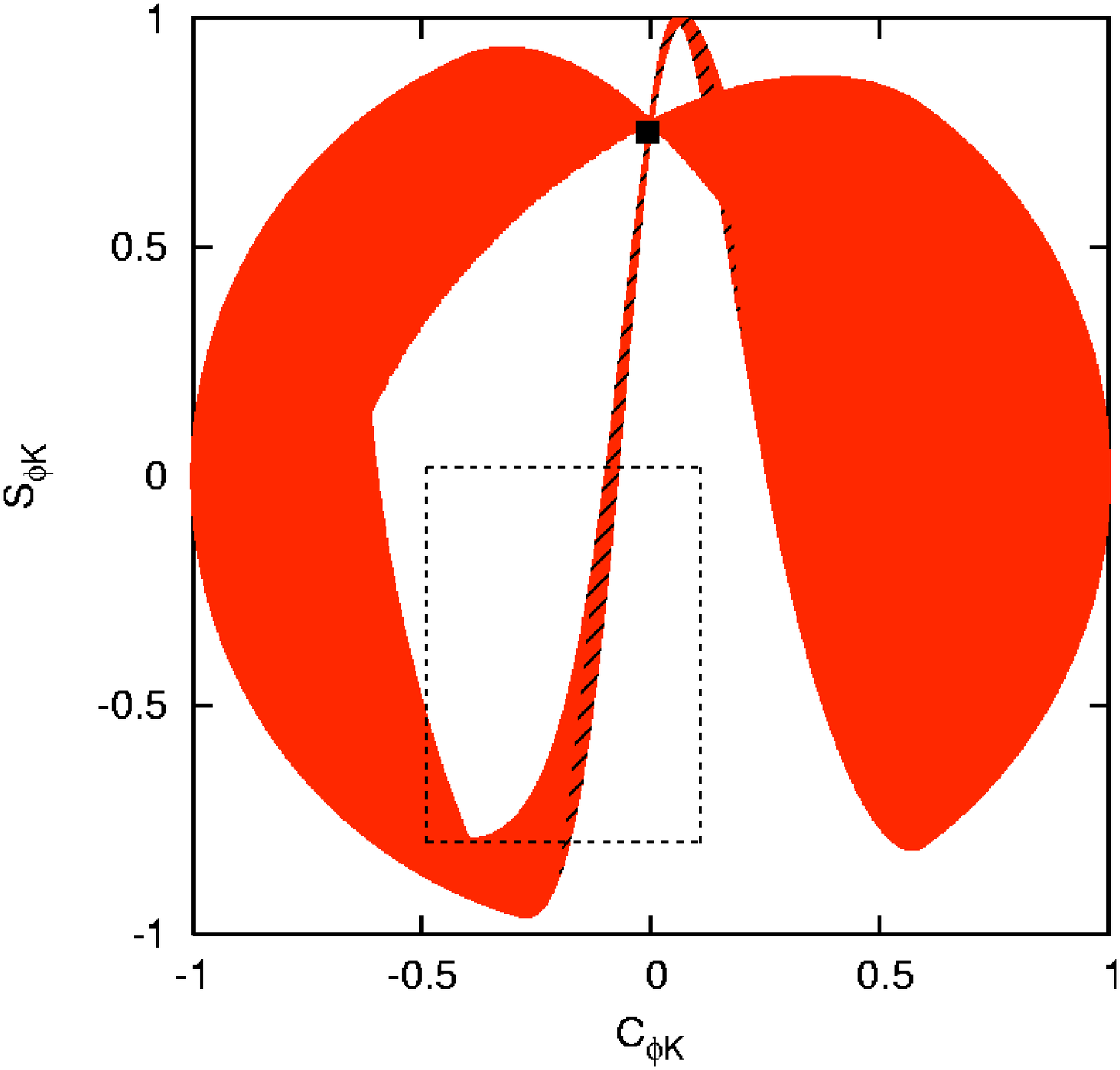}}
 \caption{
(a) The allowed region in the $( {\rm Re}\,( \delta_{RL}^d )_{23}, 
     {\rm Im}\,( \delta_{RL}^d )_{23} )$ plane in the
special case of section~\ref{RLdom}, and the correlations between
(b) $\SphiK$ and $B (\BtophiKs)$, 
(c) $\SphiK$ and $\CphiK$, and 
(d) $\SphiK$ and $\Delta M_s$ for 
$m_{\tilde{g}} = 400$ GeV and $x=1$. 
Current data at $1\,\sigma$ level is shown 
by the dotted boxes, and the hatched regions represent
the region where $B (\BtophiKs) > 1.6 \times 10^{-5}$.
}
\label{RLdomfig}
\end{figure}

Now the constraint from $B\rightarrow X_s \gamma$ plays a completely
different role in constraining our parameter space. Rather than
limiting the size of the $RL$ insertion to be close to zero, it
actually demands a finite non-zero value for the insertion in order to
reproduce the observer branching ratio. Thus, it has the effect of
constraining the complex parameter $( \delta_{RL}^d )_{23}$ to lie in
an annulus, just as the constraint from the $\BtophiKs$ branching
fraction does (though of different center and radius).
The resulting allowed parameter space is shown in
Fig.~{\ref{RLdomfig}}(a) for 
$m_{\tilde{g}} = \tilde{m} = 400$ GeV ($x=1$). 
In Fig.~{\ref{RLdomfig}}(b) and (c), 
we show the correlation of $\SphiK$ with 
$B (\BtophiKs)$ and $\CphiK$, respectively.  Note that
$\SphiK$ can take almost any value between $-1$ and $+1$ 
without conflict with the observed branching ratio for $\BtophiKs$.
Also $\CphiK$ can be positive for a negative $\SphiK$, unlike 
the $LR$ case. 
In particular, $|\CphiK| >0.2$ for $\SphiK < 0$.  
This could be a useful probe for identifying this scenario.
Since there is only one diagram contributing to $b\to s\gamma$ in 
this scenario, one has $\Acp = 0$ trivially.  

However, as
discussed in Ref.~\cite{kane}, if $( \delta_{RR}^d )_{23}$ is also 
nonzero then $\Acp$ arises as an interference between 
$\widetilde{C}_{7\gamma}$ and $\widetilde{C}_{8g}$. Then the $RR$ 
insertion can contribute more significantly to $\Delta M_s$;
we will not study this more complicated scenario here
because the observables will depend crucially on the relative size of
the various insertions. For example, the $RL$ insertion
could be induced by a large $LL$ or $RR$ insertion coupled
with an extra SU(2)-violating mass insertion.
In such a case, there could be large deviations both in 
$\SphiK$ and $\bsbsbar$ mixing. Nonetheless,
the correlations between $\SphiK$ and $\CphiK$ 
that we demonstrate 
in this paper depend primarily on the $LR$ or $RL$ insertions, 
and thus are good probes of new physics contributions to 
$\BtophiKs$.

Finally in the allowed region of $( \delta_{RL}^d )_{23}$, we find 
$\Delta M_s < 16.5 $ ps$^{-1}$, whereas the SM prediction is 
$\Delta M_s = 16.2 $ ps$^{-1}$.
If the pure $RL$-dominance 
scenario is realized in nature, $\bsbsbar$ mixing 
will be observed shortly at the Tevatron. 
On the other hand, both 
$\All$ and $\sin 2 \beta_s$ are too small to be observed.

If we lower the common squark mass to $x = 3$, the asymmetries  become 
somewhat smaller and 
$\Delta M_s$ falls in the range $[15.7, 16.7]$ ps$^{-1}$.
On the other hand, for a heavier squark $x=0.5$, the asymmetries can be
larger, but $\Delta M_s \simeq \Delta M_s^{\rm SM}$.

~~

In summary, $LR$ and $RL$ 
insertions of a size $\sim 10^{-3}$ -- $10^{-2}$ can induce large shifts in 
$\SphiK$ and $\CphiK$ 
(and $\Acp$ for $LR$ insertion or for $RL+RR$ insertion) 
but will not affect $\bsbsbar$ mixing,
whereas the $LL$ and $RR$ insertions with a size $> 0.2$ can induce 
large changes in $\bsbsbar$ mixing (both in its modulus and 
phase) but are still too small in generic models
to affect $\SphiK$, $\CphiK$ and $\Acp$ significantly. 
By understanding how each mass insertion individually affects the
various observables, it is rather straightforward to go from here to
the case in which more than one insertion is present. We will not
complete that study here. However there has been recent work that
benefits from multiple insertions that arise in well-motivated
ultraviolet physics scenarios~\cite{murayama}, and we refer the reader
there for further details.

\subsection{Gluino mass dependence of $\SphiK$}
\label{gluino}

In order to generalize our results and present a more complete picture
of SUSY contributions in $\BtophiKs$, it is necessary to study how
$\SphiK$ changes as a function of the SUSY mass scale. We have already
considered how changes in the squark mass scale, with the gluino scale
held constant, will affect the observables. Now we turn to the gluino
mass. 

\begin{figure}
\subfigure[$LR$ insertion]
{\includegraphics[width=6cm]{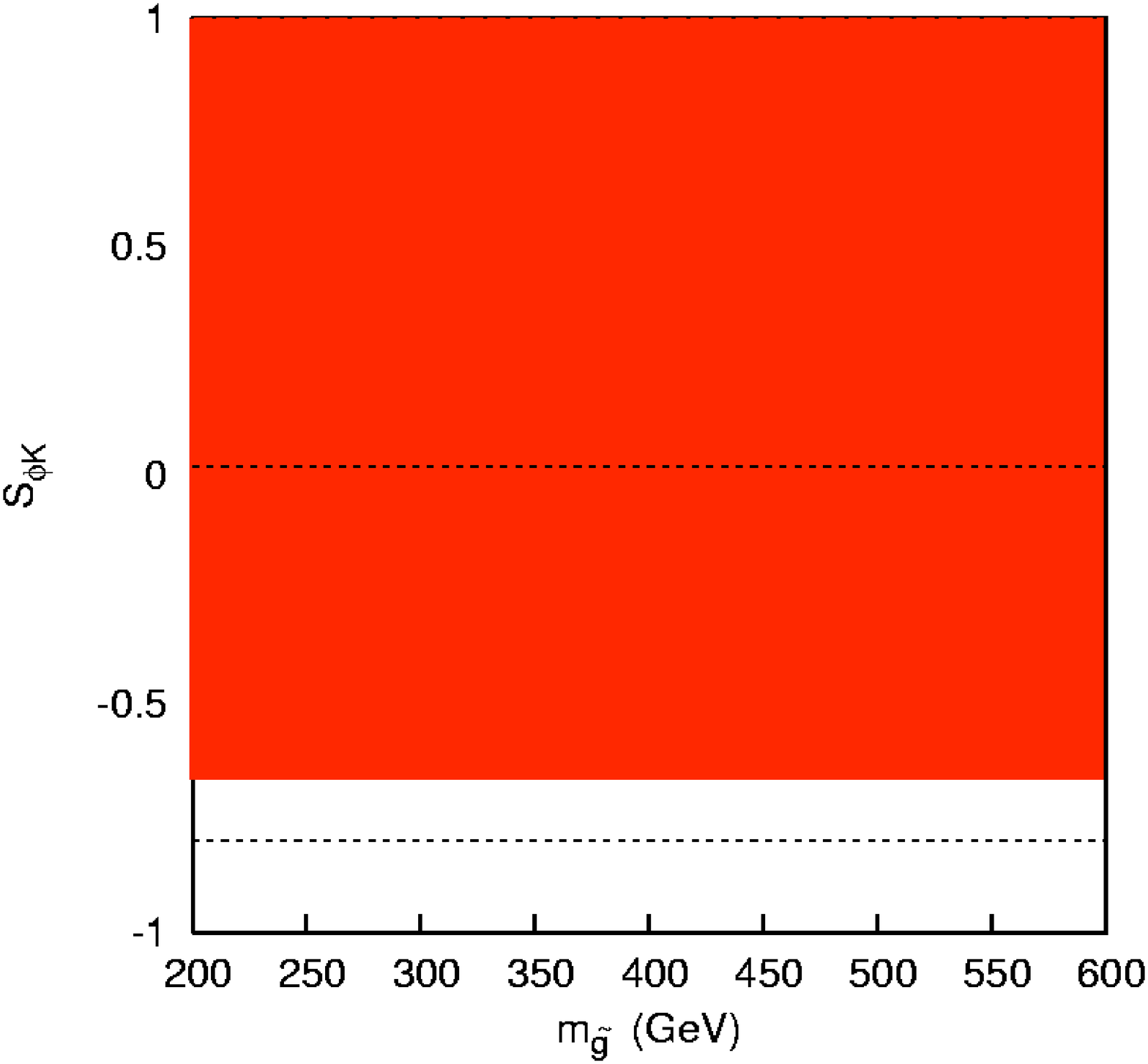}}
\subfigure[$RL$ insertion]
{\includegraphics[width=6cm]
{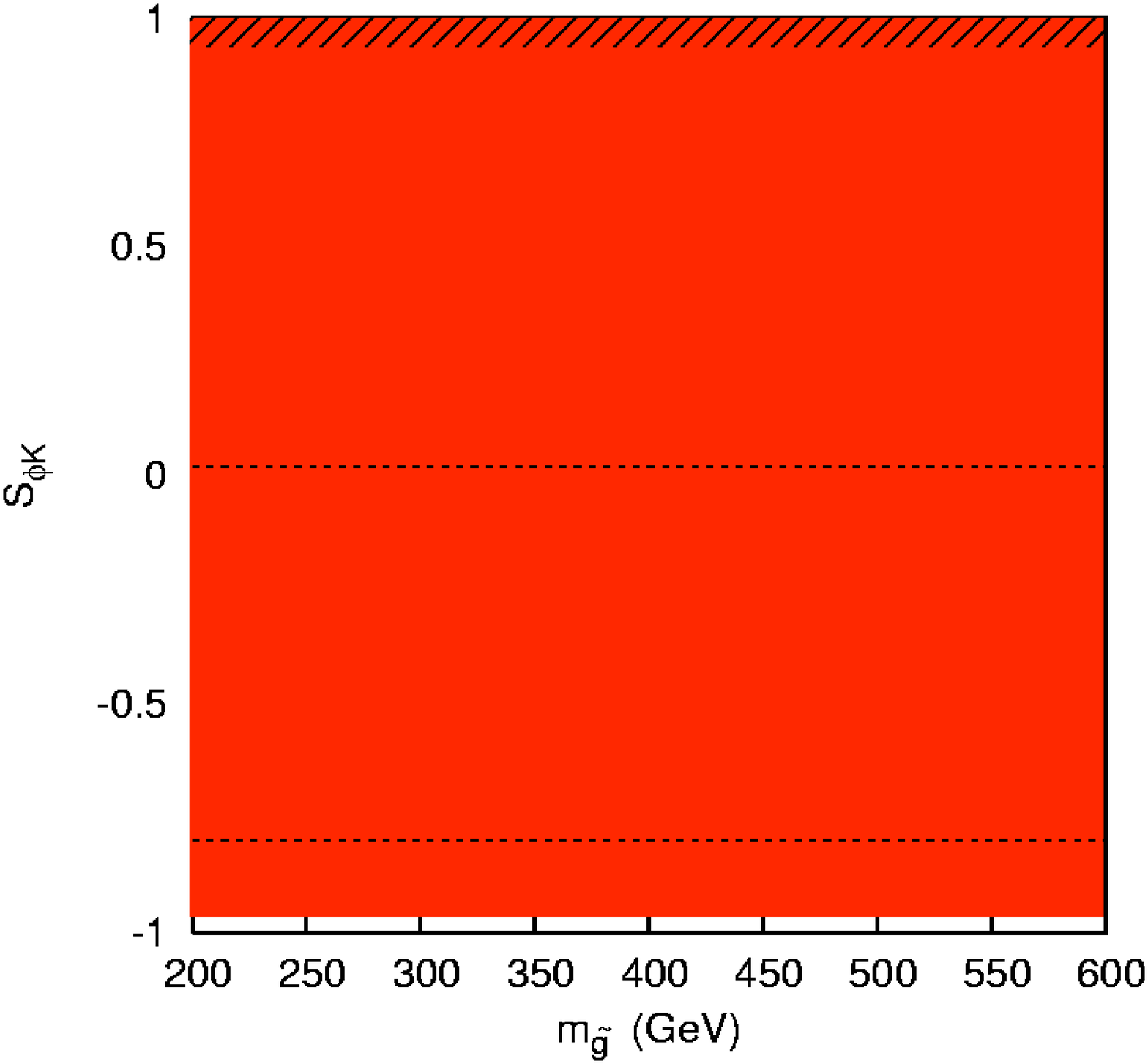}}
\subfigure[$LL$ insertion] 
{\includegraphics[width=6cm]%
{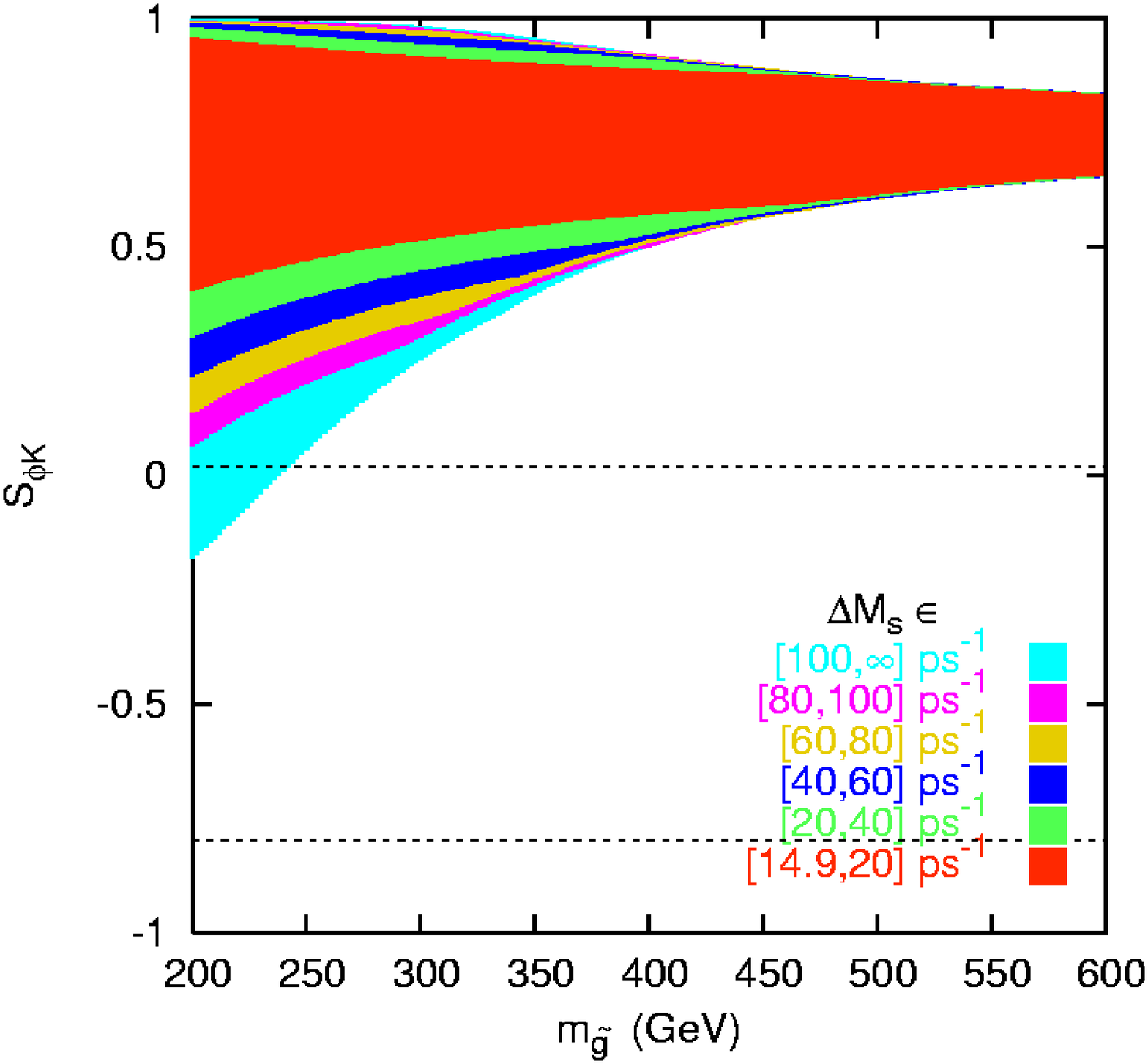}}
\subfigure[$RR$ insertion] 
{\includegraphics[width=6cm]
{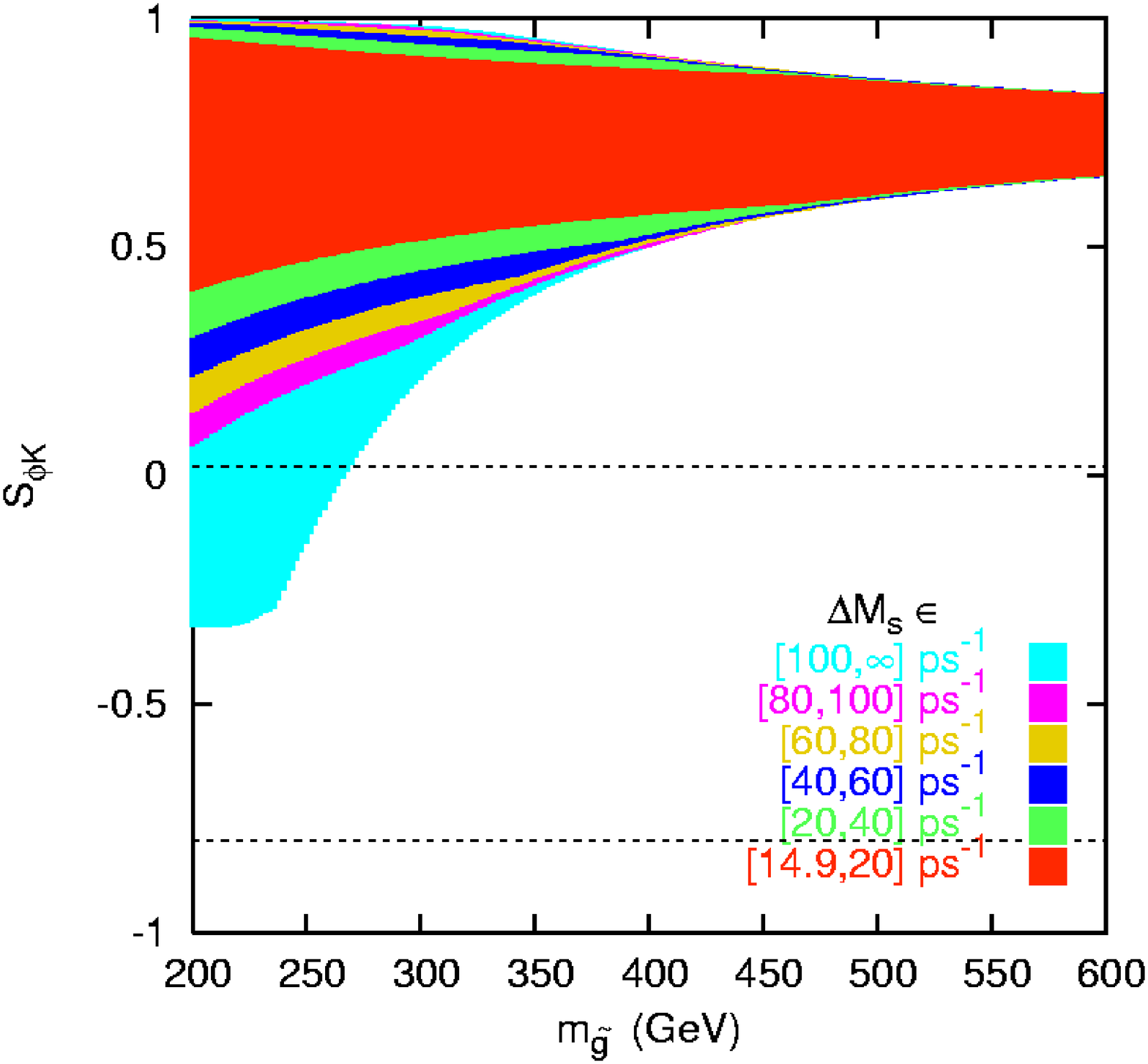}}
\caption{
$\SphiK$ as a function of $m_{\tilde g}$ for $x=1$.
The shading represents 
$\Delta M_s$ in the ranges 
[14.9,20], [20,40], [40,60], [60,80],  [80,100] and 
[100,$\infty$]$\,$ps$^{-1}$, respectively.
}
\label{gluinofig}
\end{figure}

In particular, we have calculated $\SphiK$ as a function of the gluino
mass, for $x=1$, in each of the four insertion scenarios: $LL$, $RR$,
$LR$ and $RL$. We have shown the results of our calculations in
Fig.~\ref{gluinofig}.  In each of the four figures we have shaded in
the allowed region of $\SphiK$ as a function of $m_{\gluino}$. The
different shades represent different predictions for $\Delta M_s$ as
described in the figure legends.
It is clear that $\SphiK$ can have a large negative value very easily
for the $LR$ or $RL$ insertion cases over 
a wide range of gluino and squark masses. 
On the other hand, in the $LL$ or $RR$ case $\SphiK$ can have a large 
negative value only for a very light gluino 
(close to the current bound),
and quickly reduces to the SM prediction as the gluino mass increases above 
$\sim 300$ GeV.  For heavier squarks $x=0.5$, the window for a negative 
$\SphiK$ increases slightly, but again the effect goes away quickly 
as $m_{\tilde{g}}$ becomes larger than $300\gev$. 
Furthermore, this always 
accompanies a very large $\Delta M_s$ (greater than  100~ps$^{-1}$).
Thus Run~II can rule out an $RR/LL$ explanation for $\SphiK$ if
gluinos are not found and/or $\Delta M_s<100\,\mbox{ps}^{-1}$ is
measured. 

These results speak directly to the analyses found in
Refs.~\cite{murayama,masiero02}. Both of these analyses followed the
same general lines as ours, but found that $RR$ insertions could
sizably alter $S_{\phi K}$. The reasoning behind these claims is now
clear. $RR$ insertions (and also $LL$ to a lesser extent) can indeed
produce $S_{\phi K}<0$, but only for very light
sparticles. Specifically, we find that $m_{\tilde g}$ must fall below
$300\gev$ in order to obtain $S_{\phi K}<0$ with an $RR$ or $LL$
insertion. Like us, the authors of Ref.~\cite{murayama} find that
large changes in $S_{\phi K}$ must be accompanied by large shifts in
$\Delta M_s$, making it unobservable at the
Tevatron. Ref.~\cite{masiero02} disagrees with this finding, though
the reasons probably have something to do with each groups'
estimations of the uncertainties in the BBNS factorization procedure. (See
Section~\ref{uncertainties} for more discussion of this point.)

One note of explanation is needed in order to understand the graphs,
and in particular why the $LR$ and $RL$ insertions seem to show no
sign of decoupling, while decoupling is readily apparent for the $RR$
and $LL$ insertions. The reasoning quite simple. For the $RL$ and $LR$
insertions, the leading constraint comes from $B\to X_s \gamma$, which
decouples at the same rate as the new physics contributions to 
$\SphiK$. Thus, as $m_{\gluino}$
increases, the strength of the $B\to X_s \gamma$ constraint decreases,
but the ratio of $(\delta_{LR,RL}^d)_{23}/m_{\gluino}$ remains
constant.
On the other hand, there are no strong constraints on the
$LL$ and $RR$ insertions, so even for light gluinos we have allowed
values of $(\delta_{LL,RR}^d)_{23}$ as large as 1; as the gluino mass
increases, we are not free to increase the insertion any further to
offset the mass suppression. Thus the effects on $\SphiK$ fall as 
$1/m_{\gluino}^{2}$.

\subsection{Hadronic uncertainties}
\label{uncertainties}

Finally, we should address the hadronic uncertainties within the BBNS 
approach to factorization. These uncertainties appear 
to be the origin of the qualitative differences that are found between
our results in the previous subsections and those in Ref.~\cite{masiero02}.

BBNS factorization fails two possibly important ways: it does not
properly account for the subleading power corrections coming
from annihilation diagrams, nor for those coming 
from hard scattering with a spectator quark. Each of these
contributions should be subleading, but each involves an integral that
blows up in the deep infrared ({\it i.e.}, $\int_0^1 dy/y$). Thus these
``subleading'' corrections are effectively infinite. In order to
control the size of the integral, it is necessary to cut it off. We
follow the original work of BBNS in parametrizing the integrals
as $\Delta\equiv (1+\rho e^{i\varphi})\log(m_B/\lambda_h)$ where
$\lambda_h=500\,$MeV. For the results we have presented in this paper,
we have taken $\Delta=0$; that is, we have assumed that the subleading
corrections are very small. 

\begin{figure}[htbp]
  \centering
  \subfigure[$\Delta_{\rm hard} = \Delta_{\rm annih} = 0$]
  {\includegraphics[width=6cm]{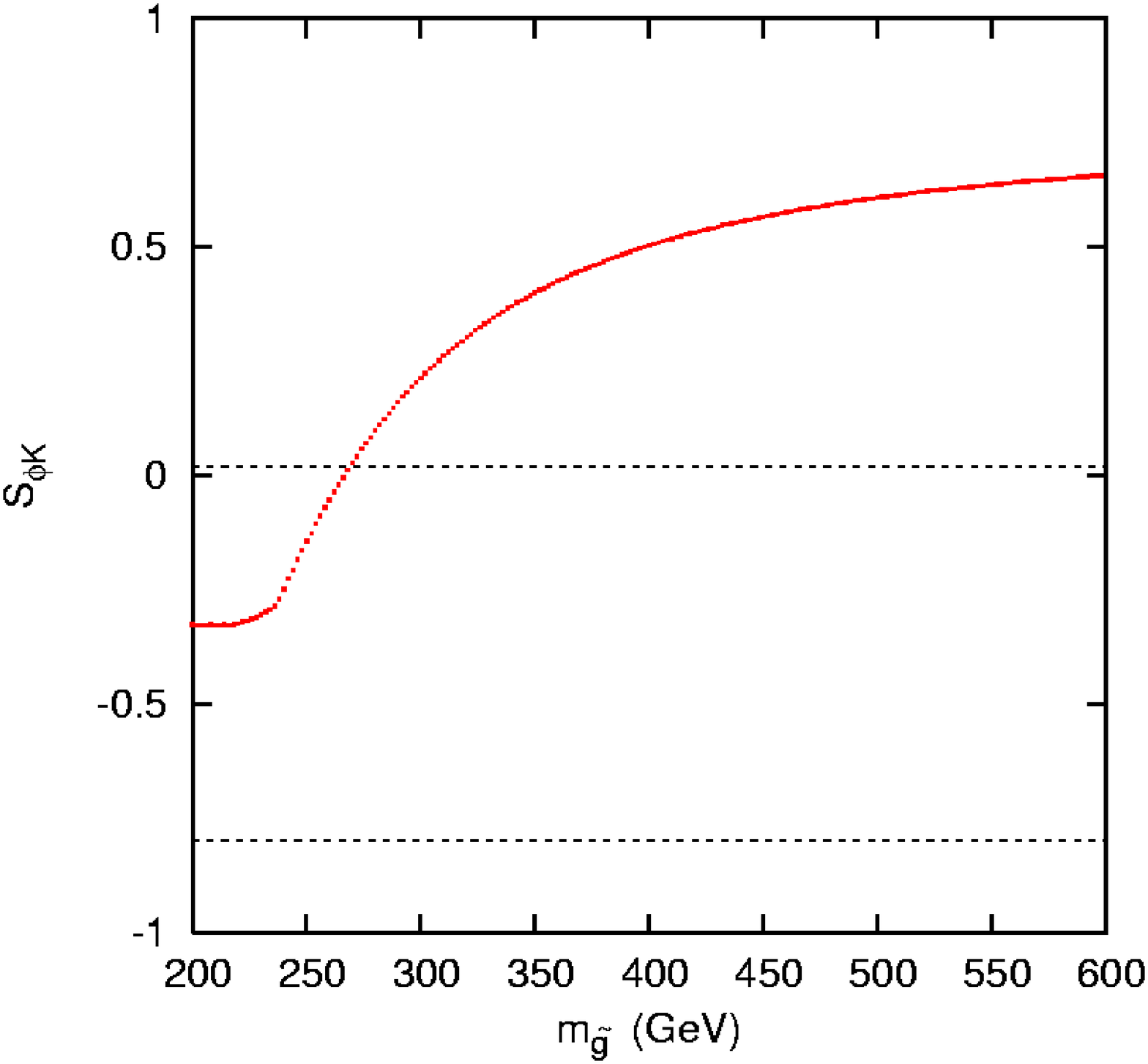}}
  \subfigure[$\rho_{\rm hard} \le 1, \Delta_{\rm annih} = 0$]
  {\includegraphics[width=6cm]{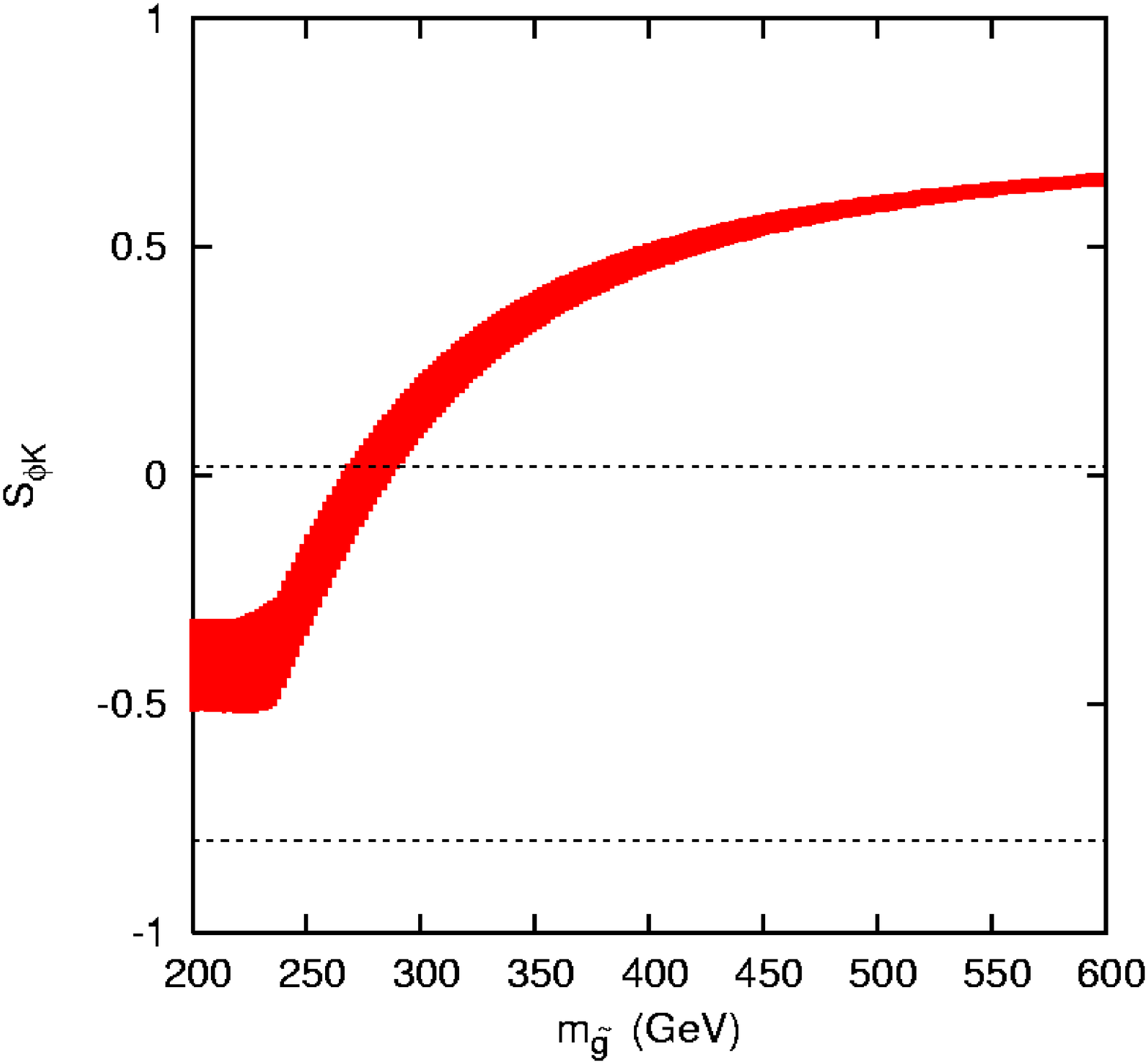}}\\
  \subfigure[$\Delta_{\rm hard} = 0, \rho_{\rm annih} \le 1$]
  {\includegraphics[width=6cm]{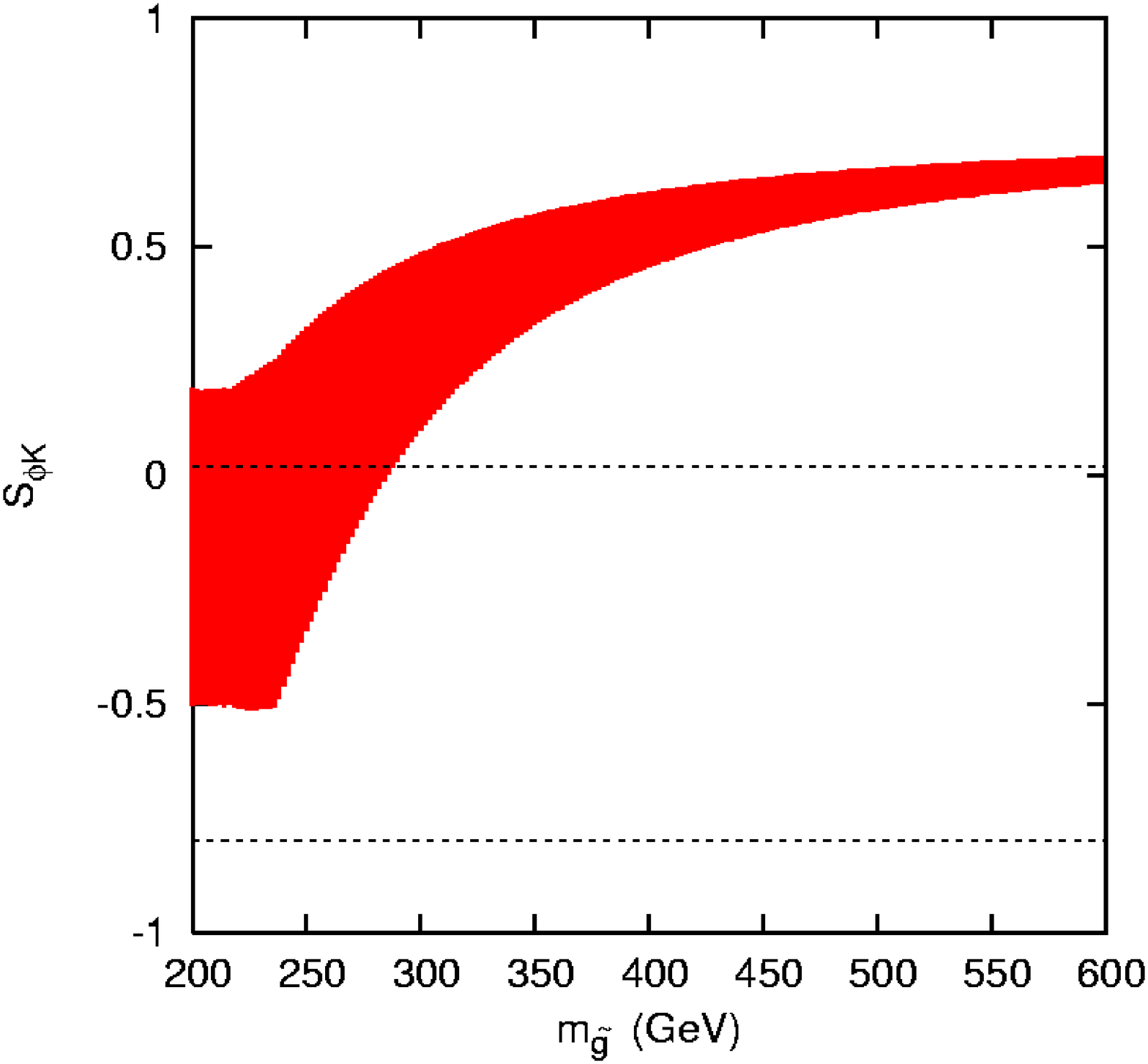}}
  \subfigure[$\rho_{\rm hard} \le 1, \rho_{\rm annih} \le 1$]
  {\includegraphics[width=6cm]{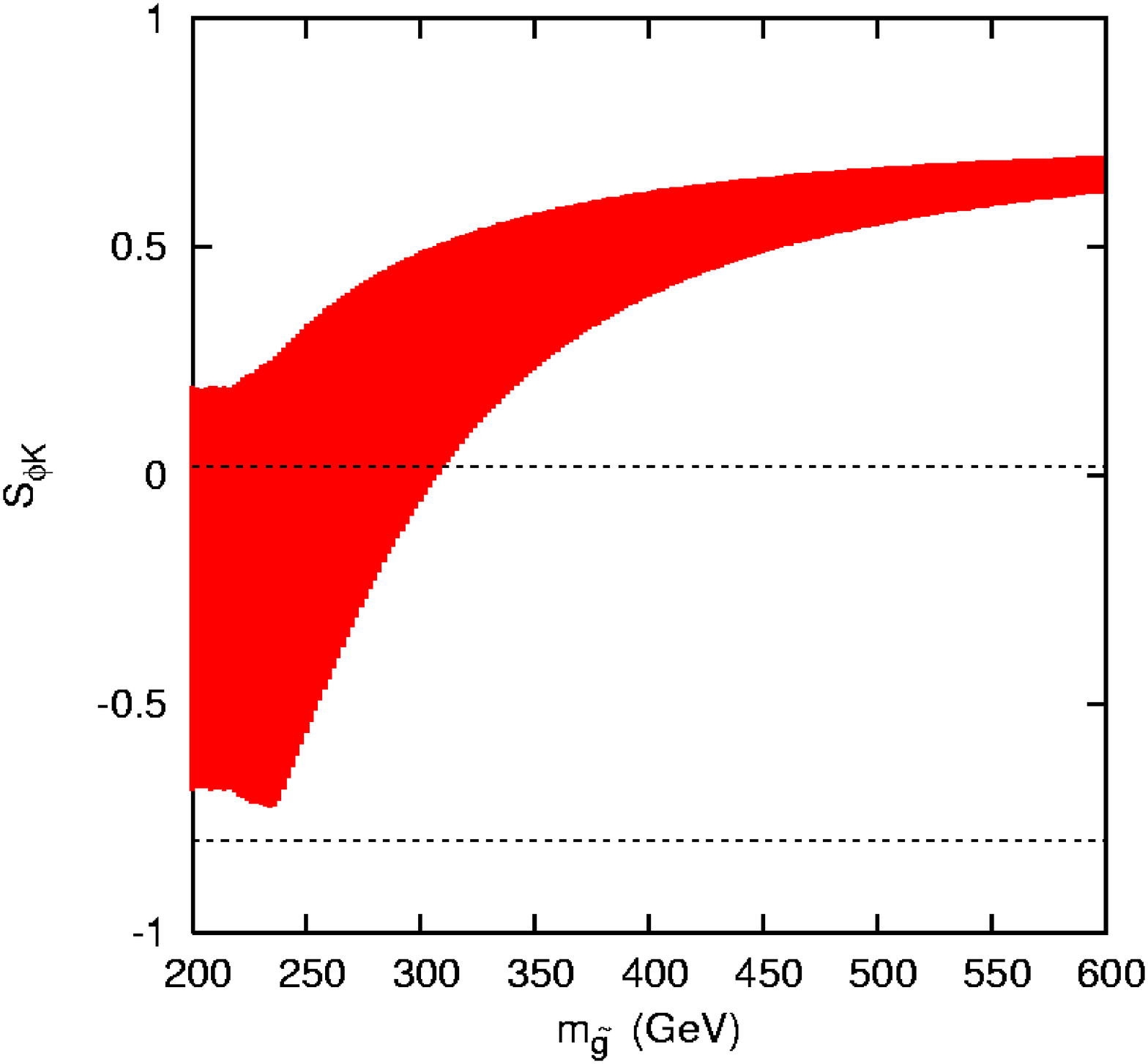}}
  \caption{Range of $\SphiK$ predicted in the BBNS approach for
several parametrizations of the subleading uncertainties, as a
function of $m_{\tilde{g}}$. In (a), all subleading uncertainties are
taken to be zero. In (b) and (c), hard scattering and annihilation
corrections are each included. In (d), both are.
Plots take $(\delta^d_{23})_{RR}$ to generate most negative $\SphiK$,
    for $m_{\tilde{g}} = 250$ GeV and $x = 1$.
}
\label{varyrhos}
\end{figure}

However, we now wish to see how our results depend on the size and
phase of $\Delta_{\rm hard}$ and $\Delta_{\rm annih}$.
Again following BBNS, we vary $\rho$ and
$\varphi$ in the ranges $0\leq\rho\leq 1$ and $0\leq\varphi\leq
2\pi$. (Each $\Delta_{\rm hard,annih}$ has its own $\rho,\varphi$
which are varied independently.) In Fig.~\ref{varyrhos} we 
consider the $RR$ 
insertion, since most recent works have emphasized this. For this case, 
the most negative value of $\SphiK$ occurs at 
$(\delta^d_{23})_{RR} = 1 - i $,
for fixed $m_{\tilde{g}} = 250$ GeV and $x = 1$.
With this value of $(\delta^d_{23})_{RR}$, we plot $\SphiK$
as a function of $m_{\tilde{g}}$. In Fig.~\ref{varyrhos}(a), who show
the result without the subleading corrections. In (b), we turn on the
hard scattering corrections only; in (c) we do likewise for the
annihilations corrections. Finally in (d) we allow both $\Delta_{\rm
hard}$ and $\Delta_{\rm annih}$ to vary over their entire ``allowed'' 
ranges. Note that the 
annihilation diagrams generate much 
larger hadronic uncertainties than do the
hard scattering amplitudes. Theoretical uncertainties decrease
quickly for heavier gluinos, and there is no possibility to get a negative
$\SphiK$ for $m_{\tilde{g}} = \tilde{m} = 400\,$GeV within the mass 
insertion approximation. 

In the work of Ref.~\cite{masiero02}, the authors did claim to find
negative $\SphiK$ without generating a large $\bsbsbar$ mass
difference. In order to avoid a large $\Delta M_s$, a smaller value
for the $RR$ insertion was used. Yet our results would indicate that a
small $RR$ insertion will not generate highly negative $\SphiK$. The
solutions to this paradox appears to lie in how the authors of
Ref.~\cite{masiero02} treated the hadronic uncertainties in the BBNS
prescription. In particular, they allowed $\rho_{\rm hard}$ and
$\rho_{\rm annih}$ to take on values as large as 8. It is not surprising
that this 8-fold increase in the uncertainties allows for a much wider
range for $\SphiK$. On the other hand, at these extremes, the
``subleading'' corrections are no longer subleading, but are every bit
as large as the leading terms in the BBNS scheme. For our analysis we
have felt it important to keep our theoretical uncertainties at the
level of those proposed by BBNS. It seems to us that doing otherwise
it tantamount to claiming that the BBNS prescription is not a 
good one. We have taken the opposite view in this paper.

\section{Higgs-mediated FCNCs}
\label{higgssec}

At large $\tan\beta$, FCNCs can also be mediated by exchange of neutral 
Higgs bosons, as found in 
$B\rightarrow X_s l^+ l^-$~\cite{huang}, 
$B_{s,d}\rightarrow \mu^+ \mu^-$~\cite{kolda} and 
$\tau \rightarrow 3 \mu$~\cite{Babu:2002et}, etc.
In this paper we consider the same mechanism leading to $b\rightarrow 
s s \bar{s}$. Unlike the gluino mediated amplitudes, which generate
flavor independent amplitudes for $b\rightarrow s q \bar{q}$,  
the Higgs exchange diagrams generate flavor-dependent 
amplitudes since the Higgs coupling is proportional to the Yukawa couplings.
Therefore $b\rightarrow s s \bar{s}$ can be modified by a significant
amount, whereas $b\rightarrow s u \bar{u} $ or $b\rightarrow s d \bar{d}$ 
remain essentially unchanged. Thus decays such as 
$\BtophiKs, \eta^{(}{}'{}^{)} K$ can be affected by  
Higgs exchange, whereas $B\rightarrow \pi K$, 
$\pi\pi$ modes are not.

In the presence of Higgs-mediated FCNCs (Fig.~\ref{figone}(c)), 
new operators 
$O_{\{R,L\},S}$ and $O_{\{R,L\},P}$ 
appear in the $\Delta B =1$ effective Hamiltonian:
\begin{equation}
H_{\rm eff} \rightarrow H_{\rm eff} 
- \frac{G_F}{\sqrt{2}} V_{ts}^* V_{tb}~ \left[ C_{R,S} O_{R,S} + 
C_{R,P} O_{R,P} + (R\rightarrow L) \right] 
\end{equation}
where 
\begin{equation}
O_{R,S} = ( \overline{s}_L b_R ) 
~( \overline{s} s ), ~~~
O_{R,P} = ( \overline{s}_L b_R ) 
~( \overline{s} \gamma_5 s )
\end{equation}
where
the first index $R$ or $L$ denotes the chirality of the initial $b$
quark in the $B$ meson, and color is conserved pairwise.
In the MSSM with minimal flavor violation, 
the Wilson coefficients $C_{L,S}$ and $C_{L,P}$ are suppressed 
relative to $C_{R,S}$ and $C_{R,P}$ by $m_s / m_b$, and will be ignored 
in the following.
The explicit form of $C_{R,S(P)}$ due to the neutral Higgs exchange is 
given by
\begin{equation}
C_{R,S} \simeq - C_{R,P} \simeq {\alpha \over \pi}\,
\left( { \tan^3 \beta \over 4 \sin^2 \theta_W } \right)\,
\left( { m_b m_\mu m_t \mu \over M_W^2 M_A^2 } \right) \,
{\sin 2\theta_{\tilde{t}} \over 2} \,
f ( m_{\tilde{t_1}}^2, m_{\tilde{t_2}}^2, \mu^2 ).
\end{equation}
where the loop function $f(a,b,c)$ can be found in Refs.~\cite{kolda}, 
for example.
Note that the $O_{R,S} - O_{R,P}$ operator is equivalent to $-O_6$ after 
Fierz transformation (but without summing over all flavors,  only with 
$\bar{s} s$).

In a model with minimal flavor violation (as defined in
Ref.~\cite{mfv}) the phases of the SUSY contributions would match
those of the SM and so no new source of CP violation would be
generated. However, if we extend minimal flavor violation to allow
arbitrary phases (a definition used by other authors), we can allow
our $C_{R,S}$ and $C_{R,P}$ operators to be complex and contribute to $S_{\phi
  K}$. Therefore we will assume that the new operators have complex
coefficients with $O(1)$ phases.

However the Wilson coefficients $C_{R,S}$ and $C_{R,P}$ also generate 
$B_s \rightarrow \mu^+ \mu^-$, which has been and is being searched for 
actively at the Tevatron.  
The upper limit  on this decay from CDF~\cite{cdf} during the 
Tevatron's Run I is 
\begin{equation}
B( B_s \rightarrow\mu^+ \mu^- ) < 2.6 \times 10^{-6}.
\label{eq:cdf}
\end{equation}
The relevant effective Hamiltonian is 
\begin{equation}
H_{\rm eff} ( B_s \rightarrow \mu^+ \mu^- ) = - {G_F \over \sqrt{2}}~V_{tb}
V_{ts}^* \left[ C_{R,S} ( \bar{s}_L^\alpha b_R^\alpha )~(\bar{\mu} \mu )
+   C_{R,P} ( \bar{s}_L^\alpha b_R^\alpha )~(\bar{\mu} \gamma_5 \mu )
+ (R \rightarrow L) \right].
\end{equation}
Note that the Wilson coefficients $ C_{R,S(P)}$ here are essentially the same
as those that appear in the effective Hamiltonian for 
$b\rightarrow s s \bar{s}$, up to a small difference between the muon and 
the strange quark masses (or their Yukawa couplings).
We find that the current upper limit (\ref{eq:cdf}) puts rather 
a strong constraint on $C_{R,S(P)}$:
\[
| C_{R,S} |, | C_{R,P} | \lesssim 0.005,
\]
since 
\beq
\mbox{B}(B_s \rightarrow \mu\mu) \propto
\left| C_{R,S} \right|^2 + \left| C_{R,P} \right|^2.
\label{eq:higgs}
\eeq
We then find that $\SphiK$ cannot be smaller than 0.71 for such 
small values of $C_{R,S(P)}$ within the naive factorization approach.
Thus it appears impossible to explain the large deviation in 
$\SphiK$ with Higgs-mediated $b\rightarrow s s \bar{s}$  alone.

Recently the authors of Ref.~\cite{cheng} have re-examined this issue
in models with non-minimal flavor violation. Among other things, they
allow a large $(\bar b_L s_R)(\bar ss)$ operator, though one would
normally expect such an operator to be suppressed by $m_s/m_b$ with
respect to the $(\bar b_R s_L)(\bar ss)$ operator.
The authors of Ref.~\cite{cheng} argue that one can explain
the data on $\SphiK$ without violating the CDF bound on $B_s\to\mu\mu$
if one allows the original operator and its parity conjugate to both be
large and roughly the same size.

Since our explanation for $S_{\phi K}$ involves non-minimal flavor violation, 
we should in principle allow similar violation in the Higgs
operators. In the general case
the branching fraction for $B_s\to\mu\mu$ depends on
\beq
\mbox{B}(B_s\to\mu\mu)\propto
\left|C_{R,S} - C_{L,S}\right|^2 + \left|C_{R,P} - C_{L,P}\right|^2
\label{eq:higgs2}
\eeq
rather than Eq.~(\ref{eq:higgs}) above. It would then 
appear that our limit applies to this difference
rather than the coefficients individually~\cite{isidori}. 
The authors of Ref.~\cite{cheng} make use of this to find a region of
parameter space consistent with negative $S_{\phi K}$.

However, this is not the complete story. Because the couplings and
masses in the MSSM Higgs sector are highly constrained ({\it e.g.},
$m_H\simeq m_A$), the scalar and pseudoscalar coefficients are tied to
one another. In particular, one finds that
\begin{eqnarray*}
C_{R,S} &\simeq& - C_{R,P} \\
C_{L,S} &\simeq& + C_{L,P}
\end{eqnarray*}
where the approximate equality becomes exact for $m_A\gg m_Z$. Thus
Eq.~(\ref{eq:higgs2}) can be rewritten as:
\beq
\mbox{B}(B_s\to\mu\mu)\propto
\left|C_{R,S} - C_{L,S}\right|^2 + \left|C_{R,S} + C_{L,S}\right|^2
=2\left(\left|C_{R,S}\right|^2 + \left|C_{L,S}\right|^2\right).
\eeq
Our argument from the CDF bound applies as before, and we
see no way to use Higgs mediation to generate $S_{\phi K}<0$, at least
in this way.

\section{Motivating $( \delta_{LR,RL}^d )_{23}$ of $O ( 10^{-2} )$}

In this section, we want to provide some motivation for 
$( \delta_{LR}^d )_{23} \sim  10^{-2}$ that we find phenomenologically 
is needed to generate $\SphiK$. 
Conversely, considerations such as those in this section show 
how one might use data to point toward classes of string--based theories. 
At this stage, we are not advocating any model of 
$( \delta_{LR}^d )_{23} \sim 10^{-2}$, but rather showing that 
$( \delta_{LR}^d )_{23}$ of the needed size can be 
theoretically plausible. 

Before going to a specific construction, we write down the generic
expression for the flavor parameters $(\delta)_{LR}$ in a supergravity
Lagrangian (which is the low energy effective field theory derived from
a string theory model). 
Under the assumption that either the Yukawa coupling does not depend on
the moduli fields, or it is a function of moduli times some constant
proportional to the Yukawa coupling itself, we can write the trilinear
couplings as~\cite{kobayashi} 
\begin{equation}
\label{trilinear}
\tilde{A}^f_{ij} \propto (A_L \cdot Y^f + Y^f \cdot A_R
)_{ij},  
\end{equation}
where $A_L$ and $A_R$ are $3 \times 3$ diagonal matrices, and
$f=u,d$. We also neglect a term which is proportional to a constant
times the Yukawa matrix which will not affect the prediction of
$(\delta)_{ij}$. The mass insertions $\delta_{LR}^f$ are defined as the ratio
between the off diagonal elements and the diagonal ones of the squark
mass matrices in the superCKM basis. The superCKM basis is defined by
rotating the quark superfields by $V_L^f$ and $V_R^f$, 
where $V_L^f Y^f V_R^{f \dagger} = Y_{\mathtt{diag}}$. 
Then we obtain the expression for the relevant LR mass insertions 
in the general MSSM as   
\begin{eqnarray}
\label{deltaLR}
(\delta_{LR})_{ij, i \neq j} &\propto& m^f_{j} 
(A_2^L - A_1^L) (V_L)_{i 2}(V_L^*)_{j 2} +  m^f_{j} (A_3^L -
A_1^L) (V_L)_{i 3}(V_L^*)_{j 3} \nonumber \\
&+& m^f_{i} (A_2^R -
A_1^R) (V_R)_{i 2}(V_R^*)_{j 2} +  m^f_{i} (A_3^R -
A_1^R) (V_R)_{i 3}(V_R^*)_{j 3}, \nonumber \\
\end{eqnarray}
where $A_{i}^{L,R}$ are model dependent parameters of order
the gravitino mass,
$m_{3/2}$. The most important feature of this expression is that the
sizes of various terms are proportional to SM fermion masses 
$m^f_{i,j}$. Some of the immediate conclusions one can draw
from this observation are
\begin{itemize} 
\item $(\delta_{LR})_{12} \ll
( \delta_{LR} )_{13}, ( \delta_{LR} )_{23} $.
\item $\delta^u/\delta^d \sim m^u/m^d$. In
particular, $(\delta^u_{LR})_{23}/(\delta^d_{LR})_{23} \sim
m_t/m_b$.    
\end{itemize}
Note that these are relations obtained at some high energy
scale. However, the dominant RGE runnings of the trilinears are
diagonal in the superCKM basis. Therefore, those relation also hold
approximately at low energy scale. 

We now turn to a possible string theory motivated scenario
where those features can arise. For definiteness, we focus on the
models constructed from a intersecting D5-brane setup.  
Suppose we have two stacks of D5-branes intersecting at 
$90^{\circ}$. There are six compact dimensions which are grouped
into three complex pairs labeled by $i=1,2,3$. D-branes are objects on
which open strings can end. Open string sectors in this type of
scenario are classified by (i) which D-brane it ends on, and (ii) which
complex dimension it moves in. We will have the following open string
states~\cite{Ibanez:1998rf} 
\begin{itemize}
\item $C^{5_i}_{j}$: open strings which end on the $5_i$ brane and move along
the $j$th complex compact dimension.
\item $C^{5_i 5_j}$: open strings which start from the $5_i$ brane and end
on the $5_j$ brane.
\end{itemize}
The string theory selection rules permit two type of large couplings
(provided they are gauge invariant): 
$C^{5_i}_{i}C^{5_i}_{j}C^{5_i}_{k}$ and $C^{5_i}_{j}C^{5_i 5_k}C^{5_i
5_k}$, where $i \neq j \neq k$.

The next question is how to embed the MSSM matter content into the
this scenario. Since we need some large mixing between the last two
generations, one obvious choice is to embed them identically, {\it i.e.},
assign them to the same open string sector. However, this type of
construction has a problem because it also predicts
$A_2^L=A_3^L$ which is not desirable. Therefore, we have to embed the
last two generations into different open string sectors and yet give
them a large mixing. 

Suppose we have the following embedding of the MSSM matter content into 
open string sectors, adopting the notation that $F_i$ and $\overline{F}_i$ 
are the $i$th generation left and right handed quarks, respectively. 
\begin{itemize} 
\item The Higgs field: $h=\frac{1}{\sqrt{2}}(C^{5_1}_1 + C^{5_1}_2)$.
\item The quarks: 
\begin{eqnarray}
F_3 = C^{5_1}_2, &\quad\quad &\overline{F}_3 = C^{5_1}_3 \nonumber \\
F_2 = C^{5_1}_1, &&\overline{F}_2 = C^{5_1}_3 \nonumber \\
F_1 = C^{5_1 5_2}, && \overline{F}_1 = C^{5_1 5_2}.
\end{eqnarray}
\end{itemize}
From the string theory selection rules, we can then determine the Yukawa
coupling matrix to be 
\begin{equation}
Y \propto \left(\begin{array}{ccc} 
\cdot & \cdot & \cdot \\
\cdot &  1 & 1 \\
\cdot & 1 & 1 
\end{array} \right). 
\end{equation}

We can also work out the soft supersymmetry breaking terms
\cite{Brignole:1997dp}. In particular, the trilinear couplings are 
\begin{eqnarray}
\frac{\tilde{A}}{-\sqrt{3} m_{3/2}} &=& -\frac{1}{\sqrt{2}} (X_0+X_3)
\cdot Y + \left(\begin{array}{ccc} 
\frac{1}{2}X_0 + X_1 + \frac{1}{2} X_3 & & \\
 & X_1 + X_3 & \\
 & & X_0 + X_1
\end{array} \right) \cdot Y \nonumber \\
&+& Y \cdot \left(\begin{array}{ccc} 
-\frac{1}{2}X_0 + X_2 - \frac{1}{2} X_3 & & \\
 & 1 & \\
 & & 1
\end{array} \right),
\end{eqnarray} 
where $X_i$'s are {\it complex} parameters satisfying $\sum_{i=0}^3
|X_i|^2 =1$. Notice now we have written the trilinear couplings in the 
form of Eq.~(\ref{trilinear}). 
We have singled out the part which is the Yukawa matrix multiplied by a
diagonal matrix proportional to the identity since this term is
diagonal in the super-CKM basis and does not contribute to flavor
violating processes. We also have the following expressions for soft
masses 
\begin{eqnarray}
m^2_{F_3} &=& m^2_{3/2} (1-3 |X_3|^2) \nonumber \\
m^2_{F_2} &=& m^2_{3/2} (1-3 |X_0|^2) \nonumber \\
m^2_{\bar{F}_3} &=& m^2_{3/2} (1-3 |X_2|^2) \nonumber \\
m^2_{\bar{F}_2} &=& m^2_{3/2} (1-3 |X_2|^2).
\end{eqnarray}

Using Eq.~(\ref{deltaLR}), our ansatz for the Yukawa textures and the 
definition of the mass insertion  parameter, we then have the following 
estimate for the size of $(\delta^d_{LR})_{23}$:
\begin{equation}
(\delta^d_{LR})_{23} \sim  \frac{\sqrt{3} m_{b}}{m_{3/2}} \frac{(X_0 -
X_3)}{\sqrt{(1-3 |X_0|^2)(1-3|X_2|^2)}}. 
\end{equation} 
Clearly, this can be as large as of order $\sim 0.01$. It can also have a
quite nontrivial phase structure since $X_0$ and $X_3$ are in general
complex parameters. 

Notice however in this particular implementation, we will generically
have $(\delta^d_{RR})_{23} \sim 0$ since
$m^2_{\bar{F}_3}=m^2_{\bar{F}_2}$. We can also choose 
$(\delta^d_{LL})_{23}$ to be close to zero by setting  $|X_0| \approx |X_3|$. 

As remarked above, this model suggests 
$( \delta_{LR}^u )_{23} \sim {m_t \over m_b }~(\delta^d_{LR})_{23} $, 
so $(\delta^d_{LR})_{23} \sim 0.008$ suggests 
$( \delta_{LR}^u )_{23} \sim 0.3$. Such a large value is not in conflict 
with any data on $B\rightarrow X_s \gamma$ or $B\rightarrow X_s l^+ l^-$, 
which give essentially no constraint on 
$( \delta_{LR}^u )_{23}$~\cite{Buras:1997ij}.
Also $( \delta_{LR}^u )_{23} \sim 0.3$ can generate a flavor changing top 
decay $t \rightarrow c g$ only up to $(6.0, 8.5, 9.2) \times 10^{-6}$ for 
$x = (0.5, 1, 3)$ and $m_{\tilde{g}} = 400$ GeV, which are still below the 
threshold for observability ($\sim 10^{-5}$) at future colliders such as LHC 
and NLC.  Another prediction of this model is 
$( \delta_{LR}^d )_{12} \sim 5 \times 10^{-5}$, which could  dominate
$\epsilon^{'} / \epsilon_K$ through the chromomagnetic $s\rightarrow d g$ 
transition.

Let us also comment on the size of $(\delta^d_{RL})_{23}$ in our current
embedding. From our formula for the trilinears, we see that
$(\delta^d_{RL})_{23} = (\delta^{d*}_{LR})_{32}$ is determined by
$(A^R_{2,3}-A^R_1) V_R V_R^*$. In our current embedding, we have
$A^R_2=A^R_3$ since the last two generations of the right-handed (s)quarks
have the same embedding. We can then use the unitarity of $V_R$ to simplify
the expression of $(\delta^d_{RL})_{23}$ to $(\delta^d_{RL})_{23}\propto
\frac{m_b}{m_{3/2}} (V_R)_{31} (V_R)^*_{21}$ which is suppressed by higher
powers of $\lambda$. However, we can realize a larger 
$(\delta^d_{RL})_{23}$ by simply switching 
the open string embedding of the last two generations of left-handed and 
right-handed (s)quarks, $F_{2,3}$ and $\overline{F}_{2,3}$. 
Then we can have a sizable  $(\delta^d_{RL})_{23} \sim 0.01$ 
while $(\delta^d_{LR})_{23}$ is suppressed by the higher power of $\lambda$.

Before closing this section, let us comment on the mass mixings/insertions
that would be generated by the RG running from the high energy 
scale to electroweak scale. If we assume a universal boundary condition for 
scalar masses at the reduced Planck scale $M_* \simeq 2.4 \times 10^{18}$ GeV, 
and all $( \delta_{AB} )_{ij}$ are zero at the high energy scale, 
nonnegligible values are induced by renormalization group (RG) running.  
The approximate solutions to the RG equation for the left squark mass squared
becomes
\[
( m_{LL}^2 )_{ij} ( \mu = M_{weak} ) \simeq
-{1 \over 8 \pi^2}\,Y_t^2 \left( V_{\rm CKM} \right)_{3i}  
\left( V_{\rm CKM}^* \right)_{3j}\,\left( 3 m_0^2 + a_0^2 \right)\,
\log ( { M_{*} \over M_{\rm weak} } ),
\]
and one can estimate $( \delta_{LL}^d )_{23} \simeq 9 \times 10^{-3}$. 
Note that this parameter is real however, since $V_{23}$ and $V_{33}$ are
real in the Wolfenstein parameterization. 
Also other mass insertions are further suppressed by additional factors of 
$Y_d$ and $Y_d^2$, so that $| ( \delta_{LR,RL}^d )_{23} | \sim 10^{-5}$ and 
$| ( \delta_{RR}^d )_{23} | \sim 10^{-7}$. These are too small to explain 
a large shift of $\SphiK$ relative to $\SpsiK$. 

In some SUSY GUT scenarios, the large mixing in the neutrino
sector induces a large mixing in the right squark sector through 
the analogous RG running. The resulting mixing is about~\cite{moroi} 
\[
( \delta_{RR}^d )_{23} \simeq 2 \times 10^{-2}\,\left( { M_{\nu_R} \over 
10^{14}~{\rm GeV} } \right).  
\]
Therefore the $RR$ mixing is generically larger than $LL$ mixing in 
SUSY GUTs where the see-saw mechanism is generating the nonvanishing 
neutrino masses. Still it cannot affect $\SphiK$ by a large amount
as discussed in the previous section, although 
the size of $( \delta_{RR}^d )_{23}$ could be $\sim O(0.5)$ with a new 
complex phase so that $\bsbsbar$ mixing could be significantly shifted.

On the other hand, if the SUSY flavor problem is resolved by the alignment 
mechanism using 
some spontaneously broken flavor symmetries, or decoupling (the effective 
SUSY models),  the resulting $LL$ or $RR$ mixings in the $23$ sector could
be easily order of $\sim \lambda^2$ ($\lambda \equiv \sin\theta_c \approx 
0.22$) with either $LL >> RR$,  $LL << RR$ or $LL \approx RR$~\cite{align,
decoupling}, whereas the mixing in the 13 sector is further suppressed by 
additional power(s) of $\lambda$('s) in order to avoid large 
contributions to $\bbbar$ mixing.  These parameters will carry 
CP violating phases in general and can contribute 
significantly to $\Delta M_s$ and $\sin 2 \beta_s$.

Further, in the presence of an $LL$ or $RR$ insertion, there can be 
an induced $LR$ or $RL$ insertion when $\mu \tan\beta$ is large with
respect to the scalar masses, 
due to the double mass insertions~\cite{ko01,bjkp}:
\[
( \delta_{LR}^d )_{23}^{\rm ind} =  ( \delta_{LL}^d )_{23} \times 
{ m_b ( A_b - \mu \tan\beta ) \over \tilde{m}^2 }.
\]
In case $ ( \delta_{LL,RR}^d )_{23} \sim 10^{-2}$, 
one can achieve $( \delta_{LR,RL}^d )_{23}^{\rm ind} \sim 10^{-2}$,
if $\mu \tan\beta \sim 30\,$TeV. This could be natural if $\tan\beta$ is 
large $\sim 40$. For larger $LL,RR$ mixing, even smaller $\mu \tan\beta$ 
would  suffice to induce the needed $LR,RL$ mass insertions of a size 
$10^{-2} - 10^{-3}$.
Since  the $\delta_{LL,RR}$'s in SUSY flavor models are generically 
complex, the induced $( \delta_{LR}^d )_{23}^{\rm ind}$ could carry a new CP 
violating phase~\cite{bjkp}, which can again explain a large negative 
$\SphiK$.  
Also, the $RR$ insertion can have a phase inherited from the neutrino mixing 
matrix in a SUSY GUT model, which would be transferred to the induced $RL$ 
insertion~\cite{moroi}. Therefore the induced $RL$ insertion can explain 
the observed $\SphiK$ in SUSY GUT models.   On the other hand, 
the phase of $( \delta_{LL}^d )_{23}$ for the 
case of minimal SUGRA is given by 
the CKM matrix elements, and it is real for the 2-3 mixing. Therefore the 
induced $( \delta_{LR}^d )_{23}^{\rm ind}$ in SUGRA-like models
is incapable of explaining the negative $\SphiK$.

\section{Conclusions}

Recent data has given us hope that there may be new physics lurking in
rare $B$-decays, particularly $\BtophiKs$.
In this paper, we considered several potentially important SUSY 
contributions  to this process in order to 
see if a significant deviation in its time-dependent CP asymmetry 
$\SphiK$ could arise from 
SUSY effects. In particular, we considered the SUSY gluino
contributions in models with non-minimal flavor violation, and the
Higgs-mediated contributions in models with minimal flavor
violation. While the latter has the advantage of only affected the
$b\to ss\bar s$ transition, current bound on $B_s\to\mu^+ \mu^-$ at
Fermilab constrain the relevant operators to be too small.
Models based on the $LL$ and $RR$ insertions in gluino interaction vertices
also give contributions too small to alter $\SphiK$ very much,
unless gluinos are very light (close to the experimental bounds).
However, in an $LL$ or $RR$ scenario one expects
an impending observation of gluinos and squarks, and absence of
$\bsbsbar$ oscillations at the Tevatron, since $\Delta M_s$ increases
dramatically in the $LL$ ($RR$) case when $\SphiK<0$.

On the other hand, gluino-mediated $LR$ or $RL$
contributions can generate sizable effects 
in $\SphiK$ as long as $| ( \delta_{LR,RL}^d )_{23} | 
\sim 10^{-3}$ -- $10^{-2}$.
As a by-product, we found that nonleptonic $B$ decays such as $\BtophiKs$
are beginning to constrain $| ( \delta_{LR,RL}^d )_{23} | $ 
as strongly as $B\rightarrow X_s \gamma$.  
We also studied various correlations among $\SphiK$, 
$\CphiK$, the direct CP asymmetry in $B\rightarrow X_s \gamma$, 
$\Delta M_{s}$, $\All$ and $\sin 2\beta_s$. 
Using the $LR$ or $RL$ insertions, 
it is easy to obtain a negative $\SphiK$ without conflict with any other
observables. Furthermore, there are definite 
correlations among $\SphiK$, $\CphiK$ and 
$\Acp$, and our explanation for the negative 
$\SphiK$ can be easily tested by measuring these other correlated 
observables. In particular $\CphiK$ can be positive only for the $RL$ 
insertion. For this case, the direct CP asymmetry in 
$B\rightarrow X_s \gamma$ vanishes (assuming no $RR$ insertion is present), 
unlike the case of the $LR$ or $LL$ insertions.  In a scenario with
both $RL$ and  
$RR$ insertions, the resulting direct CP asymmetry in 
$B\rightarrow X_s \gamma$ could be as large as in the $LR$ mixing case, and
there could be a large, complex contributions to $\bsbsbar$
mixing, leading to significant effects in
$\Delta M_s$ and $\sin2 \beta_s$. The effects of the (pure)  
$LR$ and $RL$ insertions on $\bsbsbar$ are rather small
and it would be difficult to distinguish our model from the SM by the 
$\bsbsbar$ mixing, considering 
various theoretical and experimental uncertainties. 

Thus, we have found that three classes of supersymmetric models 
could not naturally explain
a negative time dependent CP asymmetry $\SphiK$ in 
$\BtophiKs$ if the initial hints of such an asymmetry 
should persist when data improves: $LL$, $RR$ insertions, and Higgs-dominated 
decays. Two other classes, $LR$ and $RL$ insertions, can explain the data
and can be distinguished to some extent    
with more and better data. Insertions of the size and kind needed can
be generated naturally in simple
string-motivated models and SUSY flavor models.

~

\begin{acknowledgments}
We are grateful to Seungwon Baek for useful communications on 
clarifying the signs of the gluino induced QCD penguins. 
We appreciate helpful comments
fron H.~Davoudiasl, B.~Nelson, K.~Tobe and J.~Wells.  
We also thank Alex Kagan for discussions and useful communications about 
references and pointing out the sign error in the correlation between 
$\SphiK$ and $A_{\rm CP}^{b\rightarrow s\gamma}$.
PK and JP are grateful to the Michigan Center for Theoretical Physics 
for the hospitality extended during their stays.
This work is supported in part by the BK21 Haeksim Program, 
KRF grant KRF-2002-070-C00022 and KOSEF through CHEP at 
Kyungpook National University, by a KOSEF Sundo Grant (PK and JP), 
by the National Science Foundation under grant PHY00-98791 (CK), 
by the Department of Energy (GK and HW) under
grant DE-FG02-95ER40896, and by the Wisconsin Alumni
Research Foundation (LW). 
\end{acknowledgments}

%\vspace{0.5in}

\end{document}